\documentclass[12pt]{article}

\makeatletter
\newcommand{\lambdabar}{{\mathchoice
  {\smash@bar\textfont\displaystyle{0.25}{1.2}\lambda}
  {\smash@bar\textfont\textstyle{0.25}{1.2}\lambda}
  {\smash@bar\scriptfont\scriptstyle{0.25}{1.2}\lambda}
  {\smash@bar\scriptscriptfont\scriptscriptstyle{0.25}{1.2}\lambda}
}}
\newcommand{\smash@bar}[4]{%
  \smash{\rlap{\raisebox{-#3\fontdimen5#10}{$\m@th#2\mkern#4mu\mathchar'26$}}}%
}
\makeatother

\RequirePackage[a4paper,top=30.6mm,bottom=38.6mm,left=26mm,right=26mm,footskip=1.3cm]{geometry}
\usepackage{setspace}
\usepackage{amsmath}
\usepackage{amsfonts} 
\usepackage{amssymb}
\usepackage{graphicx}
\usepackage{hyperref}
\usepackage{tensor}
\usepackage{siunitx}
\usepackage{float}
\usepackage{bm}
\usepackage{slashed}
\usepackage{booktabs}
\usepackage{cancel}
\usepackage{multirow}
\usepackage{comment}
\usepackage{bbold}
\usepackage{caption}
\usepackage{amsmath}
\usepackage{enumerate}
\usepackage{cite}
\usepackage{tensor}
\usepackage{slashed}
\usepackage{inputenc}
\usepackage{rotating}
\usepackage{bigfoot}
\usepackage{cancel}
\usepackage{mathtools}
\usepackage{xcolor}
\usepackage{color, colortbl}

\usepackage[textsize=footnotesize]{todonotes}

\numberwithin{equation}{section}

\def \< {\left<}
\def \> {\right>}

\newcommand{\be}{\begin{equation}} \newcommand{\ee}{\end{equation}}
\newcommand{\bea}{\begin{eqnarray}}  \newcommand{\eea}{\end{eqnarray}}
\newcommand{\nn}{\nonumber}

\usepackage{amsfonts, amsthm}
\usepackage[english]{babel}
\usepackage{slashed}
\usepackage{mathrsfs}
\usepackage{amssymb}
\usepackage{color}

\def\Im{{\mathrm{Im}}}
\def\Re{{\mathrm{Re}}}

\begin{document}
		
	\begin{center}        
		\Large On Type II$_0$ Loci in Moduli Space 
		\end{center}
	
	\vspace{0.4cm}
	\begin{center}        
		{\large Jarod Hattab \;and\;  Eran Palti}
	\end{center}
	
	\vspace{0.15cm}
	\begin{center}  
		\emph{Department of Physics, Ben-Gurion University of the Negev,}\\
		\emph{Be'er-Sheva 84105, Israel}\\[.3cm]

	\end{center}
	
	\vspace{1cm}
	
	
	\begin{abstract}
	\noindent  
    We study type II$_0$ loci in the moduli space of type IIB string theory compactified on Calabi-Yau manifolds. We show that around these infinite distance singular loci the leading order behaviour of the gauge kinetic matrix, and of the prepotential, can always be written in the form of a threshold correction from integrating out a BPS state, but one with an effectively complex charge. In order to understand the physical meaning of this, we carefully identify the splitting in the effective supergravity between the graviphoton direction and matter vector multiplets. Within a specific two-parameter example of a Calabi-Yau, we use this to identify a strongly-coupled matter sector involving both light electric and light magnetic states. We propose that the leading gauge kinetic matrix arises as a threshold correction from integrating out this non-perturbative sector, and that the sector has an effective weakly-coupled infrared description in terms of the complex-charged state. The region in moduli space has a Heterotic string dual microscopic description. The light magnetic state in this description corresponds to a Kaluza-Klein monopole, which becomes lighter than the fundamental Heterotic string, leading to the non-perturbative sector. Assuming this picture is correct, it implies the existence of infrared emergent infinite distance loci in moduli spaces of quantum gravity.
    \end{abstract}
	
	\thispagestyle{empty}
	\clearpage
	
	\tableofcontents

\section{Introduction}
\label{sec:int}

Compactifications of type IIB string theory on Calabi-Yau manifolds lead to effective four-dimensional ${\cal N}=2$ supergravity theories. The vector multiplet sector of those theories can be solved for exactly (see the pioneering work \cite{Candelas:1990rm}). That is, the moduli space can be mapped exactly, and the gauge kinetic matrix can be calculated to all orders. More precisely, the infinite infrared limit of the sector is known exactly. In this limit, all charged finite-energy BPS states have been integrated out. Any states that become massless manifest as singularities in the moduli space, while massive states manifest as threshold corrections to the effective action, for example, in the gauge kinetic matrix. 

While the infrared physics anywhere in the moduli space can be described by the four-dimensional supergravity, the finite energy physics can be very exotic, depending on the region of moduli space of interest. Each controlled region in moduli space is an expansion around a singular locus. All the possible singular loci, and some properties of the expansions around them, were classified in \cite{Grimm:2018ohb,Grimm:2018cpv}.\footnote{The topic of such loci has been thoroughly studied, we refer to \cite{Palti:2019pca,vanBeest:2021lhn,Harlow:2022ich} for reviews and \cite{Cota:2022maf,Friedrich:2025gvs,Rudelius:2023odg,Gendler:2022ztv,Kaufmann:2024gqo,Marchesano:2022avb,Castellano:2022bvr,vandeHeisteeg:2022btw,Cribiori:2022nke,Marchesano:2022axe,Blumenhagen:2023tev,Castellano:2023qhp,Cribiori:2023ffn,Blumenhagen:2023yws,Blumenhagen:2023xmk,Seo:2023xsb,Calderon-Infante:2023ler,Calderon-Infante:2023uhz,Castellano:2023aum,Castellano:2023jjt,Castellano:2023stg,Cota:2023uir,Casas:2024ttx,Blumenhagen:2024ydy,Blumenhagen:2024lmo,Artime:2025egu,Hattab:2023moj,Hattab:2024thi,Hattab:2024chf,Hattab:2024yol,Hattab:2024ewk,Hattab:2024ssg,Monnee:2025ynn} for a small selection of more recent work.} Some regions, for example around the large complex-structure locus, can have very standard finite energy physics, where the low-energy degrees of freedom are that of the fundamental type II string. Other loci, such as type II loci (in the classification of \cite{Grimm:2018ohb,Grimm:2018cpv}), have more exotic and interesting finite energy physics. Indeed, such loci are more naturally associated to a Heterotic string description \cite{Hassfeld:2025uoy,Monnee:2025ynn}. However, even a perturbative Heterotic string description fails for some such loci \cite{Hattab:2025aok,Monnee:2025msf}. Overall, it is likely that fully understanding the microscopic finite energy physics of all regions in the moduli space will likely require a substantial research program. 

One, of many, motivations for such a research direction is to understand aspects of the Swampland program \cite{Vafa:2005ui} (see \cite{Palti:2019pca,vanBeest:2021lhn} for reviews). One of the central themes in this program is universal behaviour when approaching infinite distances in moduli space. The most well-known such proposal is the Distance Conjecture \cite{Ooguri:2006in}. This proposes that there is a tower of weakly-coupled fundamental states that become exponentially light in the distance. An interesting refinement of it is the Emergent String Conjecture \cite{Lee:2019wij,Lee:2019xtm}, which proposes that this tower are either Kaluza-Klein states or oscillator modes of a critical string. Both of these conjectures are statements about the ultraviolet nature of physics, specifically a tower of weakly-coupled states, but are formulated in terms of the infrared data of the moduli space. It is certainly likely that any weakly-coupled completion to quantum gravity in the ultraviolet will require a tower of states along the lines of the conjectures. Indeed, that string theory is the unique parametrically weakly-coupled ultraviolet completion of Einstein gravity was already proposed from scattering amplitudes earlier, see \cite{Camanho:2014apa}. Recent studies along this direction from scattering amplitudes only seem to strengthen this proposal (see, for example, \cite{Albert:2024yap}). 

The more challengeable aspect of the conjectures is the proposal that infinite distances in moduli space correspond to weak-coupling limits in the ultraviolet. One could imagine a situation where the infinite distance emerges in the infrared, as threshold corrections from integrating out light states, so that it does not correspond to a weak-coupling limit of the ultraviolet physics. From the perspective of the conjectures, the ultraviolet corresponds to the scale of the critical string. So we would require an emergent infinite distance below the string scale. In other words, we could consider the possibility that the finite energy physics about infinite distances in the infrared moduli space involves a strongly-coupled transition such that while the infrared degrees of freedom are parametrically weakly-coupled, the ultraviolet stays strongly-coupled, or at least not infinitely weakly-coupled. In quantum field theory, this is the type of physics in Seiberg-Witten theory \cite{Seiberg:1994rs}. It is also the physics of the conifold locus in string theory \cite{Strominger:1995cz,Lerche:1996xu}. There is a weakly-coupled description in the infrared, arising from integrating out a monopole state, while the ultraviolet coupling stays fixed. However, such loci are at finite distance in moduli space. An emergent infinite distance locus, has yet to be established. The first concrete candidate for such a locus was presented in \cite{Hattab:2025aok}, associated to K-points in one-parameter moduli spaces. 

Note that emergent aspects of moduli spaces in string theory have been studied as part of the Emergence Proposal, which was stated in this context in \cite{Grimm:2018ohb}, related to ideas in \cite{Harlow:2015lma,Heidenreich:2017sim,Heidenreich:2018kpg}, and formulated generally in \cite{Palti:2019pca}.\footnote{We refer to \cite{Palti:2020tsy,Marchesano:2022avb,Castellano:2022bvr,vandeHeisteeg:2022btw,Cribiori:2022nke,Marchesano:2022axe,Blumenhagen:2023tev,Castellano:2023qhp,Cribiori:2023ffn,Blumenhagen:2023yws,Blumenhagen:2023xmk,Seo:2023xsb,Calderon-Infante:2023ler,Calderon-Infante:2023uhz, Castellano:2023aum,Castellano:2023jjt,Castellano:2023stg,Cota:2023uir,Hattab:2023moj,Casas:2024ttx,Blumenhagen:2024ydy,Hattab:2024thi,Blumenhagen:2024lmo,Hattab:2024ewk,Artime:2025egu,Hattab:2024ssg,Hattab:2024chf} for a selection of works studying these ideas in various settings in string theory.} It is important however to distinguish emergence from scales above the fundamental string scale, and emergence below the string scale. Indeed, following the work in \cite{Blumenhagen:2023tev,Hattab:2023moj,Hattab:2024thi,Hattab:2024ewk,Hattab:2024chf,Hattab:2024ssg,Hattab:2024yol,Blumenhagen:2025zgf}, the proposal is that loci such as large complex-structure should be understood as emergent from integrating out physics above the fundamental string scale. This was termed Dual Emergence in \cite{Hattab:2024chf}. The notion of emergence at the K-point, and as studied in this work, is one where one integrates out physics below the fundamental string scale (in this case the dual Heterotic string scale). It is this notion of infrared emergence that can challenge the Distance and Emergent string conjectures, the ultraviolet emergence is perfectly compatible with the conjectures. 

What type of physics could possibly lead to an emergent infinite distance in the infrared, which is not present at the fundamental string scale? Using the supergravity data, various aspects of this physics were proposed in \cite{Hattab:2025aok}. It was proposed that the physics should be strongly-coupled and likely non-Lagrangian, in the sense of involving both light electric and light magnetic degrees of freedom. The physics should be similar to the conifold locus in the sense of the pure graviphoton kinetic terms not diverging. However, differently from the conifold, the mixed kinetic terms of the graviphoton with the other gauge fields do diverge. We therefore expect a light exotic sector, which mixes with the graviphoton, such that integrating out the sector induces a threshold correction to the graviphoton (mixed) kinetic terms. For this threshold correction to dominate in the infrared, the matter sector should be extremely light, doubly exponentially light in the distance. For the K-point, this could only be matched onto a threshold correction which behaves as if a BPS state with complex charges was integrated out \cite{Hattab:2024chf}. The effectively complex charges were proposed to be a manifestation of the absence of an electric-magnetic splitting, both in the matter sector and in the relation of the graviphoton to the gauge fields. Specifically, the graviphoton is a mixture of electric and magnetic fields.

In order to investigate this type of possibility, it is useful to move away from the single-parameter K-points, to multi-parameter moduli spaces. The additional parameters allow improved control over the physics. That is the aim of this work. In such cases the analogue of the K-point is a type II$_0$ locus \cite{Grimm:2018ohb}. A first question to address is whether multi-parameter type II$_0$ loci could in general exhibit similar physics to the K-point? We answer this with a partial yes. We show that for any type II$_0$ locus, much of the supergravity features of K-points can be reproduced. Specifically, it is possible to write the leading behaviour of the prepotential and the gauge kinetic matrix as if they arose from a threshold correction from integrating out a complex-charged BPS state.

After establishing that type II$_0$ loci can shed light on K-points, and more generally, on emergent infinite distances, we go on to study a two-parameter example in detail. The example we study is very well-known and studied. It was the original example of four-dimensional ${\cal N}=2$ type II - Heterotic duality \cite{Kachru:1995wm,Kachru:1995fv}. More recently, this type of duality was studied in further detail, and in more generality, in \cite{Blumenhagen:2024ydy,Artime:2025egu,Blumenhagen:2025zgf,Hassfeld:2025uoy,Monnee:2025ynn,Hattab:2025aok,Monnee:2025msf}. Indeed, this example was studied in \cite{Monnee:2025msf} precisely in order to shed light on K-point physics. Our analysis is therefore very similar in nature to that work.

The loci in the moduli space which have a perturbative Heterotic dual are type II$_1$ loci. We first establish the physics of those loci. We then study their connection to type II$_0$ loci limits. In particular, it is possible for type II$_1$ loci and type II$_0$ loci to intersect in moduli space. While type II$_0$ loci can never intersect the large complex-structure point \cite{Grimm:2018cpv}. Therefore, this suggests that the Heterotic perspective is better suited to understanding the microscopic physics of type II$_0$ loci than the type IIB perspective. Indeed, this was the general proposal of \cite{Hassfeld:2025uoy,Monnee:2025ynn,Monnee:2025msf}. 

We find that, as indeed proposed in \cite{Hattab:2025aok,Monnee:2025msf}, that type II$_0$ loci are non-perturbative from the Heterotic string perspective. Specifically, a Kaluza-Klein monopole state, which is non-perturbative with respect to the Heterotic string coupling, becomes lighter than a perturbative Kaluza-Klein (and winding) state. We identify in more detail the region in moduli space where this non-perturbative physics becomes important.\footnote{We propose that its onset is exponentially earlier, with respect to the distance, than in \cite{Monnee:2025msf}.} The non-perturbative sector has properties matching, at least qualitatively, the physics of a sector that can lead to emergent infinite distances: it is doubly exponentially light in the distance approaching the type II$_0$ locus, and it involves light electric and magnetic degrees of freedom.   

It is therefore natural to consider whether the non-perturbative sector could yield an emergent infinite distance upon integrating out. More specifically, whether integrating out the non-perturbative sector could manifest as a light complex-charged BPS state? We do not know how to microscopically integrate out a sector which has both light electric and magnetic states simultaneously, so is not Lagrangian, and so cannot answer this definitively. However, we do find various pieces of evidence towards the validity of such an interpretation of the physics, which we summarise below. 

A first crucial aspect of the physics is identifying the splitting between the graviphoton and the matter vector multiplets. Such a distinction between gravitational and matter sectors has featured in recent studies of type II loci \cite{Monnee:2025ynn,Hattab:2025aok,Monnee:2025msf}. It has also been studied in the context of the curvature of moduli spaces \cite{Marchesano:2023thx,Marchesano:2024tod,Castellano:2024gwi,Blanco:2025qom,Castellano:2026bnx}.\footnote{In particular, while this work was under completion, the example setting of section \ref{sec:modspace} was investigated from the curvature context in \cite{Castellano:2026bnx}. Some of those results overlap with ours.} We develop a new approach to this splitting: we identify certain projection matrices that project the gauge fields into directions that are orthogonal to a certain projection to the graviphoton. So these project onto the pure matter sectors in the theory. There exists such a projection for each vector multiplet, or modulus parameter. The projections are complex in nature, rather than symplectic. That is, they act naturally on the self-dual parts of the gauge fields, rather than their electric or magnetic components. This means that, in general, pure matter sectors are not embeddable purely into the electric gauge fields. Rather, they are a combination of both electric and magnetic fields. 

In the two-parameter example, we have two pure matter sectors. We show that indeed one of them coincides, at leading order, with the light electric state. However, the light magnetic state has components along both matter sectors. The leading component along the second matter sector is precisely captured by the complex charge. The complex charge corresponds to the fact that this matter sector is embedded into a combination of electric and magnetic directions.

Indeed, this mixing between the light states and two, rather than one, matter sectors coincides with an absence of factorization of the gauge group into a $U(1)$ with light states and a $U(1) \times U(1)$ with heavy states. We show that there is a general $\mathbb{Z}_2$ obstruction to such a splitting of the gauge group (even though the gauge algebra does factorize). 

Because the second matter sector is embedded into both the electric and magnetic fields, the pull back of the gauge kinetic matrix to it does not behave in a physically simple way. However, it can be made completely standard, so with positive kinetic terms and periodic real parts, by rotating one of the gauge fields by a factor of $i$. This rotation, acting on the complex charge, then makes it real and integral. In other words, the projection to the matter sector can be written as a standard matter sector, with an integrally charged state, after a certain complex rotation on the gauge fields. The same rotation also makes the embedding of the graviphoton purely electric. In other words, the fact that the matter sector is embedded into a combination of the electric and magnetic fields can be seen directly from the fact that the graviphoton is also embedded into such a combination. 

These results suggest that approaching type II$_0$ limits there is a light non-perturbative sector which in the infrared has a weakly-coupled description, similarly to Seiberg-Witten physics \cite{Seiberg:1994rs}, except that it is rotated in the complex plane of the gauge fields. This rotation is capturing the fact that it is embedded into a combination of the electric and magnetic gauge fields, and there is no symplectic frame where it can be made purely electric (similarly to Argyres-Dougles theories \cite{Argyres:1995jj}). Integrating out this sector then induces an infinite distance singularity in the infrared moduli space. This is the physics proposed also for the one-parameter K-point in \cite{Hattab:2025aok}. 

The paper is set out as follows. In section \ref{sec:sugra} we introduce the relevant notions in ${\cal N}=2$ supergravity. In section \ref{sec:ii0loci} we analyse the supergravity data of type II$_0$ loci in generality. This is where we show the similarity to the supergravity physics of the K-point. In section \ref{sec:modspace} we introduce and analyse the two-parameter example model. In particular we give an integral representation of the periods around the different loci in moduli space. We give more details of these results in appendices \ref{app:lcs} and \ref{app:ii}. In section \ref{sec:hetdual} we study the duality with the Heterotic string. In section \ref{sec:hetdualii0} we discuss the physics of type II$_0$ loci, in particular with regards to the graviphoton embeddings and mixing of the matter and gravitational sectors. We summarise our results in section \ref{sec:sum}. 

\section{The supergravity}
\label{sec:sugra} 

For comprehensive reviews of ${\cal N}=2$ supergravity theories, see for example \cite{Andrianopoli:1996cm,Freedman_VanProeyen_2012}. We mostly follow the conventions in \cite{Dedushenko:2014nya,Hattab:2024ssg,Hattab:2025aok}. 

Compactifying type IIB string theory on a Calabi-Yau manifold yields a four-dimensional ${\cal N}=2$ supergravity (see, for example, \cite{Gurrieri:2003st,Grimm:2005fa,Palti:2006yz,Palti:2008mg}).
There is a gravitational multiplet whose bosonic fields are comprised of the graviton $h_{\mu\nu}$ and the graviphoton $V_{\mu}$. There are also $n_V =h^{(2,1)}$ vector multiplets, where $h^{(2,1)}$ is the Hodge number of the Calabi-Yau, labelled by an index  $i=1,...,h^{(2,1)}$. 

It is convenient to describe these multiplets with a redundancy as $h^{(2,1)}+1$ multiplets labelled by an index $I=0,i$. The bosonic components of these are labelled as $X^I$, and the vector bosons as $F^{I}_{\mu\nu}$. 
The redundancy corresponds to a rescaling symmetry $X^I \rightarrow \lambda X^I$, with $\lambda \in \mathbb{C}^*$. It is often common to use the symmetry to fix $X^0=1$, though we will not do so in this work. The gauge field of the additional vector multiplet is identified with the graviphoton.  

The action for the vector multiplet fields takes the following form\footnote{This is coupled to the gravitational Lagrangian as ${\cal L} = \frac{M_p^2}{2} R + {\cal L}_F$, with $R$ the Ricci scalar and $M_p$ the Planck scale.}
\be 
{\cal L}_{F} =  + \frac{1}{4}\text{Im}\,\mathcal{N}_{IJ} F^{I}_{\mu\nu} F^{J,\mu\nu} -\frac{1}{8}  \text{Re}\,\mathcal{N}_{IJ} F^{I}_{\mu\nu} F^{J}_{\rho\sigma} \epsilon^{\mu\nu\rho\sigma}\;,
\label{n2lfac}
\ee 
where 
\bea 
\mathcal{N}_{IJ}&=& \overline{F}_{IJ}+2i\frac{\text{Im}\,F_{IL}X^{L} \text{Im}\,F_{JK}X^{K}}{\text{Im}\,F_{MN}X^{M}X^{N}} \;.
\label{genNexo}
\eea 
Here we utilise the notation
\be 
F_{IJ}=\partial_{X^I}\partial_{X^J}F\left(X\right) \;,\;\; F_{I}=\partial_{X^I}F\left(X\right)\;.
\label{magper}
\ee 
The (homogeneous degree-two) function $F\left(X\right)$ is known as the supergravity prepotential. 
The matrix ${\cal N}$ relates electric and magnetic quantities, for example a useful relation is
\be 
F_I = {\cal N}_{IJ} X^J \;\;.
\label{FNXfpr}
\ee
From this also follows the useful
\be 
2 F = {\cal N}_{IJ} X^I X^J \;.
\label{NtoF} 
\ee 

The physical moduli space is parameterised by gauge invariant combinations 
\be 
z^I = \frac{X^I}{X^0} \;.
\label{zidef}
\ee 
The associated kinetic terms in the action are then 
\be
{\cal L}_z = -g_{ij} \partial_{\mu}z^i\partial^{\mu}\overline{z}^j \;.
\label{supkint}
\ee 
The moduli space metric is given in terms of the non-homogeneous coordinates as
\be 
g_{ij} = -\frac{\partial^2}{\partial z^{i}\partial \overline{z}^j}\log\left(-2\left(\mathrm{Im\;}F\right)_{KL}z^{K}\overline{z}^{L}\right) \;.
\label{Kahlmet}
\ee 
The metric on the moduli space $g_{ij}$ has an associated Kahler potential $K$, so $g_{ij} = \partial_i \bar{\partial}_j K$, given by 
\be
\left|X^0\right|^2e^{-K} = i\left(F_I \overline{X}^I - \overline{F}_I X^I \right)\;.
\label{kahlpot}
\ee

\subsubsection*{Electric and magnetic field strengths}

The supergravity is a local Lagrangian theory, and therefore has either electric $F^I_{\mu\nu}$ or magnetic $G_{I,\mu\nu}$ field-strengths as dynamical. In order to write the action in terms of only one of these, we need a relation between them. This comes from a variation of the action, but has a compact expression in the supergravity. To write this expression, it is useful to work with self-dual $F^+_{\mu\nu}$ and anti self-dual $F^-_{\mu\nu}$ field-strengths. We define $F^{\pm}_{\mu\nu}$ as 
\be 
F^{I,\pm}_{\mu\nu} = \frac12 \left( F^I_{\mu\nu} \mp i\tilde{F}_{\mu\nu}^I\right) \;,
\label{fpmmink}
\ee 
with the (Hodge) dual electric field strength $\tilde{F}_{\mu\nu}^I$ defined as\footnote{We use metric conventions of $\left(-,+,+,+\right)$ and take $\epsilon_{0123}=1$.}
\be 
\tilde{F}_{\mu\nu}^I = \frac{1}{2} \epsilon_{\mu\nu\rho\lambda} F^{I,\rho\lambda} \;.
\label{tildFidefm}
\ee 
We therefore have
\be 
F^I_{\mu\nu} = F^{I,+}_{\mu\nu} + F^{I,-}_{\mu\nu} \;,\;\; F^{I,\pm}_{\mu\nu} = \left( F^{I,\mp}_{\mu\nu}\right)^*\;.
\ee 

The relation between the electric and magnetic field strengths is then given in the holomorphic form
\be 
G_{I,\mu\nu}^+ = {\cal N}_{IJ}F^{J,+}_{\mu\nu} \;\;,\;\;\; G_{I,\mu\nu}^- = \overline{{\cal N}}_{IJ}F^{J,-}_{\mu\nu} \;.
\label{GNFminrel}
\ee

The quantized electric $q^I$ and magnetic $p^I$ charges of states are given by the integrals of the field-strengths over the sphere at infinity
\be 
\int_{S^2_{\infty}} F^I_{\mu\nu} \;dx^{\mu}\wedge dx^{\nu} = -2\sqrt{2\pi} \;p^I \;\;,\;\; \int_{S^2_{\infty}} G_{I,{\mu\nu}} \;dx^{\mu}\wedge dx^{\nu} = -2\sqrt{2\pi} \; q_I \;\;.
\label{chargedefFint}
\ee 
This quantization then leads to an integral electric-magnetic duality group, to which we now turn. 

\subsubsection*{The electromagnetic symplectic frame}

An important role in our analysis is played by the symplectic electric-magnetic duality group of the supergravity $\mathrm{Sp}\left(2\left( n_V+1\right),\mathbb{Z}\right)$. We can write a general elements $S$ of this group as an integral matrix of the form
\be 
S = \left( \begin{array}{cc} {\bf A} & {\bf C} \\ {\bf D} & {\bf E} \end{array}\right) \;.
\label{Sformabde}
\ee 
The matrices $S$ are integer matrices satisfying
\be 
S^T \cdot \eta \cdot S = \eta \;,
\label{symconst}
\ee 
where the symplectic inner product $\eta$ can, in generality, be taken as 
\be 
\eta = \left( \begin{array}{cc} 0 & \mathbb{1} \\ -\mathbb{1} & 0 \end{array}\right) \;.
\label{symplecticform}
\ee 
Therefore, the block entries of $S$ satisfy the relations
\bea 
{\bf D}^T \cdot {\bf A} = {\bf A}^T \cdot {\bf D} \;\;,\;\; {\bf E}^T \cdot {\bf C} = {\bf C}^T \cdot {\bf E}\;\;,\;\; {\bf A}^T \cdot {\bf E} - {\bf D}^T \cdot {\bf C} = \mathbb{1} \;.
\eea 

The symplectic action acts on the electric and magnetic field strengths, and therefore also on their superpartner scalars. We can form symplectic vectors as
\be 
\Gamma_{\mu\nu} = \left( \begin{array}{c} {\bf F}_{\mu\nu} \\ {\bf G}_{\mu\nu} \end{array}\right) \;\;,\;\; \Pi = \left( \begin{array}{c} {\bf X} \\ {\bf F} \end{array}\right) \;,
\label{PiasXF}
\ee
with the vectors ${\bf F}_{\mu\nu}$, ${\bf G}_{\mu\nu}$, ${\bf X}$ and ${\bf F}$ defined as 
\be 
{\bf F}^I_{\mu\nu} = F^I_{\mu\nu}\;\;,\;\; {\bf G}_{I,\mu\nu} = G_{I,\mu\nu}\;\;,\;\;{\bf X}^I = X^I \;\;,\;\; {\bf F}_I = F_I \;.
\label{fieldstrengvectors}
\ee
Then the symplectic action is
\be 
\Gamma_{\mu\nu} \rightarrow S\cdot \Gamma_{\mu\nu}  \;\;,\;\;\Pi \rightarrow S\cdot \Pi \;.
\label{SactionP}
\ee

The action on the scalars gives in particular 
\be 
{\bf X} \rightarrow {\bf A} \cdot {\bf X} + {\bf C} \cdot {\bf F}  \;\;,\;\; 
{\bf F} \rightarrow {\bf D} \cdot {\bf X} + {\bf E} \cdot {\bf F}  \;\;,\;\; {\cal N} \rightarrow \left({\bf E} \cdot {\cal N} + {\bf D}\right)\cdot\left({\bf C} \cdot {\cal N} + {\bf A}\right)^{-1}\;.
\ee 
Note that the transformation of ${\cal N}$ ensures that the relations (\ref{FNXfpr}) and (\ref{GNFminrel}) are maintained under the symplectic action.

It is useful to note that the prepotential is not invariant under general symplectic transformations. It transforms as
\be 
F \rightarrow F + \frac12\; {\bf X}^T \cdot\Big(\;{\bf A}^T \cdot {\bf D} + 2\; {\cal N}\cdot  {\bf C}^T \cdot {\bf D} + {\cal N}\cdot {\bf C}^T \cdot {\bf E} \cdot {\cal N} \;\Big) \cdot{\bf X} \;.
\label{prepottrans}
\ee

\subsubsection*{The graviphoton and BPS masses}

The four-dimensional graviphoton $V_{\mu}$ is defined as the vector superpartner to the graviton. We denote its field-strength as $W_{\mu\nu}$, so
\be 
W_{\mu\nu} = \partial_{\mu} V_{\nu} - \partial_{\nu} V_{\mu} \;.
\ee 


The anti self-dual part of the four-dimensional graviphoton field is then given by \cite{Freedman_VanProeyen_2012}
\be
X^0\;W_{\mu\nu}^- = 2\;e^{\frac{K}{2}}\;\Gamma_{\mu\nu} \cdot \eta \cdot \Pi = 2\;e^{\frac{K}{2}}\left(-X^I G_{I,\mu\nu} + F_I F^I_{\mu\nu} \right) = 4ie^{\frac{K}{2}}X^I\mathrm{Im\;}{\mathcal{N}}_{IJ}F^{J,-}_{\mu\nu}\;\;.
\label{gravWdef}
\ee
Note that this is manifestly invariant under the symplectic action (\ref{SactionP}). An important property of the graviphoton embedding (\ref{gravWdef}) into the gauge field strengths $G_{I,\mu\nu}$ and $F^I_{\mu\nu}$ is that it is holomorphic rather than symplectic. That is, it is given by a holomorphic combination of the field strengths, and therefore cannot always be embedded into purely electric or magnetic field strengths by an appropriate real integral symplectic rotation. 

The anti self-dual part of the graviphoton is related to the mass of BPS states, which is given by the integral of it over the sphere at infinity $S^2_{\infty}$. We can define the central charge $Z(q)$ as
\be 
Z(q)=-\frac{i}{2}\int_{S^2_{\infty}} W^-_{\mu\nu} \;dx^{\mu}\wedge dx^{\nu} = -i\;2\sqrt{2\pi}\left(X^0\right)^{-1}e^{\frac{K}{2}}\left( q_I X^I - p^I F_I\right)  \;.
\label{defcenhargint} 
\ee 
The mass $M(q)$ of charged BPS states is then given by
\be 
M(q) = \left|Z(q) \right| M_p \;.
\label{massBPSqg} 
\ee 
We sometimes set $M_p=1$, but sometimes keep it explicitly manifest when it is informative to do so.  

Because the natural formulation of the graviphoton is in a self-dual basis, it is useful to write also the kinetic terms of the fields in this basis. The action (\ref{n2lfac}) can be written as
\be
{\cal L}_{F} = \frac{i}{4}\; \mathcal{\overline{N}}_{IJ}\; F^{I,-}_{\mu\nu} \left(F^{J,-}\right)^{\mu\nu}+\; \mathrm{h.c.} \;\;\;.
\label{n2lfacasd}
\ee

\section{Generalities of type II loci}
\label{sec:ii0loci}

In this section we discuss type II loci in general, and then specifically type II$_0$ loci. The analysis is performed around any generic point on a type II locus in the moduli space. In the case where the point is at the intersection with other singular loci there is additional structure, but the analysis presented still holds. We utilize the results in \cite{Grimm:2018ohb,Grimm:2018cpv}, and refer to those works for further details. We also note that a general analysis of one-parameter type II$_0$ loci, K-points, was performed \cite{Bastian:2023shf}, and in \cite{Bastian:2021eom} a general analysis of all types of singular loci was performed. Our results have some overlap with those works.

\subsection{The supergravity data}
\label{sec:sugraii0gen}

The starting point is the Nilpotent orbit theorem which states that the period vector $\Pi$ can be written locally around any type II locus in the form \cite{Schmid:1973cdo,Grimm:2018cpv}
\be 
\Pi = e^{\frac{\log s}{2\pi i} N} \left(a_0 + s \;a_1 + ... \right) \;.
\label{NOTF}
\ee 
Here $s$ is the local holomorphic coordinate such that the type II locus is at $s=0$. The $a_i$ are independent of $s$. In the case where the moduli space is one-parameter they would be constants, while if there are other parameters then they are functions of them. 
The matrix $N$ in (\ref{NOTF}) is constant with integer entries. Importantly, for type II$_n$ loci, it is nilpotent and has rank $2+n$ \cite{Grimm:2018cpv}
\be 
\text{Type II$_n$} \;\;:\;\; N^2 =0 \;\;,\;\; \text{Rank\;} N = 2+n \;.
\ee 

We can further constrain the form (\ref{NOTF}) by using the fact that there is a symplectic symmetry group. 
The period vector can be acted on by symplectic transformations $S \in \mathrm{Sp}\left(2\left( n_V+1\right),\mathbb{Z}\right)$ as in (\ref{SactionP}).
The form (\ref{NOTF}) is maintained by transforming
\be 
N \rightarrow S \cdot N \cdot S^{-1} \;.
\ee 

The monodromy matrix must respect the symplectic structure, and so 
\be 
N^T \cdot \eta = - \;\eta \cdot N \;.
\label{Netacom}
\ee 
This, combined with $N^2=0$, implies that its image subspace is isotropic. That is, 
\be 
\left(N\cdot v\right)^T \cdot \eta \cdot \left(N \cdot u\right) =0 \;, 
\ee 
for any vectors $u,v$. A nilpotent matrix with an isotropic image can always be brought by a symplectic transformation to the form 
\be 
N = \left( \begin{array}{cc} 0 & 0 \\ -{\bf B} & 0 \end{array} \right) \;,
\label{monN}
\ee 
with ${\bf B}$ an integer matrix. From (\ref{Netacom}) we also deduce ${\bf B}^T = {\bf B}$. Also, for type II$_n$ loci, ${\bf B}$ has two positive eigenvalues and $n$ negative ones \cite{Grimm:2018cpv}. We therefore have that
\be 
{\bf B}^T = {\bf B} \;\;,\;\; \lambda^1_{\bf B} > 0 \;\;,\;\lambda^2_{\bf B} > 0\;,\;\lambda^i_{\bf B} < 0 \;,
\label{Bprop} 
\ee 
where $\lambda^1_{\bf B}$ and $\lambda^2_{\bf B}$ are the two non-vanishing eigenvalues of ${\bf B}$ and $i=3,...,n+2$.

The supergravity period vector takes the general form (\ref{PiasXF}). The symplectic group is then the electric-magnetic group of the supergravity. Choosing a symplectic gauge where $N$ takes the form (\ref{monN}) is then a choice of electric-magnetic splitting such that all the logarithmic dependence on $s$ is in the $F_I$, and the $X^I$ are holomorphic. We call such a symplectic gauge an electric gauge. 

Usually, such a logarithmic splitting implies that the electric particles, which couple to the  $X^I$ are light, and the magnetic particles which couple to the $F_I$ are heavy. However, this need not always be the case: it is possible, and indeed is true for the example studied in section \ref{sec:modspace}, that some magnetic particles could become lighter than some electric ones. This happens when the logarithmic terms in $s$ are multiplied by powers of $s$ in the magnetic periods.

Let us write the coefficients $a_n$ in (\ref{NOTF}) as 
\be 
a_n = \left( \begin{array}{c} {\bf X}^{(n)} \\ {\bf F}^{(n)} \end{array} \right) \;,
\ee  
with the ${\bf X}^{(n)}$ and ${\bf F}^{(n)}$ independent of $s$. We then have from (\ref{NOTF}) and (\ref{PiasXF}) that, in the chosen symplectic gauge,
\bea 
{\bf X} &=& {\bf X}^{(0)} + {\bf X}^{(1)} \; s+ ... \nn \\
{\bf F} &=& - {\bf B} \cdot {\bf X} \;\frac{\log s}{2 \pi i} +  {\bf F}^{(0)} + {\bf F}^{(1)} \; s+ ... \;,
\label{expXFB}
\eea 
where the ellipses denotes additional powers in $s$ but no more $\log s$ terms. 

We now note some further important properties of ${\bf B}$. These follow from the polarized mixed Hodge structure at the type II locus.\footnote{They also appropriately generalize to type III loci, but one has to use the full $N$ rather than ${\bf B}$.} We refer to \cite{Grimm:2018ohb,Grimm:2018cpv} for details, and here just present the necessary data for the conclusions. Approaching any type II locus, there is a vector space which has the following properties. There are subspaces labelled as $P^{p,q}$ which satisfy polarized inner product relations
\bea 
\left(v^{p,q}\right)^T \cdot \eta \cdot N \cdot u^{r,s} &=& 0 \;\;,\;\; \text{unless}\; p=s\;,q=r\;,\;p+q=4 \nn \;, \\
-i^{p-q} \left(v^{p,q}\right)^T \cdot \eta \cdot N^{p+q-3} \cdot \bar{u}^{p,q} &>& 0 \;,
\label{genpolarrel}
\eea 
for any integers $p,q$ and elements $v^{p,q}\;,\;u^{p,q} \;\in \; P^{p,q}$.
Now, for type II loci we have
\be 
a_0 \in P^{3,1} \;,\;\; \bar{a}_0 \in P^{1,3} \oplus P^{0,2} \;.
\ee 
Therefore, using (\ref{genpolarrel}) we find the important constraints
\bea 
{\bf B}_{IJ} X^{(0),J} &\neq& 0 \;,
\label{BXContr1} \\
{\bf B}_{IJ} X^{(0),I} X^{(0),J} \;&=& \; 0 \;,
\label{BXContr2}
\eea 
where we define $X^{(0),I} = {\bf X}^{(0),I}$.

We can then assign a {\it rank} to a type II locus according to the vanishing order of ${\bf B}_{IJ} X^I X^J$. So we say that a type II locus is of rank $n$ if 
\be 
{\bf B}_{IJ} X^{I} X^{J} \sim s^n \;.
\label{rankntypeIIgen}
\ee 
All type II loci therefore have rank $n \geq 1$.

%

Comparing (\ref{expXFB}) and the supergravity formulas (\ref{FNXfpr}) and (\ref{NtoF}) we see that we have
\be 
{\cal N}_{IJ} = -\frac{1}{2 \pi i}\Big(  {\bf B}_{IJ} + {\bf M}_{IJ}\Big) \log s + ... \;,
\label{NBformpre}
\ee 
where the ellipses denote terms that are polynomial in the real and imaginary parts of $s$. The matrix ${\bf M}$ satisfies the constraint
\be 
{\bf M} \cdot  {\bf X} = 0 \;. 
\label{MXcond0full}
\ee 
It is important to note that this constraint is {\it exact} , so holds at any perturbative order in $s$. 
Using (\ref{NtoF}), and the constraint (\ref{MXcond0full}), we see that the logarithmic part of the prepotential takes the form
\bea 
F = -\frac{1}{4 \pi i}{\bf B}_{IJ} X^I X^J \log s + ... \;.
\label{FBform}
\eea 

The matrix ${\bf M}$ has a certain physical role associated to the compatibility with the eignenvalues of ${\bf B}$ with the physical requirement of the negativity of the imaginary part of ${\cal N}$ which follows from the action (\ref{n2lfac}). 
We can expand ${\bf M}$ as\footnote{In fact, ${\bf M}$ also has logarithmic terms in it, so powers of $\log s$. However, they appear at high orders in the examples we use, and so we do not include them in the general expansion for ease of notation.}  
\be 
{\bf M} = \sum_{i,j} {\bf M}^{(i,j)}\left(\mathrm{Re}\;s\right)^i \left(\mathrm{Im}\;s\right)^j  \;,
\ee 
where the ${\bf M}^{(i,j)}$ have no dependence on $s$. The constraint (\ref{MXcond0full}) in particular implies
\be 
{\bf M}^{(0,0)}_{IJ} X^{(0),J} = 0 \;. 
\label{MXcond0}
\ee
The divergent part of the gauge kinetic matrix then takes the form
\be 
{\cal N}_{IJ} = -\frac{1}{2 \pi i} \Big( {\bf B}_{IJ} + {\bf M}^{(0,0)}_{IJ}\Big) \log |s| + ... \;.
\label{NBform}
\ee 

From (\ref{NBform}) we see that $\mathrm{Im\;} {\cal N}$ takes the leading form 
\be 
\mathrm{Im\;} {\cal N}_{IJ} = \frac{1}{2 \pi}\Big(  {\bf B}_{IJ} + \mathrm{Re\;}{\bf M}^{(0,0)}_{IJ}\Big) \log |s| + ... \;.
\label{ImNBform}
\ee 
Positivity of the gauge kinetic terms demands that $\mathrm{Im\;} {\cal N}_{IJ} $ always has negative definite eigenvalues. Since $\log |s|$ is negative, we therefore require that ${\bf B}_{IJ} + \mathrm{Re\;}{\bf M}^{(0,0)}_{IJ}$ must be positive definite. However, for a type II$_n$ locus, ${\bf B}_{IJ}$ has two positive eigenvalues and $n$ negative ones. The role of $\mathrm{Re\;}{\bf M}^{(0,0)}_{IJ}$ is then to make sure that the negative eigenvalues are made positive. It therefore can only vanish for type II$_0$ loci. Indeed, we conjecture that it always does vanish for type II$_0$ loci \footnote{We expect that there is a simple proof of this.}
\be 
\text{(Conjecture) \;\;Type II$_0$\;\;:\;\;} {\bf M}^{(0,0)}_{IJ} = 0 \;.
\label{typeii0m0vani}
\ee 

\subsubsection{Type II$_0$ loci}
\label{sec:typeii0logens}

In the specific cases of type II$_0$ loci ${\bf B}$ is positive definite, symmetric and rank two, and so can be written in the form of an outer product of a constant complex vector ${\bf q}^{(b)}$ as
\be 
{\bf B} = {\bf q}^{(b)} \left({\bf \bar{q}}^{(b)}\right)^T+ {\bf \bar{q}}^{(b)} \left({\bf q}^{(b)}\right)^T \;,
\label{Bvf}
\ee 
with 
\be 
{\bf q}^{(b)}_I \in \mathbb{C} \;.
\ee
In terms of components we can write (\ref{Bvf}) as
\be 
{\bf B}_{IJ} = {\bf q}_I^{(b)}\; \bar{{\bf q}}_J^{(b)} + \bar{{\bf q}}_I^{(b)}\; {\bf q}_J^{(b)}  \;.
\label{Bintermsofb}
\ee 
The constraints (\ref{BXContr1}) and (\ref{BXContr2}) therefore imply
\be 
{\bf q}^{(b)} \cdot {\bf X}^{(0)} \neq 0 \;\;,\;\; \bar{{\bf q}}^{(b)} \cdot {\bf X}^{(0)} = 0 \;.
\label{qconst}
\ee  

From the definition of the rank of type II loci (\ref{rankntypeIIgen}), we have that a type II$_0$ locus is of rank $n$ if
\be 
\text{Type II$_0$ of rank n}\;:\; \bar{{\bf q}}^{(b)} \cdot {\bf X}^{(n)} \neq 0 \;\;,\;\; \bar{{\bf q}}^{(b)}  \cdot {\bf X}^{(i)} = 0 \;\;\; \forall\; i < n \;.
\label{ranknii0def}
\ee 
Examples of a rank one type II$_0$ loci are one-parameter K-points, as studied in \cite{Hattab:2025aok}. An example of a type II$_0$ locus of rank two is the II$_0$ locus in the $\mathbb{P}_{[1,1,2,2,6]}$ Calabi-Yau, as discussed in section \ref{sec:modspace}. It is interesting to note that if the moduli space has (complex) dimension $n$, so equal to the rank of the II$_0$ locus, as in the example cases above, then the vector ${\bf q}^{(b)}$ is completely fixed (up to an overall rescaling). If the rank of the type II$_0$ locus is lower than the dimension of the moduli space, then ${\bf q}^{(b)}$ would not be fixed.\footnote{The rank cannot be larger than the dimension of the moduli space.} It is not clear if such a situation is realised in Calabi-Yau compactifications.

We can write the prepotential (\ref{FBform}) in the form
\bea 
F &=& -\frac{1}{2 \pi i} \Big({\bf q}^{(b)} \cdot {\bf X} \Big) \Big( \bar{{\bf q}}^{(b)} \cdot {\bf X}\Big)  \log s + ...  \nn \\
& =& -\frac{1}{2 \pi i} s \Big({\bf q}^{(b)} \cdot {\bf X}^{(0)} + s \;{\bf q}^{(b)} \cdot {\bf X}^{(1)}\Big) \Big( \bar{{\bf q}}^{(b)} \cdot {\bf X}^{(1)} + s\; \bar{{\bf q}}^{(b)} \cdot {\bf X}^{(2)} \Big)  \log s + ... \;.
\label{FBformb2}
\eea 
We therefore see that the leading form of the prepotential is 
\be 
F =  -\frac{1}{2 \pi i} \left({\bf q}^{(b)} \cdot {\bf X}^{(0)}\right)\left( \bar{{\bf q}}^{(b)} \cdot {\bf X}^{(n)}\right) s^n \log s + ... \;,
\ee 
for a rank $n$ type II$_0$ locus. 


The gauge kinetic matrix (\ref{NBform}) can also be written using ${\bf q}^{(b)}$ and ${\bf M}$ as
\be 
{\cal N}_{IJ} = -\frac{1}{2 \pi i} \Big( {\bf q}^{(b)}_I \;\bar{{\bf q}}^{(b)}_J + \bar{{\bf q}}^{(b)}_I\; {\bf q}^{(b)}_J + {\bf M}^{(0,0)}_{IJ}\Big) \log s + ... \;.
\label{NBformb2}
\ee 
As noted above, type II$_0$ loci are unique in that they allow for ${\bf M}^{(0,0)}=0$, and we expect that this indeed always holds (\ref{typeii0m0vani}). 

The Kahler potential of the vector multiplet moduli space $K$, is given by the formula
\be 
\left|X^0\right|^{2}e^{-K} = -i\;\Pi^T \cdot \eta  
\cdot \overline{\Pi}\;.
\label{KahlpPiGen}
\ee 
Using (\ref{NOTF}) and (\ref{Netacom}), as well as the expansion (\ref{expXFB}), we can read off the leading logarithmic behaviour as
\be
\left|X^0\right|^{2}e^{-K} 
= - \kappa_0 \log |s| + ... \;.
\label{kapotgenii0}
\ee 
with
\be 
\kappa_0 = \frac{1}{\pi}\;{\bf B}_{IJ} X^{(0),I}\bar{X}^{(0),J} = \frac{1}{\pi} \left| {\bf q}^{(b)} \cdot {\bf X}^{(0)} \right|^2\;.
\ee 
This Kahler potential is divergent and implies that the locus is at infinite distance.

\subsection{BPS states}
\label{sec:intbpsgen}

A central role in our understanding of the moduli space is played by charged BPS states. All such states have already been fully integrated out from the theory  down to zero momentum, they are therefore not dynamical states. Their BPS nature is crucial for two aspects of the theory. First, their energy, or mass if they are single-particle states, is determined completely from their charge. Second, they are the only states that can contribute threshold corrections, upon integrating out, to the prepotential (and therefore to the vector-multiplet two-derivative action). This is because the vector multiplets action is an F-term in ${\cal N}=2$ superspace, and only (half) BPS states can contribute to F-terms. 

BPS states, with electric charges $q_I$ and magnetic charges $p^I$, defined as in (\ref{chargedefFint}), have an associated central charge (\ref{defcenhargint}). Their mass is then given by the magnitude of the central charge (\ref{massBPSqg}). 

The central charge (\ref{defcenhargint}) can be written in terms of the period vector and the symplectic structure as 
\be 
Z\left(q\right) = -i\;2\sqrt{2\pi}\;\left(X^0\right)^{-1}e^{\frac{K}{2}} \;q \cdot \eta \cdot \Pi \;,
\ee 
where we define the symplectic charge vector 
\be 
q = \left( \begin{array}{c} {\bf p}  \\ {\bf q}\end{array} \right) \;, \;\; {\bf q}_I = q_I \;,\;\; {\bf p}^I = p^I\;.
\label{symcharvec}
\ee 

Around type II loci, we can use the expressions (\ref{expXFB}) to therefore write the central charge as
\be 
Z\left(q\right) = -i\;2\sqrt{2\pi}\;\left(X^0\right)^{-1}e^{\frac{K}{2}} \left[  \;{\bf q} \cdot \left(\; {\bf X}^{(0)} + {\bf X}^{(1)} \; s + ...\right) - {\bf p}\cdot \left({\bf F}^{(0)} - {\bf B} \cdot {\bf X}^{(0)} \;\frac{\log s}{2 \pi i} \;+ ...\right)  \right]  \;.
\label{gencentypii}
\ee 

We are interested in the light spectrum of states. In an electric gauge (\ref{monN}), the lightest state is an electric state. We can classify electric states as being of type $n$ by the definition
\be 
\text{Type $n$} \;:\; {\bf q} \cdot {\bf X}^{(n)} \neq 0  \;\;\text{and} \;\;{\bf q} \cdot {\bf X}^{(i)} = 0 \;\;,\; \forall \; i < n \;.
\label{typendefq}
\ee 
We therefore have that for a type $n$ electric state the central charge $Z$ behaves as
\be 
\text{Type $n$} \;:\; Z = -i \;2\sqrt{2\pi}\;\left(X^0\right)^{-1}e^{\frac{K}{2}} \;{\bf q} \cdot {\bf X}^{(n)} \;s^{n} + ... \;.
\label{typencench}
\ee 
Note that a state where $n=0$ was denoted a type G (or type I) state in \cite{Grimm:2018ohb,Palti:2019pca}, while if $n>0$ it was denoted a type F (or type II) state. 

\subsubsection*{Complex charges and chiral central charges}

Following the results in \cite{Hattab:2025aok}, it is informative to introduce the notion of complex electric charges, and the associated notion of chiral central charges. This allows to write the prepotential in a certain form which closely resembles the form induced by threshold corrections from BPS states. At the moment it is a purely mathematical formulation, and our aim in this section is to show that it holds for type II$_0$ loci generally. We return to the physical meaning of this in section \ref{sec:hetdualii0}. 

Threshold corrections to the prepotential are usually evaluated as an integrating out calculation in a supersymmetric gauge field background. The seminal such calculation was performed in the work of Gopakumar and Vafa \cite{Gopakumar:1998jq}, and developed in more detail in \cite{Dedushenko:2014nya,Hattab:2024ewk,Hattab:2024ssg}. The gauge background for such a calculation is uniquely fixed by the requirement of preserving the full ${\cal N}=2$ supersymmetry in flat space. As emphasised in \cite{Dedushenko:2014nya,Hattab:2024ssg,Hattab:2025aok}, such a background is not possible to realise in flat Minkowski space with real gauge fields. However, for the purpose of the mathematical formulation, we only need to use it as a motivation for an important projection from the anti self-dual parts of the gauge field-strengths to the graviphoton field strength, which takes the form
\be 
F^{I,-}_{\mu\nu} = \frac{i}{2} \left(\bar{X}^0\right)^{-1}e^{\frac{K}{2}} \;\overline{X}^I W^-_{\mu\nu} \;.
\label{gravproj}
\ee

One way to think of this projection is as the statement that the prepotential $F$ controls the coupling of the graviphoton. That is, starting from the form of the action in (\ref{n2lfacasd}), and inserting the projection (\ref{gravproj}), we have
\bea
{\cal L}_{F} &=& -\frac{i}{16} \left(\bar{X}^0\right)^{-2}e^K \left(\mathcal{N}_{IJ} X^I X^J\right)^* W^{-}_{\mu\nu} \left(W^{-}\right)^{\mu\nu} + \mathrm{h.c.} \;\;\; \nn \\
&=& -\frac{i}{8} \left(\bar{X}^0\right)^{-2}e^K \;\overline{F} \; W^{-}_{\mu\nu} \left(W^{-}\right)^{\mu\nu} + \mathrm{h.c.} \;\;\;.
\label{n2lfacasdW}
\eea
where we utilized the identity (\ref{NtoF}). This is a manifestation of the fact that the vector multiplet action is an F-term in ${\cal N}=2$ superspace. 

The projection (\ref{gravproj}) is closely related to the central charge. We see from (\ref{defcenhargint}) that the central charge is the charge which sources the anti self-dual component of the graviphoton field strength. But in writing a local action for the electric degrees of freedom, we would like to see that this is how the graviphoton couples in the action. This coupling follows from the projection (\ref{gravproj}), by contracting the electric charge with the electric field strengths
\be 
q_I F^{I}_{\mu\nu} = \frac{i}{2} \left(\bar{X}^0\right)^{-1}e^{\frac{K}{2}} \;q_I\overline{X}^I W^-_{\mu\nu}  -\frac{i}{2} \left(X^0\right)^{-1}e^{\frac{K}{2}} \;q_I X^I W^+_{\mu\nu} \;.
\label{qiwpwmbot}
\ee 
The coefficients in front of $W^{\pm}_{\mu\nu}$ are the central charge and its conjugate for electric states.

In \cite{Hattab:2025aok}, it was noted that it is informative to relax the condition on the charges to integer, and allow them to be complex 
\be 
q_I \in \mathbb{C} \;.
\ee 
One immediate consequence is that there is not one but two central charges now associated to electric charges because the coefficients in front of $W^{\pm}_{\mu\nu}$ in (\ref{qiwpwmbot}) are no longer conjugates. We therefore distinguish them, and define chiral central charges \cite{Hattab:2025aok}, as
\be 
\sqrt{2\pi}\;q_I F^{I}_{\mu\nu} = \frac14 \overline{Z}_- W^-_{\mu\nu} + \frac14 Z_+ W^+_{\mu\nu} \;.
\ee 
Their form is then 
\be 
Z_{-} = -i \;2\sqrt{2\pi}\;\left(X^0\right)^{-1}e^{\frac{K}{2}} \;\bar{q}_I X^I \;,\;\; Z_{+} = -i\;2\sqrt{2\pi}\; \left(X^0\right)^{-1}e^{\frac{K}{2}} \;q_I X^I \;.
\label{chicentch}
\ee 
For integer (or real) charges $q \in \mathbb{Z}$ we have $Z_+=Z_-$, while for complex charges we have $Z_+ \neq Z_-$.

Around type II loci we can expand the chiral central charges in the same way that we expanded the central charge in (\ref{gencentypii}). We can then define type $\left(n,m\right)$ states as a generalisation of (\ref{typendefq}) through
\bea
\text{Type $\left(n,m\right)$} \;&:&\; {\bf q} \cdot {\bf X}^{(n)} \neq 0  \;\;\text{and} \;\; {\bf q} \cdot {\bf X}^{(i)} = 0 \;\;,\; \forall \; i < n \;, \nn \\
& \;&\; {\bf \overline{q}} \cdot {\bf X}^{(m)} \neq 0  \;\;\text{and} \;\; \overline{{\bf q}} \cdot {\bf X}^{(j)} = 0 \;\;,\; \forall \; j < m \;.
\label{typendefqpnm}
\eea 
The central charges of type $\left(n,m\right)$ states then behave as
\bea 
\text{Type $\left(n,m\right)$} \;:\;
Z_+ &=& -i\;2\sqrt{2\pi}\;\left(X^0\right)^{-1}e^{\frac{K}{2}} \;{\bf q} \cdot {\bf X}^{(n)} \;s^{n} + ... \;, \nn \\
Z_- &=& -i\;2\sqrt{2\pi}\;\left(X^0\right)^{-1}e^{\frac{K}{2}} \;{\bf \overline{q}} \cdot {\bf X}^{(m)} \;s^{m} + ... \;.
\label{typencenchnm}
\eea 

We can now utilise the results of section \ref{sec:typeii0logens} for type II$_0$ loci. In particular, (\ref{ranknii0def}) implies that for a rank $n$ type II$_0$ locus there is always a type $\left(0,n\right)$ state, with $n \geq 1$, of charge ${\bf q}^{(b)}$, 
\be 
\text{Type} \;\left(0,n\right) \;:\; {\bf q}^{(b)}\;.
\label{typeongenqb}
\ee  
The central charges of the state are then given by
\bea 
\text{Type $\left(0,n\right)$} \;:\;
Z_+ &=& -i\;2\sqrt{2\pi}\;\left(X^0\right)^{-1}e^{\frac{K}{2}} \; {\bf q}^{(b)} \cdot {\bf X}^{(0)}   + ... \;, \nn \\
Z_- &=& -i\;2\sqrt{2\pi}\;\left(X^0\right)^{-1}e^{\frac{K}{2}} \;{\bf \bar{q}}^{(b)} \cdot {\bf X}^{(n)}  \;s^{n} + ... \;.
\label{typencenchnspiis}
\eea 

The leading dependence of the prepotential on $s$ takes the form (\ref{FBformb2}). Using the definition of the chiral central charges (\ref{chicentch}), we see that it can be written in terms of the central charges of the $(0,n)$ state (\ref{typencenchnspiis}) as 
\be 
F = -\frac{i}{16 \pi^2} \; \left(X^0\right)^2e^{-K} Z_+ Z_- \log s + ... \;.
\label{FBformb2f}
\ee
The form of the prepotential (\ref{FBformb2f}) is the same as the one derived for the one-parameter K-point in \cite{Hattab:2025aok}. The results of this section show that it is general for type II$_0$ loci. 

In \cite{Hattab:2025aok} it was pointed out that (\ref{FBformb2f}) matches the form of a threshold correction to the prepotential. For example, as occurs around the conifold locus. More precisely, we should write (\ref{FBformb2f}) in the general form for a type $\left(m,n\right)$ state as
\be 
F =-\frac{i}{16 \pi^2}\;\frac{1}{m+n} \; \left(X^0\right)^2e^{-K} Z_+ Z_- \;\log \left( \left(X^0\right)^2e^{-K} Z_+ Z_- \right) + ... \;.
\label{FBformb2fgen}
\ee
In the case of the conifold, we have $Z_+=Z_-$ and therefore it is of type $\left(1,1\right)$. 

We have therefore shown that the threshold correction form (\ref{FBformb2f}) holds true for any type II$_0$ locus, and not only K-points. The physical meaning of this remains unclear as yet, but we return to it in section \ref{sec:hetdualii0}.


In \cite{Hattab:2025aok} also the behaviour of the gauge kinetic matrix was matched to the BPS state charges as
\be 
{\cal N}_{IJ} = -\frac{1}{2 \pi i} \Big( {\bf q}^{(b)}_I \;\bar{{\bf q}}^{(b)}_J + \bar{{\bf q}}^{(b)}_I\; {\bf q}_J^{(b)} \Big) \log s + ... \;.
\label{Ntoqide}
\ee 
Again, this is the form of a threshold correction. We see from (\ref{NBformb2}) that this identification holds generally for type II$_0$ loci if ${\bf M}^{(0,0)}=0$, as conjectured to be true in (\ref{typeii0m0vani}).

\subsection{Gravitational and matter gauge fields}
\label{sec:mattergrav}

As discussed in section \ref{sec:sugra}, the gauge fields in the supergravity split into a graviphoton, which is part of the gravitational multiplet, and matter gauge fields, that are part of the vector multiplets. This decomposition depends on the point in moduli space. Further, it is a holomorphic, rather than symplectic, decomposition. Around type II loci in moduli space, this decomposition explains nicely the physical role played by the ${\bf M}$ matrix (\ref{NBformpre}). 

The central concept is the projection to the graviphoton (\ref{gravproj}). If we wish to identify the non-gravitational gauge field directions, then they should be in the kernel of the projection (\ref{gravproj}). Define the vector of anti self-dual components of the field strengths, as in (\ref{fieldstrengvectors}), by
\be 
{\bf F}^{I,-}_{\mu\nu} = F^{I,-}_{\mu\nu} \;,
\ee 
so that the projection (\ref{gravproj}) can be written as
\be 
{\bf F}^{-}_{\mu\nu} = \frac{i}{2} \left(\bar{X}^0\right)^{-1}e^{\frac{K}{2}} \;\overline{{\bf X}} \;W^-_{\mu\nu} \;.
\label{gravbackvecfor}
\ee 
Then the vector 
\be 
{\bf F}^{-}_{{\bf M},\mu\nu} = {\bf \overline{M}} \cdot {\bf F}^{-}_{\mu\nu} \;,
\label{fmMvecdf}
\ee 
is in the kernel of the projection (\ref{gravbackvecfor}), which follows from (\ref{MXcond0full}). Note that (\ref{fmMvecdf}) is a projection, since ${\bf M}$ is not of maximal rank.\footnote{Mathematically, it is only a rank-deficient map. A projection would satisfy ${\bf M}^2 = {\bf M}$, which is not generally the case. Nonetheless, we use the term projection for the map, to make it clear that its image does not span the space.} It is a projection to a subspace which is orthogonal to the graviphoton. 

There is also the conjugate map for the self-dual components
\be 
{\bf F}^{+}_{{\bf M},\mu\nu} = {\bf M} \cdot {\bf F}^{+}_{\mu\nu} \;.
\label{fmMvecdfsd}
\ee 
However, because ${\bf M}$ is complex, there is no map which acts only on the electric field-strengths. That is, in general, the directions orthogonal to the graviphoton involve the electric and magnetic field strengths. 

We therefore see that the matrix ${\bf M}$ acts a defining the direction of the gauge fields which is purely matter, so orthogonal to the graviphoton. 

Is there a way to diagonalize the kinetic terms so as to isolate the graviphoton direction from the matter directions? Because ${\bf M}$ is not invertible, there is no simple action on the electric field strengths to do so. However, there is a close analogue to such a diagonalization. Consider the magnetic field strengths (\ref{GNFminrel}), which we can write as
\be 
{\bf G}_{\mu\nu}^+ = {\cal N} \cdot {\bf F}^{+}_{\mu\nu} \;.
\label{GNFminrelpv}
\ee
Using the form of the gauge kinetic matrix (\ref{NBformpre}), we can write this as
\be 
{\bf G}_{\mu\nu}^+ = -\frac{1}{2 \pi i}\Big( \; {\bf B} \cdot {\bf F}^{+}_{\mu\nu} + {\bf F}^{+}_{{\bf M},\mu\nu} \;\Big) \log s + ... \;.
\label{NBformpreGtFM}
\ee
We see the appearance of the matter component ${\bf F}^{+}_{{\bf M},\mu\nu}$ . 
It is therefore more natural to identify the matter sector from the magnetic field strengths. This suggests that we should extract the kinetic terms from the magnetic form of the action. Using (\ref{GNFminrel}) we can write the action (\ref{n2lfacasd}) in terms of the magnetic field strengths as
\be
{\cal L}_{F} =  -\frac{i}{4} \;{\bf G}^+_{\mu\nu}\cdot\mathcal{N}^{-1} \cdot {\bf G}^{+,\mu\nu} \;+ \mathrm{h.c.} \;\;.
\label{n2lfacasdG}
\ee
Then substituting (\ref{NBformpreGtFM}) into (\ref{n2lfacasdG}), we obtain
\bea
{\cal L}_{F} 
= &-&\frac{i}{4} \left(\frac{\log s}{2 \pi i} \right)^2\;\Big[  \;\;
{\bf F}^{+}_{\mu\nu} \cdot \left({\bf B}^T \cdot\mathcal{N}^{-1} \cdot {\bf B}\right) \cdot {\bf F}^{+,\mu\nu} + 2\;{\bf F}^{+}_{\mu\nu} \cdot \left( {\bf B}^T \cdot \mathcal{N}^{-1} \cdot {\bf M} \right) \cdot {\bf F}^{+,\mu\nu} \nn \\
&+& {\bf F}^{+}_{\mu\nu} \cdot \left( {\bf M}^T \cdot \mathcal{N}^{-1} \cdot {\bf M}  \right) \cdot{\bf F}^{+,\mu\nu}
\;\;\Big] + \mathrm{h.c.} + ...
\;\;.
\label{n2lfacasdGbtFm}
\eea
The third term in (\ref{n2lfacasdGbtFm}) gives the leading kinetic term for the projected gauge fields. So we can write the associated gauge kinetic matrix as
\be 
 {\cal N}_{\bf M} = \left(\frac{\log s}{2 \pi i} \right)^2 {\bf M}^T \cdot  {\cal N}^{-1} \cdot {\bf M} + ... \;.
 \label{NMdeftra}
\ee 
This is the pull back, under the map ${\bf M}$, of the object $\left(\frac{\log s}{2 \pi i} \right)^2 {\cal N}^{-1}$. 

For type II$_0$ loci, we expect ${\bf M}^{(0,0)}=0$ , as in (\ref{typeii0m0vani}). Therefore, we have that the gauge kinetic matrix on the subspace associated to the parameter $s$ behaves as 
\be 
 {\cal N}^{\mathrm{II}_0}_M \sim {\cal O}\left(s^2\right) \;.
\label{strongcouplingii0}
\ee 
This is vanishing in the asymptotic limit, which tells us that type II$_0$ loci are associated with a strongly-coupled matter sector. 

Type II$_0$ loci also have a nice property with respect to the relation (\ref{NBformpreGtFM}). 
This is because for such loci ${\bf B}$ is integral and of rank two, and so it has an integral zero eigenvector, which we label as ${\bf p}^{(m)}$, so that
\be 
{\bf B} \cdot {\bf p}^{(m)} = 0 \;.
\ee 
We can then write (\ref{NBformpreGtFM}) as
\be 
{\bf p}^{(m)} \cdot {\bf G}_{\mu\nu}^+ = -\frac{\log s}{2 \pi i}\; {\bf p}^{(m)} \cdot{\bf F}^{+}_{{\bf M},\mu\nu}  + ... \;.
\label{NBformpreGtFMqe}
\ee
This tells us that the combination ${\bf p}^{(m)} \cdot {\bf G}_{\mu\nu}^+ $ is exactly along the matter direction. This is an important result, which is utilized in section \ref{sec:hetdualii0}. 
%


\section{A two-parameter example: the $\mathbb{P}_{[1,1,2,2,6]}$ Calabi-Yau}
\label{sec:modspace}

In this section we study aspects of a two-parameter Calabi-Yau with a type II$_0$ locus in moduli space. The example is one of the most studied in the literature, starting from the pioneering work in \cite{Candelas:1993dm,Hosono:1994ax}. It plays a central role in our understanding of duality with the Heterotic string \cite{Kachru:1995wm,Kachru:1995fv,Klemm:1995tj,Kaplunovsky:1995tm,Antoniadis:1995zn}, a topic that we return to in section \ref{sec:hetdual}. Recently it was revisited in the context of the Emergent String Conjecture \cite{Lee:2019wij} and more generally the Swampland \cite{Blumenhagen:2018nts,Friedrich:2025gvs,Monnee:2025ynn,Monnee:2025msf,Blumenhagen:2023tev,Blumenhagen:2025zgf,Castellano:2026bnx}. The existence of a type II$_0$ locus was pointed out for the very similar  $\mathbb{P}_{[1,1,2,2,2]}$ model already in \cite{Grimm:2018cpv} and studied in detail in \cite{Grimm:2018cpv,Bastian:2021eom}. 
Even though the example has been studied extensively, in this section we present a number of new results and perspectives. In particular, the gauge kinetic matrix around the type II locus is calculated including higher order corrections. These play an important role in the physics discussion of section \ref{sec:hetdualii0}. While our primary focus is on the type II$_0$ locus, we find that already at large complex-structure there is interesting physics to understand.

\begin{figure}
\centering
 \includegraphics[width=1.0\textwidth]{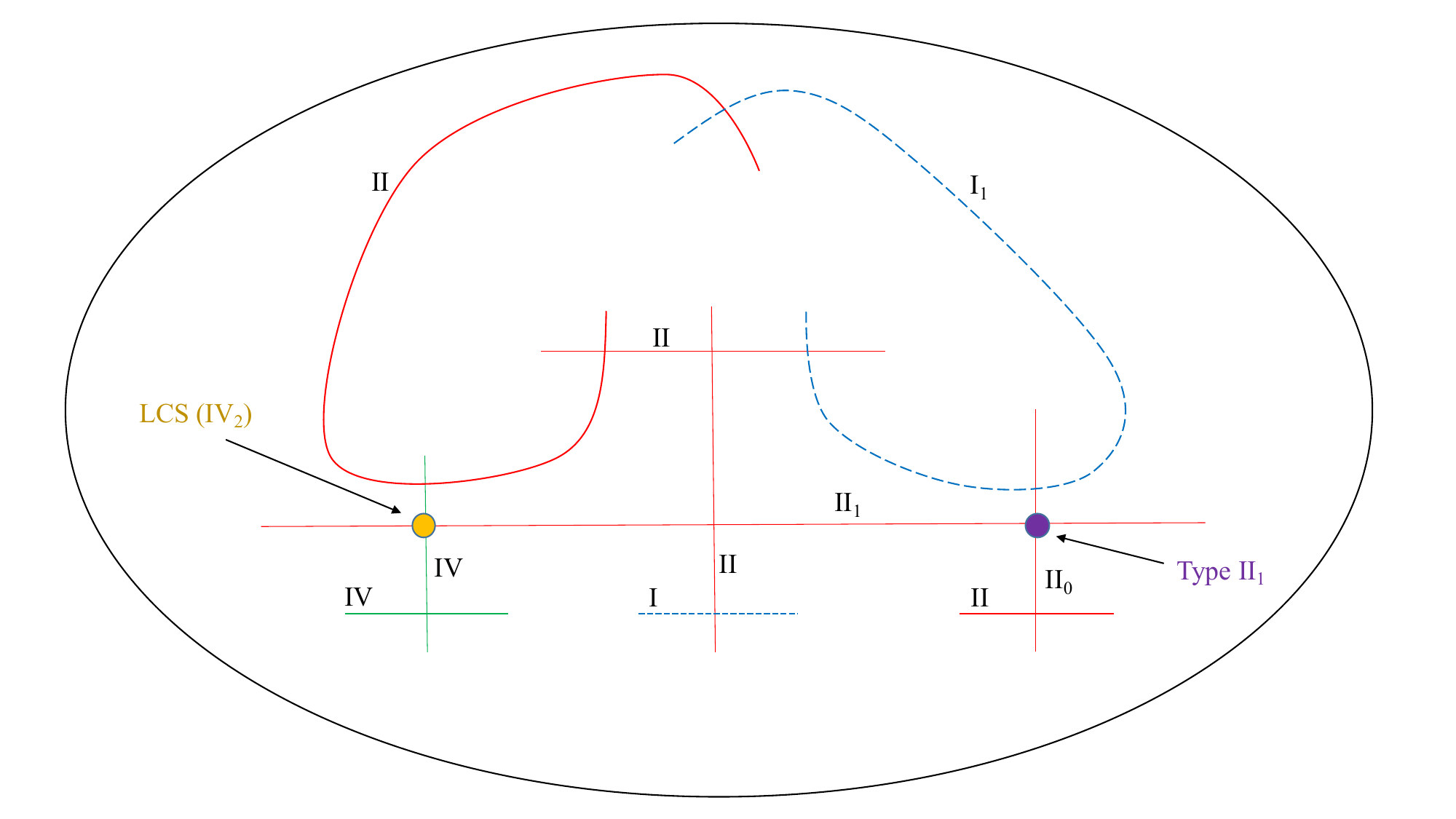}
\caption{Figure showing the complex-structure moduli space (same as in \cite{Grimm:2018cpv}). We consider local expansions about two points: the LCS point and the type II point.}
\label{fig:modulispace}
\end{figure}

We consider type IIB string theory compactified on the Calabi-Yau manifold given by the hypersurface
\be 
x_1^{12} +x_2^{12} + x_3^{6} + x_4^6 + x_5^2 - 12\; \psi \; x_1 x_2 x_3 x_4 x_5 - 2\; \phi \; x_1^6 x_2^6 = 0 \;,
\ee
in $\mathbb{P}_{[1,1,2,2,6]}$. The complex-structure moduli space is globally spanned by the complex parameters $\psi$ and $\phi$. There are two local patches in the moduli space which are of interest to us. They are labelled as the LCS (large complex-structure) point and the type II point. Figure \ref{fig:modulispace} shows a description of the moduli space, with the various singular loci and their types according to the classification of \cite{Grimm:2018ohb,Grimm:2018cpv}. 


In specifying the moduli space it is useful to work with the period vector $\Pi$. 
The period vector is specified up to symplectic transformations $S \in Sp\left(6,\mathbb{Z}\right)$. Period vectors $\Pi$ which are related by an $Sp\left(6,\mathbb{Z}\right)$ transformation are said to both be in an integral basis. When moving to different regions in moduli space, it is useful to perform $Sp\left(6,\mathbb{Z}\right)$ transformations to a particular basis. The overall integral basis is fixed in the large complex-structure region, which has a simple type IIA geometric mirror in which the quantization condition can be deduced. Once this is fixed, all other regions are in an integral basis if the period vector is related by $Sp\left(6,\mathbb{Z}\right)$ transforms from the large complex-structure vector. The $Sp\left(6,\mathbb{Z}\right)$ group is identified with the supergravity electric-magnetic duality group, as discussed extensively in sections \ref{sec:sugra} and \ref{sec:ii0loci}. 

\subsection{The large complex-structure region}
\label{sec:lcsp}

The large complex-structure (LCS) patch is given by local coordinates
\be 
z_1 = -\frac{2 \phi}{\left(1728\right)^2\psi^6}\;\;,\;\; z_2 = \frac{1}{4 \phi^2}\;\;.
\label{loclcs}
\ee 
The patch is controlled in the limit $z_1 \rightarrow 0$ and $z_2 \rightarrow 0$. The period vector for the patch is calculated in detail in appendix \ref{app:lcs}. In this section we mostly summarize the results.  

The physics at large complex-structure depends on the relative magnitudes of $z_1$ and $z_2$. There are two phases, and for each phase there is an appropriate $Sp\left(6,\mathbb{Z}\right)$ basis. At large complex-structure, there is a related, but slightly different, notion of an electric basis relative to the definition for type II loci in (\ref{monN}). The natural electric basis does have logarithmic terms in the electric periods, but such that any electrically charged state is lighter than any magnetically charged state. That is, at large complex-structure we have an electric basis where
\be 
\text{Electric basis at LCS\;\;\;:\;\;\;} \frac{q_I X^I}{p^J F_J} \xrightarrow[z_i \rightarrow 0]{} 0 \;\;,\;\; \forall \;q_I,p^I \in \mathbb{Z} \;. 
\label{eleatlcs}
\ee 
There are then two electric bases, depending on the relative magnitudes of $z_1$ and $z_2$. They corresponds to the natural frames describing a mirror microscopic type IIA string theory picture versus a microscopic Heterotic string picture. In the IIA picture, we would like D0 and wrapped D2 branes to be lighter than wrapped D4 and D6 branes, while in the Heterotic description we would like to have wrapped fundamental string states lighter than wrapped NS5 brane states. We discuss each basis in turn. 

\subsubsection*{The IIA electric basis}

We choose a symplectic basis where the electric periods are labelled as
\be
\Pi^{\mathrm{A}}_{\mathrm{LCS}} = \left( \begin{array}{c}
1 \\ 
T \\
S \\
\cdot  \\
  \cdot  \\ 
 \cdot
 \end{array}\right) \;.
\label{perveclcsA}
\ee
The two non-trivial electric periods have an expansion in terms of the local coordinates of 
\bea 
T &=& \frac{\log z_1}{2\pi i} + ... \;\;, \nn \\
S &=& \frac{\log z_2}{2\pi i} + ...\;,
\label{STlcs}
\eea 
where the ellipses denote terms that are holomorphic in $z_1$ and $z_2$. More detailed expressions are given in appendix \ref{app:lcs}.
The full periods vector can then be written as
\bea 
\Pi^{\mathrm{A}}_{\mathrm{LCS}} = \left( \begin{array}{c}
1 \\ 
T \\
S \\
S + S \;T^2 +\frac{2}{3} \;T^3 +\frac{13}{6}\; T  - \frac{63 i \zeta[3]}{2 \pi^3} + ... \\
 - 2 \;S \;T- 2\; T^2+\frac{13}{6} + ... \\ 
 -T^2  +1 + ...
 \end{array}\right) \;.
\label{perveclcsA}
\eea 
This period vector is associated with the prepotential 
\be 
F_{\mathrm{LCS}}^{\mathrm{A}} = S\;\left( 1 - T^2 \right) -\frac23 \left(T\right)^3 + \frac{13}{6} T - \frac{252i\zeta(3)}{2\left(2\pi\right)^3} + ... \;.
\label{lcsprepoA}
\ee 
The form of the prepotential (\ref{lcsprepoA}) is matched onto the infinite volume limit of the mirror Calabi-Yau, which fixes the rational coefficients in terms of topological properties of the mirror Calabi-Yau. This, in turn, fixes the integral basis of periods at large complex-structure. The integral basis of periods anywhere else in the moduli space is then related to the one at large complex-structure by $Sp\left(6,\mathbb{Z}\right)$ transformations, thereby maintaining integrability. Their exact form are given in \ref{app:pf}.

The period vector (\ref{perveclcsA}) is in an electric gauge (\ref{eleatlcs}) if we work in what we denote the IIA frame 
\be 
\text{IIA frame} \;\;:\;\; (\Im\,T)^2 \gg \Im\,S \;.
\label{IIAfrST}
\ee 
The monodromy matrices in this basis, associated to $S$ and $T$ respectively, are given by
\be 
N_T^{\mathrm{A}} = \left( \begin{array}{cccccc} 
0 & 0 & 0 & 0 & 0 & 0 \\ 
1 & 0 & 0 & 0 & 0 & 0 \\
0 & 0 & 0 & 0 & 0 & 0 \\
5 & 2 & 1 & 0 & -1 & 0 \\
-2 & -4 & -2 & 0 & 0 & 0 \\
-1 & -2 & 0 & 0 & 0 & 0 
  \end{array} \right) \;\;,\;\; N_S^{\mathrm{A}} = \left( \begin{array}{cccccc} 
0 & 0 & 0 & 0 & 0 & 0 \\ 
0 & 0 & 0 & 0 & 0 & 0 \\
1 & 0 & 0 & 0 & 0 & 0 \\
2 & 0 & 0 & 0 & 0 & -1 \\
0 & -2 & 0 & 0 & 0 & 0 \\
0 & 0 & 0 & 0 & 0 & 0 
  \end{array} \right)  \;\;.
\ee 
Since $\left(N_T^{\mathrm{A}}\right)^3 \neq 0$ and has rank four, we can identify the $T \rightarrow i\infty$ locus as type IV$_2$. The $S \rightarrow i\infty$ locus is instead of type II$_1$. The singular loci intersect at the large complex-structure point, which is of type IV$_2$. 

\subsubsection*{The Heterotic electric basis}

At $\Im\,S \sim (\Im\,T)^2$ there is a strong-coupling regime where one of the gauge couplings becomes of order one. It is simplest to see this directly from the period vector where an electric and a dual magnetic period become of the same magnitude. In the opposite regime $ \Im\,S \gg (\Im\,T)^2$, we have a new weakly-coupled description in a different electric frame. This frame we denote as the Heterotic frame 
\be 
\text{Heterotic frame} \;\;:\;\; \Im\,S \gg (\Im\,T)^2 \;.
\label{HfrST}
\ee 
The period vector in the Heterotic frame takes the leading form (including the first exponential correction in $S$) 
\bea 
\Pi^{\mathrm{H}}_{\mathrm{LCS}} = \left( \begin{array}{c}
1 \\ 
T \\
1-T^2 +\frac{1}{2\pi^2}e^{2\pi iS } + ...\\
S + S \;T^2 +\frac{2}{3} \;T^3 +\frac{13}{6}\; T  - \frac{63 i \zeta[3]}{2 \pi^3}+e^{2\pi i S}\left(\frac{1}{2i\pi^3}-\frac{S}{2\pi^2}\right) + ...\\
 - 2 \;S \;T- 2\; T^2+\frac{13}{6} + ... \\ 
 -S
 \end{array}\right) \;.
\label{perveclcsH}
\eea 
It is related to the IIA basis period vector by the symplectic transformation $S^{\mathrm{HA}}$ as
\be 
\Pi^{\mathrm{H}} = S^{\mathrm{HA}} \cdot \Pi^{\mathrm{A}} \;,
\ee 
where
\be 
\label{Hetmap}
S^{\mathrm{HA}}=
\left( \begin{array}{cccccc} 
1 & 0 & 0 & 0 & 0 & 0 \\ 
0 & 1 & 0 & 0 & 0 & 0 \\
0 & 0 & 0 & 0 & 0 & 1 \\
0 & 0 & 0 & 1 & 0 & 0 \\
0 & 0 & 0 & 0 & 1 & 0 \\
0 & 0 & -1 & 0 & 0 & 0 
  \end{array} \right) \;.
\ee
This is a simple interchange of the third and sixth periods. 

Because $S^{HA}$ is not block diagonal, it modifies the prepotential as in (\ref{prepottrans}), which now reads at leading order
\be 
F_{\mathrm{LCS}}^{\mathrm{H}} =  -\frac23 \left(T\right)^3 + \frac{13}{6} T - \frac{252i\zeta(3)}{2\left(2\pi\right)^3} + e^{2\pi i S}\left(\frac{1}{4i\pi^3}-\frac{S}{2\pi^2}\right)+... \;.
\label{lcsprepoH}
\ee 
In this frame, the variable $S$ does not appear polynomially in the prepotential, and only appears exponentially as $e^{2 \pi i S}$. 

The monodromy matrices in this basis, associated to $S$ and $T$ respectively, transform as 
\be 
N_T^{\mathrm{H}} = S^{HA} \cdot N_T^{\mathrm{A}}  \cdot \left(S^{HA}\right)^{-1} \;\;,\;\; N_S^{\mathrm{H}} = S^{HA} \cdot N_S^{\mathrm{A}}  \cdot \left(S^{HA}\right)^{-1} \;,
\ee 
and are given by
\be 
N_T^{\mathrm{H}} = \left( \begin{array}{cccccc} 
0 & 0 & 0 & 0 & 0 & 0 \\ 
1 & 0 & 0 & 0 & 0 & 0 \\
-1 & -2 & 0 & 0 & 0 & 0 \\
5 & 2 & 0 & 0 & -1 & -1 \\
-2 & -4 & 0 & 0 & 0 & 2 \\
0 & 0 & 0 & 0 & 0 & 0 
  \end{array} \right) \;\;,\;\; N_S^{\mathrm{H}} = \left( \begin{array}{cccccc} 
0 & 0 & 0 & 0 & 0 & 0 \\ 
0 & 0 & 0 & 0 & 0 & 0 \\
0 & 0 & 0 & 0 & 0 & 0 \\
2 & 0 & -1 & 0 & 0 & 0 \\
0 & -2 & 0 & 0 & 0 & 0 \\
-1 & 0 & 0 & 0 & 0 & 0 
  \end{array} \right)  \;\;.
\ee 
The matrix $N_S^{\mathrm{H}}$ is now in an electric frame form (\ref{monN}), and can be matched to the general discussions in section \ref{sec:ii0loci} on type II loci, with the matrix ${\bf B}_{\mathrm{LCS}}$ taking the form
\be 
{\bf B}_{\mathrm{LCS}}= \left( \begin{array}{ccc} 
  -2 & 0 & 1 \\ 
  0 & 2 & 0 \\ 
  1 & 0 & 0 \\ 
   \end{array} \right)\;.
\ee 
It has the expected structure for a type II$_1$ locus, being of rank 3 with two positive eigenvalues and one negative one. Note that approaching a generic point on the type II$_1$ locus corresponds to the Heterotic regime (\ref{HfrST}). The IIA regime corresponds to approaching a generic point on the type IV$_2$ locus instead.  
The leading $S$ contribution to the gauge kinetic matrix, can be extracted from \eqref{ggmHLCS} and written in the form \eqref{NBform}
\begin{equation}
    \mathcal{N}_{LCS}^H = -\left({\bf B}_{\mathrm{LCS}}+{\bf M}^{(0)} \right)S+\cdots\;.
    \label{Nhlcs}
\end{equation}
The matrix ${\bf M}^{(0)}$ takes the form, when evaluated at $\mathrm{Re\;}T=0$ , of 
\begin{equation}
   {\bf M}^{(0)} = \left( \begin{array}{ccc} 
  2+(\Im\,T)^2+\frac{1}{(\Im\,T)^2} & 0 & -1-\frac{1}{(\Im\,T)^2} \\ 
  0 & 0 & 0 \\ 
  -1-\frac{1}{(\Im\,T)^2} & 0 & \frac{1}{(\Im\,T)^2} \\ 
   \end{array} \right) \;,
\end{equation}
which indeed vanishes when contracted with $X = (1,T,1-T^2)$.

\subsection{The type II region}
\label{sec:ii0wp}

The type II patch has local coordinates
\be 
s^2 = \frac{1}{864\;\psi^6} \left(\phi + 864 \psi^6\right)\;\;,\;\; t = \frac{1}{\left(\phi + 864 \psi^6\right)^2}\;\;,
\label{lociiw}
\ee 
which in terms of the local coordinates we used in section \ref{sec:lcsp} read
\begin{equation}
    s^2 = 1-1728z_1\;\;,\;\; t = \frac{4z_2 (1728z_1)^2}{(1-1728z_1)^2}\;\;.
    \label{stdeftyII}
\end{equation}

The period vector is calculated in appendix \ref{app:ii}. It can be expressed locally in the Heterotic symplectic frame (so the same frame as the large complex-structure Heterotic frame) as
\bea 
\Pi^{\mathrm{H}}_{\mathrm{II}} = \left( \begin{array}{c}
X^0 \\ 
X^1 \\
X^2\\
F_0\\
F_1\\ 
F_2
 \end{array}\right) \;,
\label{perveclcsAX}
\eea 
with
\bea
\label{Hbasis}
    2\pi i \beta  X^0 &=& i - is\beta\left(2-\frac{t}{8}-\frac{15}{512}t^2\right)+is^2\left(\beta^2+\frac{5}{36}\right)+\cdots\;,\\ \nn
     2\pi i \beta X^1 &=& -1+s^2\left(\beta^2-\frac{5}{36}\right)+\cdots\;,\\ \nn
     2\pi i \beta X^2 &=& 2i+2is^2\left(\beta^2+\frac{5}{36}\right)+\cdots\;,\\ \nn
    2\pi i F_0 &=& (2 X^0-X^2) \log \left(t/2^6\right)+g_0\;,\\ \nn
    2\pi i F_1 &=&-2 X^1 \log \left(s^4 t\right)+g_1 \;, \\ \nn
    2\pi i F_2 &=& -\frac{X^2}{2} \log(s^4 t)-\left(X^0-\frac{X^2}{2}\right) \log \left(t/2^6\right)+g_2\;,
\eea
where 
\be 
\beta=\frac{\Gamma(\frac34)^4}{\sqrt{3}\,\pi^2} \;,
\ee
and $g_0$, $g_1$ and $g_2$ are holomorphic power series in $s$ and $t$ given in (\ref{g012}).

The map from a set of solutions to the Picard-Fuchs system to the integral large complex structure and Heterotic bases is described in detail in Appendix~\ref{app:ii}. Note that this map appeared already in \cite{Curio:2000sc,Lee:2019wij,Monnee:2025msf}. However, we find slightly different results to these, especially with regards to an integral basis.\footnote{We find better agreement with a recent analysis in \cite{Castellano:2026bnx} , which appeared while this work was in preparation.} 

This choice of integral basis leads to a prepotential $F$ of the form
\bea
     F &=&-\frac{1}{2\pi i}\left(4(X^1)^2+(X^2)^2\right)\log s \nn \\
     &-&\frac{1}{2\pi i }\left((X^1)^2-(X^0)^2+X^0X^2\right)\log t \nn \\
     &-&\frac{1}{8\pi i }\left(4(X^0)^2+(X^2)^2-4X^0X^2\right)\log 64+\frac{X^i g_i}{4\pi i }\;,
     \label{prepotentiatyepIIf}
\eea
which expressed locally gives
\bea
    (2\pi i)^3F &= &\;s^2\left(16+\frac{368 s^2}{27}+\frac{154 s^2 t}{27}+\cdots\right)\log s \nn \\
    &+& s^2\left(\frac{ t}{2}+s^2 t+\frac{13 t^2}{128}+\frac{23 s^2 t^2}{384}+\cdots\right)\log t+\cdots\;.
    \label{Fintermsst}
\eea

We can also give the leading behaviour of the Kahler potential as
\be 
e^{-K} = -\frac{4}{\pi} \log \left|s^4t \right| +... \;.
\label{Kahlerpotentypiif} 
\ee 

In this basis the monodromy matrices around $s=0$ and $t=0$ are given by
\begin{equation}
    N_s = \left( \begin{array}{cccccc} 
0&0&0&0&0&0 \\ 
0&0&0&0&0&0 \\
0&0&0&0&0&0 \\
0&0&0&0&0&0 \\
0&-8&0&0&0&0 \\
0&0&-2&0&0&0
  \end{array} \right)\;\;,\;\;  N_t = \left( \begin{array}{cccccc} 
0&0&0&0&0&0 \\ 
0&0&0&0&0&0 \\
0&0&0&0&0&0 \\
2&0&-1&0&0&0 \\
0&-2&0&0&0&0 \\
-1&0&0&0&0&0
  \end{array} \right)\;.
\end{equation}

The leading logarithmic behavior of the gauge kinetic matrix in the Heterotic frame near the Type II point, written in the form \eqref{NBform}, is
\be
    \mathcal{N}_{\mathrm{II}}^H = -\frac{\log t}{2\pi i }\;\Big( {\bf B}_t+ {\bf M}^{(0,0)}_t \Big)-\frac{\log s}{2\pi i}\;{\bf B}_s + ...\;,
    \label{gkmiiiiside}
\ee
with 
\be
\begin{array}{ccc}
     {\bf B}_t = \left( \begin{array}{ccc} 
-2&0&1 \\ 
0&2&0 \\
1&0&0 
  \end{array} \right)\;\;,&
  {\bf B}_s = \left( \begin{array}{ccc} 
0&0&0 \\ 
0&8&0 \\
0&0&2 
  \end{array} \right)\;\;, 
   & 
   {\bf M}_t^{(0,0)} = \left( \begin{array}{ccc} 
4&0&-2 \\ 
0&0&0 \\
-2&0&1 
  \end{array} \right)\;\;.
  \end{array}
    \label{bmtdef}
\ee
It is useful to note that the matrices satisfy the linear relation
\be 
4\;{\bf B}_t + 2\;{\bf M}^{(0,0)}_t = {\bf B}_s \;.    
\label{linearrel}
\ee 
The eigenvalues of the matrices are 
\be
\def\arraystretch{1.5}
\begin{array}{|c|c|c|c|}
\hline
\text{Matrix} & \multicolumn{3}{c|}{\text{Eigenvalues}} \\
\hline
{\bf B}_t &  -1-\sqrt{2} &  2 &  -1+\sqrt{2} \\
\hline
{\bf M}_t^{(0,0)} &  5 & 0 & 0 \\
\hline
{\bf B}_t + {\bf M}_t^{(0,0)} &  \frac12\left(3+\sqrt{5} \right) & 2 & \frac12\left(3-\sqrt{5} \right)  \\
\hline
{\bf B}_s &  8 & 2 & 0 \\
\hline
\end{array}
\label{mateign}
\ee 

We can also write the higher order corrections to the gauge kinetic matrix in the form (\ref{NBformpre}) as
\be
    \mathcal{N}_{\mathrm{II}}^H = -\frac{\log t}{2\pi i }\;\Big( {\bf B}_t+ {\bf M}_t  \Big) -\frac{\log s}{2\pi i}\;\Big( {\bf B}_s+ {\bf M}_s  \Big) + ...  \;,
    \label{NHtypeIIstbasis}
\ee 
where we define an expansion of ${\bf M}_t$ and ${\bf M}_s$ of the form\footnote{For ease of notation we do not include logarithmic terms, going as $\log s$ and $\log t$ in this expansion, even though they are present. The first such term behaves as $s^2t\left(\log t\right)\left(\log s\right)$, and so is not important for our purposes.}
\bea
{\bf M}_t &=& \sum_{i,j,k,l} {\bf M}_t^{\left(i,j,k,l\right)} \left(\mathrm{\Re}\;t\right)^i \left(\mathrm{\Re}\;s\right)^j\left(\mathrm{\Im}\;t\right)^k\left(\mathrm{\Im}\;s\right)^l \;,
\nonumber \\ 
{\bf M}_s &=& \sum_{i,j} {\bf M}_s^{\left(i,j,k,l\right)} \left(\mathrm{\Re}\;t\right)^i \left(\mathrm{\Re}\;s\right)^j\left(\mathrm{\Im}\;t\right)^k\left(\mathrm{\Im}\;s\right)^l \;.
\label{MsMtExpa}
\eea
Note that this is a two-parameter expansion, as opposed to the one-parameter expansion as in section \ref{sec:ii0loci}. 
Explicitly, the constant matrices in (\ref{MsMtExpa}) are given by
\be
\begin{array}{cc}
{\bf M}_t^{(0,1,0,0)} = 4\beta \left( \begin{array}{ccc} 
0 & 0 & 1\\ 
0 & 0 & 0
\\
1  & 0 & -1
  \end{array} \right) 
   \;\;, 
   & 
{\bf M}_t^{(0,0,0,1)} = 4\beta \left( \begin{array}{ccc} 
0 & 2 & 0\\ 
2 & 0 & -1
\\
0  & -1 & 0
  \end{array} \right)  \;,
\\ 
  {\bf M}_t^{(1,1,0,0)} = -\frac{1}{4}\beta \left( \begin{array}{ccc} 
0 & 2 i & 2\\ 
2i & 0 & -i
\\
2  & -i & -2
  \end{array} \right) 
   \;\;,
   &
{\bf M}_t^{(0,2,0,0)} = 8\beta^2 \left( \begin{array}{ccc} 
2 & 0 & -1\\ 
0 & 0 & 0
\\
-1  & 0 & 1
  \end{array} \right)  \;,
\\
  {\bf M}_t^{(0,0,0,2)} = 8\beta^2 \left( \begin{array}{ccc} 
2 & 0 & -1\\ 
0 & 2 & 0
\\
-1  & 0 & 1/2
  \end{array} \right) 
   \;\;, 
   &
   {\bf M}_t^{(0,0,1,1)} = \frac{1}{2}\beta \left( \begin{array}{ccc} 
0 & i & 1\\ 
i & 0 & -i/2
\\
1  & -i/2 & -1
  \end{array} \right) \;, 
  \\
    {\bf M}_t^{(1,0,0,1)} =\frac{i}{2}\beta \left( \begin{array}{ccc} 
0 & 2i & 1/2\\ 
2i & 0 & -i
\\
1/2  & -i & -1/2
  \end{array} \right) 
   \;\;,
   &
    {\bf M}_t^{(0,1,1,0)} = \frac{1}{4}\beta \left( \begin{array}{ccc} 
0 & -4 & i\\ 
-4 & 0 & 2
\\
i  & 2 & -i
  \end{array} \right) \;,  
   \\
  {\bf M}_t^{(0,1,0,1)} =8\beta^2 \left( \begin{array}{ccc} 
0 & 0 &0\\ 
0 & 0 & 1
\\
0  & 1 & 0
  \end{array} \right) 
   \;\;. &  
\end{array} 
\label{Mtexpan}
\ee

\be
\begin{array}{cc} 
{\bf M}_s^{(0,1,0,0)} = 8\beta \left( \begin{array}{ccc} 
0 & 2i & 1\\ 
2i & 0 & -i
\\
1  & -i & -1
  \end{array} \right) 
   \;\;, 
   &
   {\bf M}_s^{(0,0,0,1)} = 8\beta \left( \begin{array}{ccc} 
0 & 2 & -i\\ 
2 & 0 & -1
\\
-i  & -1 & i
  \end{array} \right)  \;, 
  \\ 
  {\bf M}_s^{(1,1,0,0)} = \frac{3}{2}\beta \left( \begin{array}{ccc} 
0 & -2 i & -1\\ 
-2i & 0 & i
\\
-1  & i & 1
  \end{array} \right) 
   \;\;, 
   &
   {\bf M}_s^{(0,2,0,0)} = 4\beta^2 \left( \begin{array}{ccc} 
0 & 0 & 0\\ 
0 & 4 & 2i
\\
0  & 2i & 3
  \end{array} \right)  \;,
  \\
  {\bf M}_s^{(0,0,0,2)} = 4\beta^2 \left( \begin{array}{ccc} 
0& 0 & 0\\ 
0 & 12 & -2 i
\\
0  & -2i & 1
  \end{array} \right) 
   \;\;, 
   &
    {\bf M}_s^{(0,0,1,1)} = \frac{3}{2}\beta \left( \begin{array}{ccc} 
0 & 2i & 1\\ 
2i & 0 & -i
\\
1  & -i & -1
  \end{array} \right)  \;, 
  \\
    {\bf M}_s^{(1,0,0,1)} =\frac{3i}{2}\beta \left( \begin{array}{ccc} 
0 & 2i & 1\\ 
2i & 0 & -i
\\
1  & -i & -1
  \end{array} \right) 
   \;\;, 
   &
   {\bf M}_s^{(0,1,1,0)} = \frac{3i}{2}\beta \left( \begin{array}{ccc} 
0 & 2i & 1\\ 
2i & 0 & -i
\\
1  & -i & -1
  \end{array} \right)  \;, 
  \\
  {\bf M}_s^{(0,1,0,1)} =8\beta^2 \left( \begin{array}{ccc} 
0 & 0 &0\\ 
0 & 4i & 2
\\
0  & 2 & -i
  \end{array} \right) 
   \;\;. & 
   \end{array}
   \label{Msexpan}
\ee

This completes our computation of the relevant quantities for the supergravity around both the large complex-structure and the type II regions. We utilize these results in section \ref{sec:hetdualii0}. However, before proceeding to this, there is another important piece of the picture that should established in more detail, and that is the duality with the Heterotic string, to which we now turn. 

\section{Duality with the Heterotic string}
\label{sec:hetdual}

Type II loci in type IIB complex-structure moduli space have a natural duality map to the Heterotic string. When the type II locus is in the large complex-structure region it can be mapped to the geometric type IIA regime, and then to the Heterotic string through type IIA - Heterotic duality. Indeed, this was the first example of ${\cal N}=2$ type IIA - Heterotic duality \cite{Kachru:1995wm}. 
The duality proposed in \cite{Kachru:1995wm} was a connection between works calculating the descriptions of the type IIB/IIA side and the Heterotic side. On the type II side the relevant work is the pioneering \cite{Candelas:1993dm}, while on the Heterotic side the classic studies \cite{Antoniadis:1995ct,deWit:1995dmj}.
For type II loci that are not at large complex-structure, the duality is more naturally stated as type IIB - Heterotic duality. Such a duality featured already in \cite{Kachru:1995wm} and developed in more detail in the follow-up \cite{Kachru:1995fv} (and in many subsequent works). The duality was revisited in the context of the Swampland and the Emergent String Conjecture \cite{Lee:2019wij,Lee:2019xtm} and in \cite{Artime:2025egu,Blumenhagen:2025zgf}. More recently, in a series of works \cite{Friedrich:2025gvs,Monnee:2025ynn,Monnee:2025msf}, type IIB - Heterotic duality associated to general type II loci has been systematically studied and developed. The duality specifically for type II$_0$ loci was briefly discussed also in \cite{Hattab:2025aok}.

For our purposes, there are four types of type IIA/IIB - Heterotic dualities. The most standard, and best understood, involves at least a three-parameter moduli space and certain fibration structures for the compactification manifolds. We refer to this as $STU$ Type II-Heterotic duality, and review it in section \ref{sec:stuiihetdual}. 

The next level of complication arises in dualities which involve two parameters. The initial example in \cite{Kachru:1995wm} was of this type, and corresponds to the example Calabi-Yau studied in detail in section \ref{sec:modspace}. This type of duality then has a further split into duality at large complex-structure in IIB, and one away from large complex-structure.  The large complex-structure duality corresponds to approaching the type II$_1$ locus around the LCS point in figure \ref{fig:modulispace}. This has been proposed to be dual to the Heterotic string on a torus which has equal Kahler and complex-structure moduli \cite{Kachru:1995fv}. We discuss this duality in section \ref{sec:hetlcstypeii}. 

The duality away from large complex-structure, on the type IIB side, is the next level of complexity. Now we are considering the type II$_1$ point in figure \ref{fig:modulispace}. The duality itself is split into two types. The first is the case of approaching a generic point on the type II$_1$ locus, around the type II$_1$ point. This is the duality studied already in \cite{Kachru:1995fv}, and more recently in \cite{Monnee:2025msf}. We study this also in section \ref{sec:hetlcstypeii}. 

Finally, the most involved setting is approaching a generic point on the type II$_0$ locus, around the type II$_1$ point. This is the case of most interest to us. It was studied in \cite{Hattab:2025aok} and in \cite{Monnee:2025msf}. We discuss this case in section \ref{sec:hetdualii0}.

\subsection{The Heterotic string on $K3 \times T^2$}
\label{sec:hetonk3t2}

In this section we discuss compactifications of the Heterotic string on $K3_H \times \left(T^2\right)_H$. The subscripts $H$ are not meaningful at this point, but are useful later to distinguish from dual type IIA quantities. The discussion here is mostly review, and we refer to \cite{Antoniadis:1995ct,deWit:1995dmj} for classic original works. We consider the $E_8 \times E_8$ Heterotic string, where the gauge bundle (with field-strength $F_G$) is chosen as the standard embedding, which means it need to have 
\be 
\int_{K3_H} c_2\left(F_G\right) = 24 \;.
\label{hetbundle}
\ee 
The simplest symmetric choice is to take the instanton numbers under the two $E_8$ factors as $\left(12,12\right)$. Embedding the gauge bundle in the $E_8$ factors means that the gravitational moduli are unrestricted, and there is a universal sector in the vector multiplets denoted as $S$, $T$ and $U$. Taking the (ten-dimensional) Heterotic string coupling as $\hat{g}_{s,H}$, the Heterotic string scale as $m_s^H$, and the torus radii as $R_1$ and $R_2$, so with 
\be 
\left(T^2\right)_H = \left(S^1\right)_{R_1} \times \left(S^1\right)_{R_2} \;,
\ee 
we can identify the moduli as\footnote{For simplicity, these identifications with the torus radii are for the case where the torus factorises, so taking $\mathrm{Re\;}U=0$. This restriction is sufficient for the current analysis. The more general map is of the form $T = b + i \left(g_{11}g_{22}-g_{12}^2 \right)^{\frac12}$ and $U= \frac{g_{12}}{g_{22}} + i \frac{\left(g_{11}g_{22}-g_{12}^2 \right)^{\frac12}}{g_{22}}$, where the $g_{ij}$ denote the torus metric components and $b$ denotes the Neveu-Schwartz anti-symmetric two-form on the torus.}
\be
    \text{Im\;}S =  \frac{1}{\left(g_{s,H}\right)^2}\;,\;\;\;\;\;\;
    \text{Im\;}T =  R_1 R_2\left(2 \pi \;m^H_s\right)^2\;,\;\;\;\;\;\;
    \text{Im\;}U = \frac{R_1}{R_2} \;.
\label{STUmap}
\ee
Here $g_{s,H}$ denotes the four-dimensional Heterotic string coupling, which is related to the ten-dimensional coupling by
\be 
\left(\hat{g}_{s,H}\right)^2 = \left(g_{s,H}\right)^2 \text{Vol}\left(K3_H\right) \text{Vol}\left(T^2\right)_H\left( m_{s}^H\right)^6 \;.
\ee 
The Heterotic string scale is related to the four-dimensional Planck scale as
\be 
\left(2 \pi \;m_s^H\right)^2 = \pi \left(g_{s,H}\right)^2 M_p^2 =\frac{\pi}{\text{Im\;}S}\;M_p^2\;.
\label{mstomprel}
\ee

The infrared four-dimensional supergravity theory has a gauge (sub-)group $U(1)^4$ associated to the three vector multiplets and the graviphoton. It is simple to identify the appropriate charged BPS states at the microscopic level in an appropriate symplectic basis. In this basis, we denote the electric field strengths as $F^I$ and the magnetic ones as $G_I$. The states and their masses, in Heterotic string units, are shown in table \ref{tab:HetBPS}.  
\begin{table}
\begin{center}
\def\arraystretch{1.8}
	\begin{tabular}[t]{|c|c|c|c|c|c|}
	\hline
	 Gauge field & State & $K3_H$ & $R_1$ & $R_2$ & $\left(\frac{M}{2\pi m_s^H}\right)^2$ \\
	 \hline
	$F^0$ & F1KK &  & x & & $\frac{1}{\left(\mathrm{Im\;}T\right)\left(\mathrm{Im\;}U\right)}$ \\
	\hline
	$F^1$ & F1W  &  &  & x & $\frac{\mathrm{Im\;}T}{\mathrm{Im\;}U}$ \\
	\hline
	$F^2$ & F1KK  &  &  & x & $\frac{\mathrm{Im\;}U}{\mathrm{Im\;}T}$ \\
	\hline
	$F^3$ & F1W  &  & x & & $\left(\mathrm{Im\;}T\right)\left(\mathrm{Im\;}U\right)$ \\
	\hline
	$G_0$ & KK5 & x & & x & $\left(\mathrm{Im\;}S\right)^2 \left(\mathrm{Im\;}T\right)\left(\mathrm{Im\;}U\right)$ \\
	\hline
	$G_1$ & NS5 & x & x & & $\frac{\left(\mathrm{Im\;}S\right)^2 \left(\mathrm{Im\;}U\right)}{\mathrm{Im\;}T}$ \\
	\hline
	$G_2$ & KK5 & x & x & & $\frac{\left(\mathrm{Im\;}S\right)^2 \left(\mathrm{Im\;}T\right)}{\mathrm{Im\;}U}$ \\
	\hline
	$G_3$ & NS5 & x &  & x & $\frac{\left(\mathrm{Im\;}S\right)^2}{\left(\mathrm{Im\;}T\right)\left(\mathrm{Im\;}U\right)}$ \\
	\hline
	\end{tabular}
\end{center}
\caption{Table showing the schematics structure of BPS states in the Heterotic string theory. The x's denote directions, either wrapping or KK, associated to the states. F1KK and F1W denote KK and Winding modes of the fundamental Heterotic string, while NS5 denotes NS5-branes and KK5 denotes KK monopoles.}
\label{tab:HetBPS}
\end{table}

In the four-dimensional supergravity, there are no charged states under the gauge fields, they have been integrated out. We are therefore free to perform any $Sp\left(8,\mathbb{Z}\right)$ transformation. However, in a microscopic description there are fixed electric dynamical degrees of freedom, which partially fixes this freedom. The restriction for a microscopic description can be most simply studied by the spectrum in table \ref{tab:HetBPS}. We can obtain the four-dimensional supergravity from a ten-dimensional supergravity so that the KK modes become dynamical. That requires the restrictions 
\be
\text{Ten-dimensional supergravity} \;:\;\mathrm{Im\;}T \gg \mathrm{Im\;}U \gg 1 \;,\; \mathrm{Im\;}S \gg \left(\mathrm{Im\;}T\right)\left(\mathrm{Im\;}U\right)  \;\;.
\label{tendhetsup}
\ee 
These ensure that the lightest charged BPS states are KK modes of the torus. In this description all the non-perturbative states, whose mass goes as $\mathrm{Im\;}S$ are heavy and have been integrated out. Also, the winding modes are heavy, and are not dynamical. 

The dynamical charged states are only charged under $F^0$ and $F^2$, and therefore we still have a symmetry group acting on $F^1$, $F^3$ and their magnetic duals
\be 
G^{\mathrm{Uncharged}}=Sp\left(4,\mathbb{Z}\right) \;,
\ee 
which is unfixed by this microscopic description. In particular, this symmetry includes the exchange 
\be 
F^3 \leftrightarrow G_3\;.
\label{F3G3ex}
\ee 
If we act with this exchange, then we see that the electric degrees of freedom have masses of the form $\left(1,T,U,S\right)$, up to an overall normalizing factor of $ \left(\mathrm{Im\;}T\right)^{-\frac12}\left(\mathrm{Im\;}U\right)^{-\frac12}$. This electric frame is the famous $STU$ frame, and in the supergravity corresponds to a prepotential of the form
\be 
F = -STU + ... \;.
\label{FSTUcase}
\ee 

On the other hand, if we do not act with the symplectic exchange (\ref{F3G3ex}), then the electric states have masses that are independent of $S$ at the perturbative level. This means that the dependence on $S$ of the prepotential is purely exponential 
\be
F = Se^{iS} + ... \;.
\label{expShetgen} 
\ee 
In the early works, \cite{Ceresole:1995jg,deWit:1995dmj}, this situation of an exponential prepotential in $S$ was called a frame where the prepotential does not exist. This is somewhat of a misnomer, the prepotential does exist and plays the standard role of a prepotential. What does not exist is a purely perturbative prepotential.

We can relax one of the conditions in (\ref{tendhetsup}), and give up a ten-dimensional supergravity description. We can still consider a worldsheet fundamental string description. In this setting we can, in generality, consider $\mathrm{Im\;}T \geq 1$ and $\mathrm{Im\;}U \geq 1$. This allows us to write the required condition as
\bea 
\text{F1 Worldsheet} \;&:&\; \mathrm{Im\;}S \gg \left(\mathrm{Im\;}T\right)\left(\mathrm{Im\;}U\right) \;\;.
\label{hetsup}
\eea 
This constraint ensures that the non-perturbative degrees of freedom are all heavier than those of the fundamental string. In this microscopic description the symplectic frame is completely fixed.  More precisely, one can still perform $SL\left(4,\mathbb{Z}\right) \subset Sp\left(8,\mathbb{Z}\right)$ transformations which respect the F1 and non-perturbative splitting and only mix the each sector within itself. But the exchange (\ref{F3G3ex}) is not longer allowed. That is, a fundamental string description fixes the electric $U(1)$ choice for the full $U(1)^4$. This electric frame is then one where the prepotential is of the exponential form, as in (\ref{expShetgen}). 

\subsubsection{The four-dimensional supergravity and gauge kinetic matrix}

As explained above, there are two natural bases for the electric period vectors, of the forms
\be
X^I_{\mathrm{A}} =  \left( \begin{array}{c} 1 \\ T \\ U \\ S \end{array} \right) \;,\;\; X^I_{\mathrm{H}} =  \left( \begin{array}{c} 1 \\ T \\ U \\ -TU \end{array} \right) \;,
\label{sympframesforhe}
\ee 
which are related by the symplectic exchange (\ref{F3G3ex}). We consider here the natural Heterotic basis $X_H^I$, for which the prepotential is exponential (\ref{expShetgen}). The Kahler potential is symplectically invariant, and so takes the universal form
\be 
e^{-K} = 8\left(\mathrm{Im\;}S\right)\left(\mathrm{Im\;}T\right)\left(\mathrm{Im\;}U\right)\;.
\ee 
The mass of electric BPS states, with charges $q_I$, therefore takes the form
\be 
\left(\frac{M\left(q\right)}{M_p}\right)^2 = 8 \pi \;e^{K} \left|q_I X^I\right|^2 = \frac{\pi}{\left(\mathrm{Im\;}S\right)\left(\mathrm{Im\;}T\right)\left(\mathrm{Im\;}U\right)} \;\Big|q_0+q_1\;T + q_2 \;U -q_3 \;T\;U\Big|^2\;.
\label{BPSmasssupgrav}
\ee 
The gauge kinetic matrix, in the regime (\ref{tendhetsup}), has universal string coupling behaviour, and so takes the very simple form 
\be 
{\cal N}^{(STU)} = -i\;\mathrm{Im\;}S \left( \begin{array}{cccc} 
\left(\mathrm{Im\;}T\right)\left(\mathrm{Im\;}U\right) & 0 & 0 & 0 \\ 
0 & \frac{\mathrm{Im\;}U}{\mathrm{Im\;}T} & 0 & 0 \\
0 & 0 & \frac{\mathrm{Im\;}T}{\mathrm{Im\;}U} & 0 \\
0 & 0 & 0 & \frac{1}{\left(\mathrm{Im\;}T\right)\left(\mathrm{Im\;}U\right)} \\
\end{array}\right) + ... \;.
\label{Nstuclass}
\ee 
where we have set $\mathrm{Re\;}S=\mathrm{Re\;}T=\mathrm{Re\;}U=0$ for simplicity.


\subsubsection{Light states}

If we consider only the milder constraint (\ref{hetsup}), so a worldsheet description rather than a higher-dimensional supergravity one, there are certain loci in the $\left(T,U\right)$ moduli space where new charged states become massless.\footnote{It is possible to think of an intermediate eight-dimensional supergravity description, where the torus is integrated over. One can then compactify the eight-dimensional theory on $K3$. However, we are primarily interested in the torus physics here, and so we refer to this as an intrinsically worldsheet-based regime.} The simplest way to see this is though the perturbative Heterotic string spectrum (see, for example \cite{Giveon:1994fu,LopesCardoso:1994ik,Antoniadis:1995jv,LopesCardoso:1995zq,Fraiman:2022lwd}, for excellent accounts).  

From the worldsheet theory, the mass of the charged states can be written in terms of the left-moving and right-moving momenta $p_L$ and $p_R$, given by \cite{Blumenhagen:2013fgp,Giveon:1994fu,LopesCardoso:1994ik,Antoniadis:1995jv,LopesCardoso:1995zq}
\bea 
p_R &=& \frac{1}{\sqrt{2\left(\mathrm{Im\;}T\right)\left(\mathrm{Im\;}U\right)}}\;\Big|\;n_1+\;n_2\;U+\;w_2 \;T+w_1 \;T\;U \;\Big| \;, \nn \\
p_L &=& \frac{1}{\sqrt{2\left(\mathrm{Im\;}T\right)\left(\mathrm{Im\;}U\right)}}\;\Big|\;n_1+\;n_2\;U+\;w_2 \;\bar{T}+w_1 \;\bar{T}\;U \;\Big| \;,
\label{PlPRHet}
\eea 
where $n_1$ and $n_2$ are KK numbers along $R_1$ and $R_2$ respectively, and similarly $w_1$ and $w_2$ are winding numbers
\be 
n_1 \;,\; n_2 \;,\; w_1 \;,\; w_2 \; \in \mathbb{Z} \;. 
\ee 
The left-moving $N_L$ and right-moving $N_R$ oscillator numbers satisfy the level-matching condition
\be 
N_R - N_L = n_1 w_1 - n_2 w_2 \;.
\label{lvlmatmlnrs}
\ee 
The mass of the states can be written as
\bea 
\left(\frac{M\left(n_1,n_2,w_1,w_2\right)}{2\pi \;m_s^H}\right)^2 &=& 2\left|p_R\right|^2 +4 \left( N_R -1\right) \;.
\label{Mhetplrgen}
\eea 
We can read off which states are BPS by using the relation between the string and Planck scales (\ref{mstomprel}). The mass (\ref{Mhetplrgen}) then matches the supergravity formula (\ref{BPSmasssupgrav}), with the electric charges mapped as
\be 
\left(q_0,q_1,q_2,q_3\right) = \left(n_1,n_2,w_2,-w_1\right) \;,
\label{maptowindingstu}
\ee 
and the oscillator numbers taken as \footnote{We have restricted $N_L=0$ here to consider spacetime vectors.}
\be 
N_R = 1 \;,\; N_L=0 \;.
\label{oscnumnlnr10}
\ee 
Using the map (\ref{maptowindingstu}) we can then write the level-matching constraint (\ref{lvlmatmlnrs}) as 
\be 
q_0\; q_3 + q_1 \;q_2 = -1 \;.
\label{intselfdulatcons}
\ee

It is an interesting property of the Heterotic symplectic frame (\ref{sympframesforhe}) that the supergravity formula (\ref{BPSmasssupgrav}) can be extended unchanged from the regime $\mathrm{Im\;}T \gg \mathrm{Im\;}U \gg 1$ to the stringy regime of arbitrary $U$ and $T$. This holds in the weak-coupling limit $S \rightarrow \infty$. The reason is that the period $X^3_H = -TU$ arises from the derivative with respect to $S$ of the $STU$ term in the prepotential that is classical, and therefore is exact up to non-perturbative corrections in $S$. Quantum corrections in $T$ and $U$ only affect the magnetic periods in the Heterotic symplectic frame. This means that they appear in the gauge kinetic matrix, and in the Kahler potential, but not in the holomorphic part of the central charge for purely electric BPS states. 

From (\ref{Mhetplrgen}) and (\ref{PlPRHet}), we can read off the charged states which become massless on special loci in moduli space. These charged states are W-bosons which enhance the Abelian $U(1)^4$ gauge symmetry to a non-Abelian group. The specific states and enhancements, in the Heterotic symplectic basis, are shown in table \ref{tab:hetlightsta}. 

\begin{table}
\begin{center}
\def\arraystretch{1.5}
	\begin{tabular}[t]{|c|c|c|c|c|c|c|}
	\hline
	& $q_0$ & $q_1$ & $q_2$ & $q_3$ & Locus & Gauge enhancement\\
	 \hline
	$A$ & 0 & $\pm$ 1 & $\mp$ 1 & 0 &  $T=U$ & $\begin{array}{c}U(1)_1\times U(1)_2 \rightarrow \\ SU(2)_{12} \times U(1)_{1+2}\end{array}$ \\
	\hline
	$B$ & $\pm$ 1  & 0 & 0 & $\mp$ 1 &  $T=-\frac{1}{U}$ & $\begin{array}{c} U(1)_0\times U(1)_3 \rightarrow \\ SU(2)_{(03)} \times U(1)_{(0)+(3)}\end{array}$ \\
	\hline
	$C$ & $\mp$ 1 & $\pm$ 1 & $\mp$ 1 & 0 &  $T=U+1$ & $\begin{array}{c} U(1)_0\times U(1)_1\times U(1)_2\rightarrow \\ SU(2)_{(012)} \times U(1)_{(0)-(1)}\times U(1)_{(1)-(2)} \end{array}$\\
	\hline
	$D$ & 0 & $\pm$ 1 & $\mp$ 1 & $\pm$ 1 &  $T=\frac{U}{U+1}$ & $\begin{array}{c} U(1)_1\times U(1)_2\times U(1)_3\rightarrow \\ SU(2)_{(123)} \times U(1)_{(1)+(2)}\times U(1)_{(2)+(3)} \end{array}$ \\
	\hline
	\multicolumn{5}{|c|}{$A\cap B$} &  $T=U=i$ & $\begin{array}{c} U(1)_0 \times U(1)_1\times U(1)_2 \times U(1)_3 \rightarrow \\ SU(2)_{(12)} \times SU(2)_{(03)}  \\ \times  U(1)_{(1)+(2)}\times U(1)_{(0)+(3)}\end{array}$ \\
	\hline
	\multicolumn{5}{|c|}{ $B\cap C\cap D$}  &  $\begin{array}{c} T=e^{\frac{ \pi i}{3}} \;,\\ \;U=e^{\frac{2 \pi i}{3}} \end{array}$ & $\begin{array}{c} U(1)_0 \times U(1)_1\times U(1)_2 \times U(1)_3 \rightarrow \\ SU(3)_{(0123)}\times U(1)_{(1)+(2)} \\ \times  U(1)_{(0)+(1)-(3)} \end{array}$ \\
	\hline
	\end{tabular}
\end{center}
\caption{Table showing charges of states which become massless on special loci in moduli space, and the associated gauge enhancements. The notation is that a non-Abelian enhancement with sub-index $(ijk)$ means that it involves that $U(1)$ directions $i$, $j$, and $k$. An Abelian factor of the form $U(1)_{(i)\pm(j)}$ means that its field-strength is proportional to the combination $F^i_{\mu\nu} \pm F^j_{\mu\nu} $ , so it is the appropriate linear combination of the $U(1)$ directions.}
\label{tab:hetlightsta}
\end{table}

\subsubsection{The prepotential, gauge kinetic matrix and monodromies}
\label{sec:pregkmmm}

Around the singular loci shown in table \ref{tab:hetlightsta} the gauge kinetic matrix diverges and there are monodromies of the period vector. These arise from integrating out the states which become massless on the special loci. Unlike the type II setting, where the charged states are non-perturbative and so have automatically been integrated out into the classical description, here the states are perturbative and must be integrated out by hand into the effective infrared description. This must be performed in the appropriate electric frame, where the charged states can be treated as fundamental. In terms of the Heterotic versus type IIA frames, as in (\ref{sympframesforhe}), in the Heterotic frame all the appropriate charged states are electric. 

Integrating out the light states can be naturally expected to be captured in the effective four-dimensional quantum field theory framework. The prepotential then receives the usual one-loop contribution of the form $M^2 \log M$, with $M$ the mass of the integrated-out state. For example, for the state on the locus $A$ we should expect a contribution
\be 
F^{\text{1-loop}} \sim  \left(T-U\right)^2 \;\log \left(T-U\right) \;.
\ee 

The gauge kinetic matrix also develops a divergence of the form
\be 
{\cal N}^{\text{1-loop}}_{IJ} = -\frac{\left(-2\right)}{2 \pi i}\;{\bf M}^{(q)}_{IJ} \;\log M  \;,
\label{N1loopvecto}
\ee 
where the matrix ${\bf M}^{(q)}$ is given by 
\be 
{\bf M}^{(q)}_{IJ} = q_I q_J\;,
\ee 
with $q_I$ is the charge of the state which becomes massless, with mass $M$. Note that in the expression (\ref{N1loopvecto}) there is an explicit factor of $-2$ in the numerator. This is a factor which is associated to a vector multiplet becoming massless, with a BPS index of $-2$ .\footnote{This factor includes both the $W^+$ and $W^-$ particles in it.} If instead a hypermultiplet becomes massless, then the BPS factor is $+1$. 


\subsubsection{Gauge bundles and the $ST^2$ model}
\label{sec:st2modelhet}

On the special loci where there are enhanced non-Abelian gauge symmetries, it is possible to turn on some of the bundle (\ref{hetbundle}) in those directions. An example relevant for our purposes is where we enhance to an $SU(2)$ gauge symmetry along locus A in table \ref{tab:hetlightsta}. We can then split the bundle as $\left(10,10,4\right)$, where the bundle with instanton number 4 is turned on along the $SU(2)$. This model was considered in \cite{Kachru:1995wm}, in the context of type IIA Heterotic duality. We denote it the $ST^2$ model. Because the bundle is treated geometrically, and is turned on along the $K3$ factor, we should think of this model as starting from an effective eight-dimensional supergravity theory with an enhanced $SU(2)$ symmetry, and reducing on a large $K3$. 

Turning on a bundle along the enhanced $SU(2)$ fixes the moduli to the enhancement locus, and breaks the gauge group. The left-over Abelian gauge group is then 
\be 
ST^2\;\mathrm{model}\;:\; U(1)_0 \times U(1)_{(1)+(2)} \times U(1)_3  \;.
\ee  
Since we have only two vector multiplets now, the electric period vector takes the form
\be
{\bf X}^I_{\mathrm{A}} =  \left( \begin{array}{c} 1 \\ T \\ S \end{array} \right) \;,\;\;{\bf X}^I_{\mathrm{H}} =  \left( \begin{array}{c} 1 \\ T \\ -T^2 \end{array} \right) \;,
\label{sympframesforheST2pre}
\ee 
which is the restriction of (\ref{sympframesforhe}) to the locus $U=T$. This is associated to a prepotential (in the type IIA frame) of  
\be 
F^A = -ST^2 + ... \;.
\label{hetFst2}
\ee 
We will consider the Heterotic frame, so after the symplectic transformation (\ref{Hetmap}). However, it is convenient, for the purposes of comparing with the type II string theory dual results, to perform a further basis change by the symplectic transformation as in (\ref{Sformabde}) with ${\bf C}={\bf D}=0$ and 
\be 
{\bf A}^{(T^2\rightarrow 1-T^2 )} = \left( \begin{array}{ccc} 
1 & 0 & 0 \\
0 & 1 & 0 \\
1 & 0 & 1 \\
\end{array} \right) \;. 
\label{ttanto1mt2} 
\ee 
After this, the Heterotic basis periods take the form 
\be 
\Pi_{\mathrm{H}} =  \left( \begin{array}{c} 1 \\ T \\ 1-T^2 \\ S\;\left(1+T^2\right) \\ -2\; S\; T \\ -S \end{array} \right) \;.
\label{sympframesforheST2}
\ee

The classical part of the prepotential (\ref{hetFst2}) is supplemented by the one-loop contribution (as well as non-perturbative ones). There are two regimes of interest, the first is the large $T$ regime
\be
\mathrm{Im\;}T \gg 1 \;,
\label{largeThetst2}
\ee 
and the second around the enhanced symmetry locus B in table \ref{tab:hetlightsta}, so with 
\be 
T=i \;.
\label{ihetst2}
\ee 
Around the enhanced locus we have the behaviour of the prepotential, in the Heterotic regime where the light state is electric, of
\be 
F^H \sim \left(T-i\right)^2 \log \left(T-i\right) + ... \;.
\ee 
The monodromy about the locus is deduced by the charge of the state that becomes massless from table \ref{tab:hetlightsta}. After performing the rotation (\ref{ttanto1mt2}), the charge of the state is
\be 
{\bf q}^{\left(e\right)} = \left( \begin{array}{c} -2 \\ 0 \\ 1 \end{array} \right) \;\;,  
\label{chargetmi}
\ee 
which satisfies
\be 
{\bf q}^{\left(e\right)} \cdot  {\bf X}_{\mathrm{H}} = - \left(T+i\right)\left(T-i\right) \;,
\ee 
and yields a monodromy 
\be 
\text{Locus}\;A \cap B \;\;:\;\;{\bf M}_{IJ}^{\left(e\right)} = {\bf q}^{\left(e\right)}_I {\bf q}^{\left(e\right)}_J =  \left( \begin{array}{ccc} 
4 & 0 & -2  \\
0 & 0 & 0  \\
-2 & 0 & 1  \\
\end{array} \right)_{IJ} = \left({\bf M}^{(0,0)}_t\right)_{IJ}  \;,
\ee 
where we recall the definition of ${\bf M}^{(0,0)}_t$ is in (\ref{bmtdef}).

We also have the projection of the classical gauge coupling (\ref{Nstuclass}) to the $ST^2$ model. After performing the rotation (\ref{ttanto1mt2}), this takes the form
\be 
{\cal N}^{(ST^2)} = -S \left( \begin{array}{ccc} 
\left(\mathrm{Im\;}T\right)^2 + \frac{1}{\left(\mathrm{Im\;}T\right)^2} & 0  & -\frac{1}{\left(\mathrm{Im\;}T\right)^2} \\ 
0 & 2  & 0 \\
-\frac{1}{\left(\mathrm{Im\;}T\right)^2} & 0  & \frac{1}{\left(\mathrm{Im\;}T\right)^2} \\
\end{array}\right) + ...   \;,
\label{Nst2class}
\ee 
where we set $\mathrm{Re\;}T=0$ for simplicity. Note that the middle $2$ in (\ref{Nstuclass}) is the correct normalization for the remaining $U(1)$ combination after enhancing to the $SU(2)$ along locus A in table \ref{tab:hetlightsta}. 

As we vary $T$ towards $T=i$, the gauge coupling matrix (\ref{Nst2class}) must stay the same, since it is classical and so quantum corrections cannot modify it. Evaluating (\ref{Nst2class}) on $T=i$ yields 
\be 
\left.{\cal N}^{(ST^2)}\right|_{T=i} = -S \left( \begin{array}{ccc} 
2 & 0  & -1 \\ 
0 & 2  & 0 \\
-1 & 0  & 1 \\
\end{array}\right) + ...  = -S \;\Big( {\bf B}_t + {\bf M}_t^{(0,0)} \Big) + ... \;.
\label{Nst2classi}
\ee 
Recall that ${\bf B}_t$ and ${\bf M}_t^{(0,0)}$ are given in (\ref{bmtdef}).

We therefore have for the $ST^2$ model, using (\ref{N1loopvecto}), the behaviour of the gauge coupling matrix, around $\mathrm{Im\;}T \gg 1$ and $T=i$, of
\bea 
\mathrm{Im\;}T \gg 1 \;\;&:&\;\; {\cal N}^H = - S \left( \begin{array}{ccc} 
\left(\mathrm{Im\;}T\right)^2 + \frac{1}{\left(\mathrm{Im\;}T\right)^2} & 0  & -\frac{1}{\left(\mathrm{Im\;}T\right)^2} \\ 
0 & 2  & 0 \\
-\frac{1}{\left(\mathrm{Im\;}T\right)^2} & 0  & \frac{1}{\left(\mathrm{Im\;}T\right)^2} \\
\end{array}\right)
+ ...  \;\;,\nn \\
& &\nn \\
& &\nn \\
\mathrm{Im\;}T \sim 1 \;\;&:&\;\; {\cal N}^H = - S \;\Big( {\bf B}_t + {\bf M}^{(0,0)}_t \Big)
+\frac{1}{2\pi i}\log \left(T-i\right) \;  {\bf M}^{(0,0)}_t + ... \;.
\label{st2modelgaugeco}
\eea 

It is also informative to write the general spectrum of BPS states in the $ST^2$ model, which is the appropriate projection of table \ref{tab:HetBPS}. We consider the basis associated to the electric period vector (\ref{sympframesforheST2}), so after the symplectic transformation (\ref{ttanto1mt2}). We have a general electric-magnetic charge vector of the form (\ref{symcharvec}), with an associated central charge (\ref{gencentypii}). The masses, and nature, of the appropriately charged BPS states are shown in table \ref{tab:HetBPSST2}. 
\begin{table}
\begin{center}
\def\arraystretch{1.5}
	\begin{tabular}[t]{|c|c|c|c|c|c|c|}
	\hline
	 Charge & State & $K3_H$ & $R_1$ & $R_2$ & $\left(\frac{M}{2\pi m_s^H}\right)^2\;,\;\mathrm{Im\;}T \gg 1$ & $\left(\frac{M}{2\pi m_s^H}\right)^2\;,\;\mathrm{Im\;}T \sim 1$ \\
	 \hline
	$q_0$ & F1KK + F1W  &  & x & & $\frac{1}{\left(\mathrm{Im\;}T\right)^2}$ & $1$\\
	\hline
	$q_1$ & F1KK + F1W &  &  & x & $1$ & $1$\\
	\hline
	$q_2$ & - F1W  &  & x & & $\left(\mathrm{Im\;}T\right)^2$ & $1$\\
	\hline
	$p^0$ & KK5  & x & & x & $\left(\mathrm{Im\;}S\right)^2 \left(\mathrm{Im\;}T\right)^2$ & $\left(\mathrm{Im\;}S\right)^2\left(\left(\mathrm{Im\;}T\right)^2-1\right)^2$\\
	\hline
	$p^1$ & KK5 +  NS5  & x & x & & $4\left(\mathrm{Im\;}S\right)^2$ & $4\left(\mathrm{Im\;}S\right)^2$\\
	\hline
	$p^2$ & KK5 - NS5 & x &  & x & $\frac{\left(\mathrm{Im\;}S\right)^2}{\left(\mathrm{Im\;}T\right)^2}$ & $\left(\mathrm{Im\;}S\right)^2$\\
	\hline
	\end{tabular}
\end{center}
\caption{Table showing the BPS states in the Heterotic $ST^2$ model in the regimes $\mathrm{Im\;}T \gg 1$ and $\mathrm{Im\;}T \sim 1$, at leading order in $\left(\mathrm{Im\;}T-1\right)$ and $S$.}
\label{tab:HetBPSST2}
\end{table}

\subsection{General $STU$ duality}
\label{sec:stuiihetdual}

The most standard, and best understood, type II-Heterotic duality is of the type  
\be
\begin{array}{c}
	\frac{\rule[-0.8ex]{0pt}{2.5ex}\text{\normalsize Type IIB}}{\rule{0pt}{2.8ex}\text{\normalsize $Y$}} = \frac{\rule[-0.8ex]{0pt}{2.5ex}\text{\normalsize Type IIA}}{\rule{0pt}{3ex}\text{\normalsize $\widetilde{Y}$}} 
	= \frac{\rule[-0.8ex]{0pt}{2.5ex}\text{\normalsize Heterotic}}{\rule{0pt}{2.8ex}\text{\normalsize $K3_H\times \left(T^2\right)_H$}}\;\;,
\end{array} 
\label{chadustr}
\ee
where the IIB Calabi-Yau is denoted as $Y$ and its mirror as $\tilde{Y}$. The Heterotic $K3_H \times\left(T^2\right)_H$ is at large volume in (Heterotic) string units. 
This chain of dualities requires a certain fibration structure for the Calabi-Yau manifolds \cite{Klemm:1995tj}. This is nicely explained, for example, in \cite{Dedushenko:2014nya}. The mirror Calabi-Yau $\tilde{Y}$ must support both an elliptic fibration, with fibre $E$ and base $B$, and $K3$ fibration with base $B_S=\mathbb{P}^1$. This means that the base $B$ is a rational ruled surface, admitting a holomorphic map to a fibration with fibre $B_{S'} = \mathbb{P}^1$ over the base $B_S$. Schematically, the structure of the generic fibre is of the form
\be 
\tilde{Y} \sim \lefteqn{\overbrace{\rule{0pt}{2.5ex}\phantom{\left(\mathbb{P}^1\right)_S \times\left(\mathbb{P}^1\right)_{S'}}}^{\rule[-0.8ex]{0pt}{.5ex}\text{\normalsize Base $B$}} }\left(\mathbb{P}^1\right)_{B_S} \times
\underbrace{\rule[-1.5ex]{0pt}{.5ex}\left(\mathbb{P}^1\right)_{B_{S'}}\times \left(T^2\right)_A}_{\rule{0pt}{2.5ex}\text{\normalsize $K3_A$ Fibre}} \;.
\label{prodsctru}
\ee 
On the Heterotic side, the $K3$ is different from the $K3$ in the type IIA fibrations. But it is still an elliptic fibration, with fibre $E'$ over the base $B_{S}$. So we can write schematically for the Heterotic geometry
\be 
\mathrm{Heterotic\;}\;:\;\; \overbrace{\rule{0pt}{2.5ex}\left(\mathbb{P}^1\right)_{B_{S}} \times E'}^{\rule[-0.8ex]{0pt}{.5ex}\text{\normalsize $K3_H$}} \;\times\; \left(T^2\right)_H \;.
\label{prodhet}
\ee 
The simplest realisation of this is the case where the base $B$ is the Hirzebruch surface $\mathbb{F}_0$
\be 
B=\mathbb{F}_0=\left(\mathbb{P}^1\right)_{B_S}\times \left(\mathbb{P}^1\right)_{B_{S'}} \;,
\label{BisFo}
\ee
which is dual to the Heterotic string with instanton numbers $(12,12)$ (in the two $E_8$ groups) \cite{Morrison:1996pp,Curio:2001ae}. In this case, on the type IIA side, we have three distinguished geometric parameters, associated to the overall volumes (in type IIA string units) of $\left(\mathbb{P}^1\right)_{B_S}$, $\left(\mathbb{P}^1\right)_{B_{S'}}$ and the $\left(T^2\right)_A$. These are denoted as
\be 
    \text{Im\;}S = \text{Vol}\left(B_S\right) \left(m^A_s\right)^2,\;\;\;\;\;\;
    \text{Im\;}T = \text{Vol}\left(T^2\right)_A\;\left(m^A_s\right)^2,\;\;\;\;\;\;
    \text{Im\;}U = \text{Vol}\left(B_{S'}\right)\left(m^A_s\right)^2\;,
\label{iiageomidntom}
\ee
where $m^A_s$ denotes the type IIA string scale. 
On the heterotic side, these same vector fields $S$, $T$ and $U$ control the four-dimensional heterotic string coupling $g_{s,H}$, the (complexified) Kahler modulus and the complex-structure of the $\left(T^2\right)_H$ respectively. These are exactly the fields in (\ref{STUmap}). 

The four-dimensional Heterotic $g_{s,H}$ and type IIA $g_{s,A}$ string couplings are related to ten-dimensional couplings $\hat{g}_{s,H}$ and $\hat{g}_{s,A}$ , by 
\bea 
\left(\hat{g}_{s,A}\right)^2 &=& \left(g_{s,A}\right)^2 \left(\text{Vol}\;\tilde{Y} \right) \left( m_{s}^A\right)^6 \;,\;\; \nn \\
\left(\hat{g}_{s,H}\right)^2 &=& \left(g_{s,H}\right)^2 \text{Vol}\left(K3_H \right) \text{Vol}\left(T^2\right)_H\left( m_{s}^H\right)^6 \;.
\eea 
The respective string scales are related to the Planck scales as
\be 
m_s^A = \frac{1}{2\sqrt{\pi}}\;g_{s,A}\; M_p \;,\;\; m_s^H = \frac{1}{2\sqrt{\pi}}\;g_{s,H}\; M_p \;.
\label{stringscales}
\ee 
Therefore, in the Einstein frame we have the duality relation
\be 
\frac{m_s^A}{g_{s,A}} = \frac{m_s^H}{g_{s,H}} \;.
\label{massrelHA}
\ee 

The relation (\ref{massrelHA}) is crucial to map physical quantities from one side of the duality to the other. For example, the tension of a string which arises from an NS5 brane on the type IIA side that is wrapping the $K3_A$ is given by 
\be 
T^{\mathrm{NS5}}_{A} = \frac{2\pi\;\text{Vol}\left(K3_A \right) \left( m_{s}^A\right)^6}{\left(\hat{g}_{s,A}\right)^2} =2\pi \left(m_{s}^H\right)^2 = T^{\mathrm{F1}}_H\;,
\ee 
which identifies it as the fundamental Heterotic string. 

Similarly, (\ref{massrelHA}) is important in understanding the relation between the vector multiplets and hypermultiplets quantities. For example, the volume of the base $\text{Vol}\left(B_S\right)$ when measured in type IIA string units, is mapped to the Heterotic dilaton, which is part of the vector multiplets (\ref{STUmap}). However, it also features in the volume of the K3 on the Heterotic side (\ref{prodhet}). In that setting we should measure it in Heterotic string units, which yields
\be 
\text{Vol}\left(B_S\right) \left(m^H_s\right)^2 = \frac{1}{\left(g_{s,A}\right)^2}\;, 
\ee 
and so identifies it with the four-dimensional type IIA string coupling, that is part of the Hypermultiplet sector. 

Assuming the structures (\ref{prodsctru}) and (\ref{prodhet}), so an $STU$-type phase, one can build a rough intuition for the BPS state in the chain of dualities (\ref{chadustr}). Recalling that mirror symmetry is just three T-dualities, we can split the six dimensions of the Calabi-Yau manifolds $Y$ and $\tilde{Y}$ into a structure of the form (\ref{prodsctru}). Then the different BPS states in the IIB, IIA and Heterotic frames are shown in table \ref{tab:bpsdual}. 

\begin{table}
\begin{center}
\def\arraystretch{1.5}
	\begin{tabular}[t]{|c|c|c|c|c|c|c|}
	\hline
	 IIB & \multicolumn{2}{c}{$B_S$} & \multicolumn{2}{|c}{$B_{S'}$} & \multicolumn{2}{|c|}{$T^2$} \\
	 \hline
	D3 & x &  & x & & x & \\
	\hline
	D3 & x & & x & & & x\\
	\hline
	D3 & x & & & x & x & \\
	\hline
	\rowcolor{lightgray}
	D3 & x & & & x & & x \\
	\hline
	D3 & & x & & x & & x\\
	\hline
	D3 & & x & & x & x& \\
	\hline
	D3 & & x & x & & & x\\
	\hline
	\rowcolor{lightgray}
	D3 & & x & x & & x & \\
	\hline
	\end{tabular}
	\hfill
	\begin{tabular}[t]{|c|c|c|c|c|c|c|}
	\hline
	 IIA & \multicolumn{2}{c}{$B_S$} & \multicolumn{2}{|c}{$B_{S'}$} & \multicolumn{2}{|c|}{$T^2$} \\
	\hline
	D0 &  &  & & & & \\
	\hline
	D2 & & & & & x & x \\
	\hline
	D2 & & & x & x & & \\
	\hline
	\rowcolor{lightgray}
	D4 & & & x & x & x & x \\
	\hline
	D6 & x & x & x & x & x & x \\
	\hline
	D4 & x & x & x & x & & \\
	\hline
	D4 & x & x & & & x & x\\
	\hline
	\rowcolor{lightgray}
	D2 & x & x  &  & &  & \\
	\hline
	\end{tabular}
	\hfill
	\begin{tabular}[t]{|c|c|c|c|c|c|c|}
	\hline
	 HET & \multicolumn{2}{c}{$B_S$} & \multicolumn{2}{|c}{$E'$} & \multicolumn{2}{|c|}{$T^2$} \\
	\hline
	F1KK &  &  &  & & x & \\
	\hline
	F1W & & & & & & x \\
	\hline
	F1KK & & & & & & x \\
	\hline
	\rowcolor{lightgray}
	F1W & & & & & x & \\
	\hline
	KK5 & x & x & x & x &  & x \\
	\hline
	NS5 & x & x & x & x & x & \\
	\hline
	KK5 & x & x & x & x & x & \\
	\hline
	\rowcolor{lightgray}
	NS5 & x & x & x & x &  & x\\
	\hline
	\end{tabular}
\end{center}
\caption{Table showing the schematics structure of BPS states in type IIB, IIA and Heterotic string theories for the simple fibration geometries (\ref{prodsctru}) and (\ref{prodhet}). The x's denote directions, either wrapping or KK, associated to the states. On the Heterotic side, F1KK and F1W denote KK and Winding modes of the fundamental string, while KK5 denote KK monopoles. The Heterotic states are the same ordering as in table \ref{tab:HetBPS}. The highlighted rows are important in the sense of defining the electric frame.}
\label{tab:bpsdual}
\end{table}


In table \ref{tab:bpsdual} there are two highlighted rows, which correspond to states who mass ratio defines a natural electric frame. From the type IIA perspective, it is natural to take D2 branes as lighter than D4 branes. On the other hand, from the Heterotic perspective it is natural to have F1 strings lighter than NS5 branes. The two possibilities are related by a symplectic transformation. In fact, these are nothing but the two natural electric frames discussed in section \ref{sec:modspace}. They are universal to the Heterotic string on $K3 \times T^2$ \cite{deWit:1995dmj}. Quantitatively, we can write the masses of the two highlighted states as
\bea 
\left(M_\mathrm{F1W}\right)^2 &=& \left( 2\pi \left(m_{s}^H\right)^2 \right)^2 \left(2 \pi R_1\right)^2 = \left( 2\pi m_{s}^H \right)^2\left(\text{Im\;}T\right)\left(\text{Im\;}U\right) \;, \\ \nn
\left(M_\mathrm{NS5}\right)^2 &= & \left(\frac{2\pi\left( m_{s}^H \right)^{6} \text{Vol}\left(K3_H \right) 2\pi R_2}{\left(\hat{g}_{s,H}\right)^2}\right)^2 = \left(2\pi \;m_s^H\right)^2 \frac{\left(\text{Im\;}S\right)^2}{\left(\text{Im\;}T\right)\left(\text{Im\;}U\right)} \;.
\eea 
Therefore, their ratio is given by
\be 
\frac{M_\mathrm{F1W}}{M_\mathrm{NS5}} = \frac{\left(\text{Im\;}T\right)\left(\text{Im\;}U\right)}{\text{Im\;}S}\;.
\ee 
The two natural electric frames are then 
\bea 
\text{IIA Frame}\;&:&\; \left(\text{Im\;}T\right)\left(\text{Im\;}U\right) \gg \text{Im\;}S \;, \nn \\
\text{Heterotic Frame}\;&:&\; \left(\text{Im\;}T\right)\left(\text{Im\;}U\right) \ll \text{Im\;}S \;,
\eea 
which match the frames (\ref{IIAfrST}) and (\ref{HfrST}) under the appropriate restriction $T=U$. 

The monodromies and prepotential for the two frames are then as in the general discussion of section \ref{sec:ii0loci}. That is, Heterotic weak coupling limits $S \rightarrow \infty$ are type II limits in moduli space. The appropriate electric Heterotic frame is then one where the monodromy takes the form (\ref{monN}). In this case, as shown in generality in section \ref{sec:ii0loci},  the prepotential has no polynomial dependence on $S$, and it appears only with exponential factors. We can write this schematically as 
\bea 
\text{IIA Frame}\;&:&\; F(S) \sim S \;, \nn \\
\text{Heterotic Frame}\;&:&\; F(S) \sim S e^{-S} \;,
\eea

In the case where the base is not the Hirzebruch surface $\mathbb{F}_0$, as in (\ref{BisFo}), the relation between the moduli and the geometry is modified. For example, if the base is $\mathbb{F}_2$, then instead of (\ref{iiageomidntom}) we have the identification $\text{Vol}\left(B_{S'}\right)\left(m^A_s\right)^2=\text{Im\;}U-\text{Im\;}T$ \cite{Curio:2001ae}. Since our primary focus is on the example two-parameter model in section \ref{sec:ii0loci}, we henceforth directly restrict to this setting rather than reviewing the various other possibilities. 

\subsection{$ST^2$ duality at weak-coupling}
\label{sec:hetlcstypeii}

In this section we consider Heterotic-type II duality for the example Calabi-Yau studied in section \ref{sec:ii0loci}. This is the original example of four-dimensional ${\cal N}=2$ duality considered in \cite{Kachru:1995wm}. It is proposed to be dual to the Heterotic $ST^2$ model studied in section \ref{sec:st2modelhet}. The duality can be tracked through the transition between large complex-structure and the type II locus. That is, we identify the full type IIB periods with the Heterotic parameters, and these periods can be evaluated as an expansion about large complex-structure or about the type II locus. 

The identification is, clearly, identifying $S$ and $T$ in the heterotic $ST^2$ model, as described in section \ref{sec:st2modelhet}, with the (full) periods $S$ and $T$ in the IIB model (\ref{perveclcsA}). Note that these are the full global periods, not only their local expansions about large complex-structure. 

\subsubsection{Duality with large complex-structure}

The large complex-structure regime corresponds to 
\be 
\mathrm{Im\;}S \gg 1 \;\;,\;\; \mathrm{Im\;}T \gg 1 \;.
\ee 
In this regime one can match the Heterotic gauge kinetic matrix (\ref{st2modelgaugeco}) and the large complex-structure gauge kinetic matrix (\ref{Nhlcs}). 

Evidence for the duality is stronger than the leading behaviour, as argued in \cite{Kachru:1995wm}. One first notes that the periods $S$ and $T$ are related to the global type IIB coordinates $z_1$ and $z_2$ as \cite{Candelas:1993dm}
\be 
z_1 = \frac{1}{j\left(T\right)} + {\cal O}\left( e^{2 \pi i S}\right) \;\;\;,\;\;\; z_2 = e^{2 \pi i S} \;\Big( \;1 + {\cal O}\left( e^{2 \pi i S}\right)\;\Big) \;,
\label{z1z2stmap}
\ee 
where $j\left(T\right)$ is the elliptic j-function. This captures the modular behaviour with respect to $T$ on the Heterotic side in the weak-coupling limit. The expansion about large $\mathrm{Im\;}T$ is 
\be 
j\left(T\right) = e^{-2\pi i T} + 744 + {\cal O}\left( e^{2 \pi i T} \right) \;, 
\ee 
which shows that indeed at leading order the identification (\ref{z1z2stmap}) matches that of large complex-structure (\ref{STlcs}). 

One can even match, in the weak-coupling limit, the exponential expansions, at $\mathrm{Im\;}T \gg 1 $ of the prepotential. The match was a primary result in \cite{Kachru:1995wm}. 

\subsubsection{Duality with the type II locus}

The type II locus was proposed to be dual to the $T=i$ locus on the Heterotic side \cite{Kachru:1995wm,Kachru:1995fv}. So we are now in the regime
\be 
\mathrm{Im\;}S \gg 1 \;\;,\;\; \mathrm{Im\;}T \sim 1 \;.
\ee 
First, we should check that this indeed corresponds to the type II locus on the type IIB side. This corresponds to $s \rightarrow 0$ and $t \rightarrow 0$ in (\ref{stdeftyII}). The $s \rightarrow 0$ limit indeed matches (\ref{z1z2stmap}) upon noting that
\be 
j(i) = 1728 \;.
\ee
 
The $t \rightarrow 0$ limit is more subtle since it requires an ordering in the limits. This is because the $(s,t)=(0,0)$ point in the moduli space is singular and needs further blowing-up to reach normal crossing divisors \cite{Candelas:1993dm}. The highlighted, type II$_1$, point in figure \ref{fig:modulispace} is after this blowup. The appropriate blowup scaling is determined by the Heterotic duality \cite{Kachru:1995fv}, so let us first show the duality.

The map for $S$ and $T$ is determined precisely from the periods as
\bea 
S &=& -\frac{F_2}{X^0} = \frac{\log \left( t s^4 /1728\right) }{2 \pi i}-\frac{5+2\zeta_2\pi \beta+\log(1728)}{2\pi i } + ... \;\;,\label{STIIo} \nn \\ \nn
T-i &=&\frac{X^1}{X^0}-i = is \left(2 \beta-\frac{ t \beta}{8 }-\frac{15   t^2 \beta}{512}\right)+is^2 \left(2  \beta^2-\frac{1}{2}  t \beta^2\right) \\ 
& &+\frac{1}{54} i s^3 \beta \left(31+108 \beta^2\right)+\dots\;\;. 
\eea
A nice consistency check is that in the $t=0$ limit, it is possible to derive the expansion in $s$ for $T-i$ in (\ref{STIIo}) using the relations (\ref{stdeftyII}) and the expansion
\be 
j\left(T\right) = 1728  - \frac{1296\;\pi^4}{\Gamma\left(\frac34\right)^8}\;\left(T-i\right)^2 +{\cal O} \left( \left(T-i\right)^4 \right) \;.
\ee

In terms of these coordinates, we can write the gauge kinetic matrix on the type IIB side (\ref{NHtypeIIstbasis}) as 
\begin{align}
    \mathcal{N}_{\mathrm{II}}^H &= 
    -S\;\Big( {\bf B}_t + {\bf M}_t \Big)+\frac{2}{2\pi i}\log \left(T-i\right) \; {\bf M}^{(0,0)}_t + ...\;,
  \label{gkmiiiisideSTco}
\end{align}
where the ellipses denote terms that are finite as $s,t \rightarrow 0$. This matches precisely the form of the leading form of the gauge kinetic matrix on the Heterotic side for the $ST^2$ model (\ref{st2modelgaugeco}).

In fact, the match is precise to all order in $s$, in the $t=0$ limit. This is because the term
\begin{equation}
{\cal N}^H  \supseteq- S \left( \begin{array}{ccc} 
\left(\mathrm{Im\;}T\right)^2 + \frac{1}{\left(\mathrm{Im\;}T\right)^2} & 0  & -\frac{1}{\left(\mathrm{Im\;}T\right)^2} \\ 
0 & 2  & 0 \\
-\frac{1}{\left(\mathrm{Im\;}T\right)^2} & 0  & \frac{1}{\left(\mathrm{Im\;}T\right)^2} \\
\end{array}\right) \;\;,
\end{equation}
is classical and therefore is not modified to all order in $s$ for $t = 0$ , when going to the Type II point. Indeed, one can check using the expansion for $T$ at the Type II point (\ref{STIIo}), and the expressions for the higher order terms in ${\bf M}_t $ (\ref{Mtexpan}), that for $t = \Im\, s = 0$ we have
\begin{equation}
    \left( \begin{array}{ccc} 
\left(\mathrm{Im\;}T\right)^2 + \frac{1}{\left(\mathrm{Im\;}T\right)^2} & 0  & -\frac{1}{\left(\mathrm{Im\;}T\right)^2} \\ 
0 & 2  & 0 \\
-\frac{1}{\left(\mathrm{Im\;}T\right)^2} & 0  & \frac{1}{\left(\mathrm{Im\;}T\right)^2} \\
\end{array}\right) \stackrel{t = \Im s = 0}{=} \Big( {\bf B}_t + {\bf M}_t \Big)\;.
\label{highersmatching}
\end{equation}
The matching of the sub-leading terms in $s$ in (\ref{highersmatching}) is further strong evidence for the duality. 

The second term in (\ref{gkmiiiisideSTco}) arises as a one-loop threshold correction on the Heterotic side from integrating out a BPS state with charges (\ref{chargetmi}). It is classical from the type IIB fundamental string perspective, but should be understood as integrating out the wrapped D3 branes of charges (\ref{chargetmi}). Note that, as discussed in section \ref{sec:pregkmmm}, this state is a vector multiplet rather than a hypermultiplet.

We can now understand more precisely in what specific limit, with respect to the scaling of $s$ and $t$, there is a duality between the type II locus and the weakly-coupled perturbative Heterotic string. Certainly we require the weaker condition 
\be 
\frac{\log t}{\log s} \rightarrow \infty \;,
\ee  
which ensures that the first classical term in (\ref{gkmiiiisideSTco}) dominates over the second quantum term. However, the relation (\ref{highersmatching}) shows that the classical gauge kinetic matrix includes sub-leading terms in $s$, so which behaves as $s \log t $ in the gauge kinetic function. For a classical Heterotic string description, so an exact worldsheet description at tree-level, these corrections should still dominate over the one-loop terms. And so we require a much stronger condition
\be 
\frac{s\log t}{\log s} \rightarrow \infty \;.
\label{perthetlimisp1}
\ee 
Indeed, for an exact classical description, we really need this to hold for all powers of $s$ in the numerator. We therefore have that the appropriate limit is a pure type II$_1$ limit, that is, we should take $s$ small and fixed, and then send $t \rightarrow 0$. There is a clear physical reasoning for why a perturbative Heterotic description requires a limit of the type (\ref{perthetlimisp1}). We discuss this, and the general transition towards strongly-coupled Heterotic physics, in the next section.  

\section{Physics of the type II$_0$ locus}
\label{sec:hetdualii0}

In this section we bring together the various analyses in sections \ref{sec:ii0loci}, \ref{sec:modspace} and \ref{sec:hetdual} to understand, as much as is possible from the four-dimensional supergravity perspective, the physics of type II$_0$ loci. We focus on the specific example type II$_0$ locus studied in sections \ref{sec:modspace} and \ref{sec:hetdual}. Our approach is to consider as a starting point the type II$_1$ locus limit, which is dual to the perturbative Heterotic string, and transition towards the type II$_0$ locus. This transition can be tracked precisely from the infrared four-dimensional supergravity. Its finite energy microscopic interpretation is much more subtle and less understood. In \cite{Monnee:2025msf} a proposal for this transition was made, and we comment on this in section \ref{sec:transs2}. 

To make this section self-contained, let us recall for convenience some of the key elements in our analysis. We are considering a point in moduli space defined by two local coordinates, $s$ and $t$, such that
\be 
\text{Type II point} \;\;:\;\; s \rightarrow 0 \;\;,\;\; t \rightarrow 0 \;.
\ee 
The point is at the intersection of a type II$_1$ locus and type II$_0$ locus, defined as
\bea 
\text{Type II$_1$} \;\;&:&\;\; t \rightarrow 0 \;,\;\; s \;\mathrm{fixed} \;, \nn \\
\text{Type II$_0$} \;\;&:&\;\; t \;\mathrm{fixed} \;,\;\; s \rightarrow 0 \;. 
\label{typeii10limdef}
\eea 
In the example of section \ref{sec:modspace}, this corresponds to the intersection point in figure \ref{fig:modulispace} marked as type II. 

The relevant supergravity vector multiplet fields can be conveniently parameterised by two periods, labelled as $T$ and $S$. These take the form (\ref{STIIo})
\bea 
2 \pi i \;S &=& \log\left(ts^4\right) + 8 s \beta \log \left(s\right) + c + {\cal O}\left(s\right) \;, \nn \\
T - i&=& 2 i \beta s \;\Big(1+ s \beta  \Big) + {\cal O}\left(st\right) \;,
\label{cordtohetcha}
\eea 
with $\beta = \frac{\Gamma(3/4)^4}{\sqrt{3}\,\pi^2}$, and $c$ a constant that is given in (\ref{STIIo}). Note that the period $T$ is electric, but the period $S$ is magnetic.

\subsubsection*{The perturbative Heterotic regime}
\label{sec:hetdualii0}

The regime studied in the original Heterotic duality work \cite{Kachru:1995fv,Kachru:1995wm}, and the many follow-ups, is the limit
\be 
\left(T-i\right)S \sim s \log t  \rightarrow \infty \;.
\label{hetlimitpert}
\ee  
In this limit, there is a well-understood perturbative Heterotic dual compactified on $K3 \times T^2$. The period $S$ is identified with the Heterotic dilaton, proportional to the inverse-squared Heterotic string coupling, and $T$ is identified with the Kahler modulus of the torus (at large volume).

The divergent parts of the gauge kinetic matrix take the form
\be 
    \mathcal{N}_{\mathrm{II}}^H = -S\;\Big( {\bf B}_t + {\bf M}^{(0,0)}_t \Big)+\frac{2}{2\pi i}\log \left(T-i\right) \;{\bf M}^{(0,0)}_t + ...\;,
  \label{gkmiiiisideSTcoRR}
\ee 
with 
\be
\begin{array}{ccc}
     {\bf B}_t + {\bf M}_t^{(0,0)}= \left( \begin{array}{ccc} 
2&0&-1 \\ 
0&2&0 \\
-1&0&1 
  \end{array} \right)\;\;, 
   & 
   {\bf M}_t^{(0,0)} = \left( \begin{array}{ccc} 
4&0&-2 \\ 
0&0&0 \\
-2&0&1 
  \end{array} \right)\;\;.
  \end{array}
    \label{bmtdefRR}
\ee

From the Heterotic perspective we have a weak-coupling limit $S \rightarrow \infty$, leading to all three gauge couplings being small. The second term in (\ref{bmtdefRR}) is a one-loop correction coming from integrating out a state corresponding to the fundamental Heterotic string with appropriate KK and winding number along the first torus circle in the $T^2$. The precise state has charges (\ref{chargetmi}) and so its nature can be read off from table \ref{tab:HetBPSST2}. 
 
The perturbative heterotic regime (\ref{hetlimitpert}) is a type II$_1$ limit in (\ref{typeii10limdef}). We now would like to track how the four-dimensional supergravity behaves as we move from this regime towards a type II$_0$ limit in (\ref{typeii10limdef}). We then use this to extract whatever information we can on the finite energy microscopic physics.

\subsection{The $\mathrm{II}_1 \rightarrow \mathrm{II}_0$ transition : Stage 1}
\label{sec:transs1}

Starting from the perturbative Heterotic regime (\ref{hetlimitpert}), we now begin to make $s$ smaller, relative to $\log t$. The infrared four-dimensional supergravity remains under control irrespective of the relative magnitudes of $s$ and $t$. But the microscopic finite-energy physics can undergo significant transformations. There are two critical transition regimes, the first is discussed in this section, and the second in section \ref{sec:sdepn}. 

The first regime of interest is where
\be 
\text{Mixed II$_0$-II$_1$\;}:\;\left(T-i\right)S \sim s \log t  \rightarrow 0 \;\;,\;\; \frac{S}{\log \left(T-i\right)} \sim \frac{\log t}{\log s}  \rightarrow \infty  \;.
\label{hetstage1lim}
\ee 
This is a mixed II$_1$ and II$_0$ limit in the moduli space.

With regards to the leading form of the gauge kinetic function (\ref{gkmiiiisideSTcoRR}), there is no significant difference with respect to the perturbative Heterotic regime (\ref{hetlimitpert}). However, the graviphoton embedding into the gauge fields tells a different story. There are two aspects to this, which we describe in turn. 
	
Using the expression for the periods (\ref{Hbasis}) or (\ref{sympframesforheST2}), and the definition (\ref{gravWdef}), we can write the limiting values of the electric and dual graviphoton field-strengths as
\bea 
\text{Type II}\;\;&:&\;\;\frac{W_{\mu\nu}}{M_p} \rightarrow \frac{1}{\sqrt{2\pi}\; m_s^H} \left(  2 F^1_{\mu\nu} +\tilde{F}^2_{\mu\nu} \right) \;,\nn \\
& &\;\;\frac{\tilde{W}_{\mu\nu}}{M_p} \rightarrow \frac{1}{\sqrt{2\pi}\; m_s^H} \left( 2 \tilde{F}^1_{\mu\nu} -F^2_{\mu\nu} \right) \;.
\label{gravlimtypII}
\eea 
These can be compared with the limiting values in the large complex-structure regime $\mathrm{Im\;}T \gg 1$, which read
\bea 
\text{LCS}\;\;&:&\;\;\frac{W_{\mu\nu}}{M_p} \rightarrow \frac{1}{\sqrt{2\pi}}\;\frac{\mathrm{Im\;}T}{m_s^H}\; \tilde{F}^0_{\mu\nu} \;,\nn \\
& &\;\;\frac{\tilde{W}_{\mu\nu}}{M_p} \rightarrow \frac{1}{\sqrt{2\pi}}\;\frac{\mathrm{Im\;}T}{m_s^H} \left(  -F^0_{\mu\nu} \right) \;.
\label{gravlimlcs}
\eea
At large complex-structure, we have, up to an overall electric-magnetic flip, a natural electric embedding of the graviphoton into $F^0_{\mu\nu}$. Around the type II locus, this is no longer the case. In particular, the embedding of the electric graviphoton is into a combination of the electric and magnetic parts of the different directions. Specifically, the embedding into $F^1_{\mu\nu}$ is electric, while into the $F^2_{\mu\nu}$ directions it is magnetic. 

From the gauge kinetic matrix perspective, the variation from large complex-structure to the type II locus appears rather mild. The leading form of the matrix is just the classical large volume expression (\ref{Nst2class}) evaluated around $T=i$ rather than large $T$. So from the perturbative Heterotic perspective it is just a change in the $T^2$ circle radius. However, the embedding of the graviphoton shows that it is actually highly non-trivial. In particular, the magnetic $\tilde{W}_{\mu\nu}$ is completely rearranged, from an embedding into the electric $F^0_{\mu\nu}$ into a mixed embedding into $\tilde{F}^1_{\mu\nu}$ and $F^2_{\mu\nu}$. This suggests that the microscopic physics should also undergo a non-trivial transition between the two regions. The perturbative limit (\ref{hetlimitpert}) somewhat shields this transition, but it becomes more manifest in the regime (\ref{hetstage1lim}). 

To understand this transition, we should look at the physical magnetic field strengths which are the $G_{I,\mu\nu}$, rather than the $\tilde{F}^I_{\mu\nu}$. We can write the anti self-dual part of the graviphoton (\ref{gravWdef}) as
\be
W_{\mu\nu}^- = 2e^{\frac{K}{2}}\left( \left( -G_{0,\mu\nu} -2i \left(\mathrm{Im\;}S\right)\left(\mathrm{Im\;}T-1\right) F^0_{\mu\nu} + ... \right) + \left(X^0\right)^{-1} \sum_{i=1,2} \left( -X^i G_{i,\mu\nu} + F_i F^i_{\mu\nu} \right)  \right) \;,
\label{gravWst1}
\ee
where the ellipses denote sub-leading terms and we have set, for simplicity, $\mathrm{Re\;}T=\mathrm{Re\;}S=0$.
We see that in the perturbative Heterotic regime (\ref{hetlimitpert}) the electric field-strength $F^0_{\mu\nu}$ dominates over the magnetic field-strength $G_{0,\mu\nu}$ in the graviphoton, while in the regime (\ref{hetstage1lim}) this is inverted. This is the key aspect of the transition. Since the anti self-dual part of the graviphoton is directly related to the central charge and to the mass of BPS states, we see that this corresponds to the transition where the Heterotic Kaluza-Klein monopole associated to the $R_2$ direction, with mass (see table \ref{tab:HetBPSST2})
\be 
\left(\frac{M_{KK5}}{2 \pi \;m_s^H}\right)^2 = \left(\mathrm{Im\;}S\right)^2\left(\left(\mathrm{Im\;}T\right)^2-1\right)^2 \;,
\label{kk5nps}
\ee  
becomes lighter than the Kaluza-Klein (or Winding) mode along the same direction with mass
\be 
\left(\frac{M_{KK}}{2 \pi \;m_s^H}\right)^2 = 1 \;.
\ee 
Microscopically, the more drastic point is that the Kaluza-Klein Monopole becomes lighter than a fundamental string scale, even though it is a non-perturbative state from the Heterotic perspective. 


The transition where the KK monopole becomes lighter than the string scale is very much analogous to the conifold locus in the type IIB setting. There, a wrapped D3 brane, which is a non-perturbative state from the type IIB perspective, becomes lighter than the string scale. In that regime, there is no known microscopic description of the physics, since it would require a non-perturbative string state to be treated dynamically at the string scale. 

The Heterotic transition is even more drastic. This is because there is a different fundamental string state which is even lighter than the Kaluza-Klein monopole. This is just the electric state with charges (\ref{chargetmi}). This state is not mutually local with respect to the monopole.\footnote{In terms of the charge vector ${\bf q}^{(e)}$ and ${\bf p}^{(m)}$ defined in (\ref{chargetmi2}) and (\ref{pmch}), this is the statement that ${\bf q}^{(e)} \cdot {\bf p}^{(m)} \neq 0$.} We are therefore in a doubly exotic situation, where a non-perturbative state is lighter than the fundamental string state, and the two are also mutually non-local. In the conifold case, the dual mutually non-local state is a heavy non-perturbative state, and so it is a milder situation. 

Of course, the four-dimensional supergravity in the infrared is perfectly well-behaved and controlled, but this is because all of the states discussed here have been integrated out. A dynamical theory at the Heterotic string scale, however, would need to be exotic. 

\subsubsection*{The infrared weak-coupling}

It is interesting that even though a magnetic state becomes lighter than an electric state, the infrared four-dimensional supergravity remains weakly-coupled. This should be contrasted to what happens at large complex-structure, as discussed in section \ref{sec:lcsp},  where moving from the IIA to Heterotic regimes involves a strong coupling transition in the supergravity. To understand this, let us consider the light non-perturbative state (\ref{kk5nps}), from the supergravity perspective. This is a state with magnetic charges
\be 
\text{KK5\;\;:\;\;} p^0 = 1 \;,\;\; p^1=p^2=0 \;.
\ee 
We can these magnetic charges as a magnetic charge vector
\be 
{\bf p}^{(m)} = \left( \begin{array}{c} 1 \\ 0 \\ 0 \end{array} \right) \;.
\label{pmch}
\ee 
The BPS mass of the state can be read off from the central charge formula (\ref{defcenhargint}), and periods (\ref{Hbasis}), in Planck units, as
\be 
\left(\frac{M_{(m)}}{M_p}\right)^2 = \pi\; \mathrm{Im\;}S \left(\left(\mathrm{Im\;}T\right)^2-1\right)^2\;.
\label{massmag}
\ee 

This mass is lighter than the generic electric state, implying the non-perturbative microscopic physics. However, there is a distinguished electric state which remains lighter than the monopole state. There is no natural electric-magnetic dual to a given charged state in ${\cal N}=2$ supergravity, but one can define a close analogue as follows. We consider the relation (\ref{GNFminrel}). It is useful to write this relation in terms of the vectors of field strengths ${\bf G}_{\mu\nu}$ and ${\bf F}_{\mu\nu}$, defined in (\ref{fieldstrengvectors}). The leading order behaviour of this relation takes the form
\be 
{\bf p}^{(m)} \cdot {\bf G}^-_{\mu\nu} = -\frac{\log |t|}{2 \pi i} \;{\bf q}^{\left(e\right)}\cdot {\bf F}^-_{\mu\nu} + ... \;,
\label{ptoeemrel}
\ee 
where the charge ${\bf q}^{\left(e\right)}$ is given in (\ref{chargetmi}), which we reproduce here for convenience
\be 
{\bf q}^{\left(e\right)} = \left( \begin{array}{c} -2 \\ 0 \\ 1 \end{array} \right) \;. 
\label{chargetmi2}
\ee 
The relation (\ref{ptoeemrel}) is analogous to electric-magnetic relation in the case of a single $U(1)$. It singles out the state with charge ${\bf q}^{\left(e\right)}$ as dual to the magnetic state. To be precise in what this means, we can write the real part of (\ref{ptoeemrel}) as 
\be
G^{0}_{\mu\nu} = -\frac{\log |t|}{4 \pi} \epsilon_{\mu\nu}^{\;\;\;\;\rho\sigma}\left(-2 F^{0}_{\rho\sigma} + F^{2}_{\rho\sigma} \right) + ...  \;.
\label{G0relation}
\ee
So the magnetic state (\ref{pmch}) couples to the same combination of $U(1)$s , $-2 F^{0}_{\rho\sigma} + F^{2}_{\rho\sigma}$ , as the electric state (\ref{chargetmi2}). This is the sense in which they are electric-magnetic duals. 

The mass of this dual electric state is given as
\be 
\left(\frac{M_{(e)}}{M_p}\right)^2 = \frac{\pi}{\mathrm{Im\;}S} \Big(\left(\mathrm{Im\;}T\right)^2-1\Big)^2\;.
\label{masselect}
\ee
Because the light electric state (\ref{masselect}) is lighter than the light magnetic state (\ref{massmag}), the theory remains weakly-coupled with respect to the $\log t$ dependence. That is, the two states in (\ref{ptoeemrel}) satisfy the natural mass hierarchy, even though they are both much lighter than other electric states. This is why the theory remains weakly-coupled.

The non-perturbative aspects of the microscopic theory are therefore associated to the mixing of the different $U(1)$ directions. That is, that the magnetic state with charges (\ref{pmch}) is lighter than an electric state with say electric charges $(1,0,0)$. It is in the mixing between the light $U(1)$ directions and the other directions that the interesting physics occurs.  We therefore should investigate this mixing in order to extract information from the four-dimensional supergravity on the microscopic physics. There are two aspects to, which we study in sections \ref{sec:factoriz} and \ref{sec:sdepngravmatii0}. 

Let us summarise here concisely the crucial point, which is developed in more detail in section \ref{sec:sdepngravmatii0}. In the regime (\ref{hetstage1lim}), we have that the graviphoton is embedded into both electric and magnetic fields (\ref{gravlimtypII}). This means that the matter sectors associated to vector multiplets, that are orthogonal to it, are also not  purely electric or magnetic. Therefore, in writing the electric gauge kinetic matrix, we are not looking at pure matter sectors, but rather at combinations of them with the graviphoton and with other matter sectors. This is why everything appears weakly-coupled. We show in section \ref{sec:sdepngravmatii0} how to identify the strongly-coupled matter sub-sector. 

\subsection{No factorization of the gauge group}
\label{sec:factoriz}

In section \ref{sec:transs1} we discussed how the transition towards the type II$_0$ locus involves a magnetic state becoming lighter than some electric states. The electric state dual to the light magnetic state, as defined by (\ref{ptoeemrel}), is lighter still, and therefore along that specific $U(1)$ direction there is nothing exotic happening from the infrared supergravity perspective, and the theory stays weakly-coupled. The more non-trivial physics is therefore associated to the mixing between the $U(1)$ direction involving the light states and the other $U(1)$ directions. A first question to ask is whether, at classical tree-level, the $U(1)$ involving the light state can factor out from the other $U(1)$ directions. In this section we show that this is not possible. 

We begin by noting that the two charge vectors (\ref{chargetmi2}) and (\ref{pmch}) are not parallel, and so cannot be aligned. This arises because the gauge kinetic matrix (\ref{bmtdefRR}) is not diagonal. This is true already at the classical Heterotic level, so at leading order in $S$ or $\log t$. A decoupled $U(1)$ sector of charged states would need to be such that gauge kinetic matrix takes a block-diagonal form, where there is no mixing between the $U(1)$ direction and the other two $U(1) \times U(1)$ directions. In that case, the two charge vectors of the light states (\ref{chargetmi2}) and (\ref{pmch}) would align and define a single $U(1)$ direction, along which one would be electric and the other magnetic. 

The question we would like to answer is therefore whether there exists an integral symplectic $Sp\left(6,\mathbb{Z}\right)$ transformation that can diagonalize the gauge kinetic matrix, at least at leading order. Recall that the gauge kinetic matrix gives the relation between the electric and magnetic periods (\ref{FNXfpr}). A block-diagonal factorized form would then imply that both an electric and magnetic period would vanish simultaneously. We now prove that there is a $\mathbb{Z}_2$ obstruction to this.

It is convenient to use the Heterotic variables for legibility. The large complex-structure period vector, in the Heterotic basis, takes the form (\ref{perveclcsA}), which at leading order is 
\begin{equation}
\Pi=
\begin{pmatrix}
1\\
T\\
1-\,T^{2}\\
S \left(1+T^{2}\right)\\
-2 S T\\
- S
\end{pmatrix} \;.
\ee 
The leading order of the period vector at the type II locus is just the evaluation of this at $T=i$, which gives 
\be 
\Pi^{T=i}=
\begin{pmatrix}
1\\
i\\
2\\
0\\
-2 S i\\
- S
\end{pmatrix}\;.
\end{equation}
The dictionary to the basis (\ref{Hbasis}), ignoring vanishing subdominant contributions, is given explicitly in section \ref{sec:hetonk3t2}. 
We want to show that there exists no integer symplectic matrix
$M\in Sp(6,\mathbb Z)$, so satisfying
\begin{equation}
\label{symp}
M^{T}\,\eta\,M=\eta \quad \text{or equivalently} \quad M\;\,\eta\,M^T=\eta\;,
\end{equation}
with
\begin{equation}
\eta=
\begin{pmatrix}
0 & \mathbb{1}\\
- \mathbb{1} & 0
\end{pmatrix}\;,
\end{equation}
such that the transformed vector $M\cdot \Pi^{T=i}$ has two components vanishing for all values of $S$. 

For a given component of $M\cdot \Pi^{T=i}$ to vanish for all $S$, the
corresponding row $r\in\mathbb Z^{6}$ of $M$ must annihilate both the
$S$-independent and the $S$-dependent parts of $\Pi^{T=i}$. This requirement
forces $r$ to be of the form
\begin{equation}
r(a,b)=\left(2a,0,-a,b,0,0\right)\;,
\qquad a,b\in\mathbb Z.
\end{equation}
Therefore, for $M\cdot \Pi^{T=i}$ to have two vanishing entries, the
matrix $M$ must contain two linearly independent rows of this type,
$r(a,b)$ and $r(c,d)$. Their symplectic pairing, denoted by $\langle \;.\; ,\;.\; \rangle$ , in particular satisfies 
\begin{equation}
\label{even}
    \langle\; r(a,b)\;,\;r(c,d)\;\rangle = r(a,b)\cdot\eta\cdot r(c,d)^T = 2\left(bc-ad\right)\in2\mathbb{Z} \;.
\end{equation}
However, symplecticity of $M$ (\ref{symp}) imposes that its rows satisfy
\begin{equation}
    r_i\cdot \eta\cdot r_j^T = \eta_{ij}\in\{0,-1,1\}\;,
\end{equation}
and linear independence that 
\begin{equation}
    bc-ad\neq 0\;,
\end{equation}
meaning that no such integer symplectic transformation exists.

In fact, the obstruction to decoupling this would-be $U(1)$ does not rely on any analytic fine print of the period vector; it is a statement about the integral structure of the light charge sub-lattice. Denote by  $\Gamma\cong\mathbb{Z}^6$ the full electromagnetic charge lattice. Using the heterotic dual description, one can identify the light state sub-lattice
\begin{equation}
    \Gamma_{\{T = i\}} := \{\gamma\in \Gamma\,;\, \langle\gamma, \Pi^{T = i} \rangle= 0\}\;,
\end{equation}
with the electromagnetic charge lattice of the Cartan $U(1)$ inside the enhanced $SU(2)$ gauge sector. In the root-lattice normalization appropriate to this enhancement (only adjoint/W-boson-type charges are light), the $\eta$ pairing restricted to $\Gamma_{\{T = i\}}$ is even, i.e $\langle\Gamma_{\{T = i\}},\Gamma_{\{T = i\}}\rangle\in 2\mathbb{Z}$ which we see explicitly in (\ref{even}). But an integrally decoupled $U(1)$
factor would require a primitive rank-two symplectic summand, so
$\Lambda\simeq \mathbb Z^2\subset\Gamma$ with $\Gamma = \Lambda\oplus\Gamma/\Lambda$ admitting generators $e,m$ with
$\langle e,m\rangle=1$ , so that one could choose an integral symplectic basis in
which that $U(1)$ occupies a standard $2\times 2$ block.
The evenness of the pairing on $\Gamma_{\{T=i\}}$ forbids such a primitive
summand: any candidate pair of independent light charges has pairing in
$2\mathbb Z$ and hence can never realize the unit Dirac pairing required for an
integral symplectic block. Equivalently the vanishing-cycle/light sublattice associated with the $SU(2)$ enhancement does not embed primitively into $\Gamma$. Consequently, a decoupled $U(1)$ block can only be obtained after a half-integral change of basis, which is not allowed in $Sp(6,\mathbb{Z})$.

What does the absence of integral factorization mean physically? It means that any physics which is integrally quantized, such as the spectrum of particles or gauge flux backgrounds, cannot affect only the $U(1)$ direction with light states. Another way to think of this is that the only electric states that are mutually local with the light magnetic state (\ref{pmch}) are those with electric charges of the form
\be 
{\bf q}_{\left({\bf p}^{(m)}-\mathrm{local}\right)} = \left( \begin{array}{c} 0 \\ m \\ n \end{array}\right) \;,\;\;m,n \in \mathbb{Z} \;,
\label{mutualllocalsta}
\ee 
so that
\be 
{\bf q}_{\left({\bf p}^{(m)}-\mathrm{local}\right)} \cdot {\bf p}^{(m)} =0 \;.
\ee 
But those states are, in turn, not orthogonal electrically to the light electric state (\ref{chargetmi2}), 
\be 
{\bf q}_{\left({\bf p}^{(m)}-\mathrm{local}\right)} \cdot  \left( {\bf B}_t + {\bf M}_t^{(0,0)} \right) \cdot {\bf q}^{\left(e\right)} \neq 0 \;,
\ee 
where the inner product is taken with respect to the leading gauge kinetic matrix (\ref{bmtdefRR}). This means that we cannot construct a $U(1) \times U(1)$ theory, with a complete spectrum of states that are mutually local to the light magnetic state, so the states with charges (\ref{mutualllocalsta}), by projecting out the electric $U(1)$ direction under which the light electric states are charged. 

\subsection{A non-integral diagonal basis}
\label{sec:diag}

While a diagonal integral basis does not exist, it is, of course, possible to go to a non-integral diagonal basis. This diagonal basis is useful for understanding much of the physics.  It is simplest to reach such a basis by a block symplectic transformation, so that the prepotential is invariant. The transformation is symplectic but, as proven in section \ref{sec:factoriz}, cannot be integral. It therefore takes the form
\begin{equation}
   S_{\frac12} =  \left(\begin{array}{cccccc}
    0& 0 & 1 & 0 &0&0\\
    0& 1 & 0 & 0 &0&0\\
    2& 0 & -1 & 0 &0&0\\
    0& 0 & 0 & \frac{1}{2} &0&1\\
    0& 0 & 0 & 0 &1&0\\
    0& 0 & 0 & \frac{1}{2} &0&0
  \end{array} \right)\;.
  \label{symMtodiag}
\end{equation}
So it takes us to a half-integral basis. This is the closest to an integral symplectic diagonal basis that it is possible to reach.

After performing the change of basis, the gauge kinetic matrix, in the forms (\ref{gkmiiiisideSTcoRR}) and (\ref{NHtypeIIstbasis}), transforms into
\bea 
    \mathcal{\hat{N}}_{\mathrm{II}}^H &=& -S\;\Big( {\bf \hat{B}}_t + {\bf \hat{M}}^{(0,0)}_t \Big)+\frac{2}{2\pi i}\log \left(T-i\right) \;{\bf \hat{M}}^{(0,0)}_t + ...\;, \\
    &=& -\frac{\log t}{2\pi i }\;\Big( {\bf \hat{B}}_t+ {\bf \hat{M}}_t  \Big) -\frac{\log s}{2\pi i}\;\Big( {\bf \hat{B}}_s+ {\bf \hat{M}}_s  \Big) + ... \;.
  \label{diagnonintgkmiiiisideSTcoRR}
\eea 
The transformed matrices read
\begin{align}
     {\bf \hat{B}}_t &= \frac{1}{2}\left( \begin{array}{ccc} 
1&0&0 \\ 
0&4&0 \\
0&0&-1 
  \end{array} \right)\;,\quad
  {\bf \hat{B}}_s = 2\left( \begin{array}{ccc} 
1&0&0 \\ 
0&4&0 \\
0&0&0 
  \end{array} \right)\;\;, \nn \\
  {\bf \hat{M}}_t &= \left( \begin{array}{ccc} 
0&0&0 \\ 
0&0&0 \\
0&0&1 
  \end{array} \right)+2\beta\left( \begin{array}{ccc} 
0&0&\mathrm{Re\;}s \\ 
0&0&2  \;\mathrm{Im\;}s \nn \\
\mathrm{Re\;}s&2 \; \mathrm{Im\;}s&0 
  \end{array} \right)+\dots \;,\\
  {\bf \hat{M}}_s &= 4 \beta \bar{s} \left( \begin{array}{ccc} 
0&0&1\\ 
0&0&2 i \\
1&2 i&0
  \end{array} \right)+\dots \;.
  \label{rotamatb}
\end{align}

The periods (\ref{Hbasis}) transform to a (non-integral) basis
\bea
\label{nonintdiagHbasis}
    2\pi i \beta  \hat{X}^0 &=& 2 i + 2i s^2 \left( \frac{5}{36} + \beta^2 \right)+\cdots\;, \nn\\
     2\pi i \beta \hat{X}^1 &=& -1+s^2\left(\beta^2-\frac{5}{36}\right)+\cdots\;, \nn\\
     2\pi i \beta \hat{X}^2 &=& -4 i \beta\;s \left(1 - \frac{t}{16}\right)  +\cdots\;, \nn\\
    \left(2\pi i\right)^2 \beta \;\hat{F}_0 &=& -i\;\log \left(t s^4\right)  +... \;, \nn\\
    \left(2\pi i\right)^2 \beta \;\hat{F}_1 &=&2\; \log \left(t s^4\right) + ... \;, \nn \\
    \left(2\pi i\right)^2 \beta \;\hat{F}_2 &=& -2i\beta\;s \; \log t + ...\;.
\eea
The prepotential (\ref{prepotentiatyepIIf}) and Kahler potential (\ref{Kahlerpotentypiif}) remain invariant, since (\ref{symMtodiag}) is a block diagonal symplectic transformation. 

The charges of the light electric and magnetic states (\ref{chargetmi2}) and (\ref{pmch}) now read
\be 
{\bf \hat{q}}^{(e)} = \left( \begin{array}{c} 0 \\ 0 \\ -1 \end{array} \right) \;\;,\;\; {\bf \hat{p}}^{(m)} = \left( \begin{array}{c} 0 \\ 0 \\ 2 \end{array} \right) \;.
\label{chargetmi2trans}
\ee 

In terms of the new basis, the asymptotic graviphoton expressions (\ref{gravlimtypII}) now read
\bea 
\text{Type II}\;\;&:&\;\;\frac{W_{\mu\nu}}{M_p} \rightarrow \frac{1}{\sqrt{2\pi}\; m_s^H}  \left(  2 \hat{F}^1_{\mu\nu} +\hat{\tilde{F}}^0_{\mu\nu} \right) \;,\nn \\
& &\;\;\frac{\tilde{W}_{\mu\nu}}{M_p} \rightarrow \frac{1}{\sqrt{2\pi}\; m_s^H}  \left( 2 \hat{\tilde{F}}^1_{\mu\nu} - \hat{F}^0_{\mu\nu} \right) \;,
\label{gravlimtypIIhat}
\eea 
with the $\hat{F}^I_{\mu\nu}$ denoting the appropriately transformed gauge field-strengths, as in (\ref{SactionP}), so with 
\be 
\hat{\Gamma}_{\mu\nu} = S_{\frac12} \cdot \Gamma_{\mu\nu} \;,
\ee 
where the electromagnetic field-strengths vector $\Gamma_{\mu\nu}$ is defined in (\ref{PiasXF}). 

This choice of basis manifests more clearly the natural (almost) factorized structure, with the light states (\ref{chargetmi2trans}) now aligned along the $\hat{F}^2_{\mu\nu}$ direction, while the asymptotic form of the graviphoton is along the $\hat{F}^0_{\mu\nu}$ and $\hat{F}^1_{\mu\nu}$ directions. The leading form of the gauge kinetic matrix is now diagonal. 

The basis also manifests the mixing in the sub-leading terms (in $s$) of the gauge kinetic matrix between the $\hat{F}^0_{\mu\nu}$ and $\hat{F}^1_{\mu\nu}$ directions, and the $\hat{F}^2_{\mu\nu}$ direction. These mixing terms show that even if the leading gauge kinetic matrix factorizes (in a non-integral basis), there is residual overlap between the sectors. This residual overlap play an important role in identifying the matter sectors, as studied in section \ref{sec:sdepngravmatii0}, and in identifying the threshold effects, as discussed in section \ref{sec:sdepn}.

\subsection{Gravitational and Matter sectors}
\label{sec:sdepngravmatii0}

As discussed in section \ref{sec:mattergrav}, there is a projection onto gravitational and matter sectors around type II loci. In this example case, in the non-integral basis choice (\ref{rotamatb}), the projection is defined by the matrices ${\bf \hat{M}}_t$ and ${\bf \hat{M}}_s$. There is an informative way to write these projections, at leading order, in the form of an outer-product of vectors. Specifically, we can write 
\bea
{\bf \hat{M}}_t &=& {\bf \hat{q}}^{(e)} \left({\bf \hat{q}}^{(e)}\right)^T + ... \;, \nn \\
{\bf \hat{M}}_s &=& - \left(4 \beta \bar{s}\right)\left(  {\bf \hat{\bar{q}}}^{(b)} \left( {\bf \hat{q}}^{(e)}\right)^T + {\bf \hat{q}}^{(e)}  \left({\bf \hat{\bar{q}}}^{(b)}\right)^T  \right) + ... \;.
\label{MsMtproj}
\eea 
Here ${\bf \hat{q}}^{(e)}$ is given in (\ref{chargetmi2trans}), while ${\bf \hat{q}}^{(b)}$ is given by 
\be 
{\bf \hat{q}}^{(b)} = \left( \begin{array}{c} 1 \\ -2 i \\ 0 \end{array}\right) \;.
\label{qbexcas}
\ee 
The charge vectors ${\bf \hat{q}}^{(e)}$ and ${\bf \hat{q}}^{(b)}$ define a matter subspace orthogonal to the graviphoton, in the sense of the projection (\ref{fmMvecdf}). Each projection in itself gives a particular matter sector. 

The leading term in the projection  ${\bf \hat{M}}_t$, as shown in (\ref{MsMtproj}), is rank one, and gives the matter sector associated to the light W-boson in the Heterotic setting. This sector is weakly-coupled, as can be seen by evaluating the gauge-kinetic matrix induced by the projection (\ref{NMdeftra}), which reads 
\be
 {\cal \hat{N}}_{{\bf \hat{M}}_t} = \left(\frac{\log t}{2 \pi i} \right)^2 {\bf \hat{M}}_t^T \cdot  {\cal \hat{N}}^{-1} \cdot {\bf \hat{M}}_t 
 =   -\frac{\log t}{2 \pi i}
 \left( \begin{array}{ccc} 
0 & 0 & 0 \\ 
0 & 0 & 0 \\ 
0 & 0 & 2 \\ 
\end{array}\right) + ... \;.
 \label{NMdeftratc}
\ee 

The projection  ${\bf \hat{M}}_s$ yields another matter sector, but is much more involved. First, there is an overall factor of $\bar{s}$. This is because that matter sector is strongly-coupled, as can be seen by evaluating the gauge kinetic matrix induced by the projection (\ref{NMdeftra}), which reads
\bea
{\cal \hat{N}}_{{\bf \hat{M}}_s} &=&\left(\frac{\log s}{2 \pi i} \right)^2  {\bf \hat{M}}_s^T \cdot  {\cal \hat{N}}^{-1} \cdot {\bf \hat{M}}_s \nn \\
&=& -\frac{ \log t}{2\pi i} \left( \frac{\log s}{\log t}\right)^2 \left(4 \beta \bar{s} \right)^2 \left( \begin{array}{ccc} 
2 & 4i & 0 \\ 
4i & -8 & 0 \\ 
0 & 0 & 0 \\ 
\end{array}\right) 
+ ... \;.
\label{gaugeindNmim}
\eea 
This vanishes in the limit $s \rightarrow 0$, denoting a strong-coupling regime. However, the precise sense of strong-coupling is subtle, and is discussed in more detail in section \ref{sec:sdepn}. Further, we expect that there is a weakly-coupled dual description of the sector, as discussed in section \ref{sec:transs2}. 


The other interesting aspect is that the charge direction ${\bf \hat{q}}^{(b)}$ is complex. This tells us that the projection mixes electric and magnetic field-strengths. Recall that the projection acts on the anti self-dual parts of the field strengths as (\ref{fmMvecdf}). So explicitly, we can write for the electric field strengths  
\bea 
\label{fmMvecdfmscas}
{\bf \hat{F}}_{{\bf \hat{M}}_s,\mu\nu} &=& \left(\mathrm{Re\;} {\bf \hat{M}}_s \right)\cdot {\bf \hat{F}}_{\mu\nu} + \left(\mathrm{Im\;} {\bf \hat{M}}_s \right)\cdot {\bf \hat{\tilde{F}}}_{\mu\nu}\\ \nn 
&=& \left(\mathrm{Re\;} {\bf \hat{M}}_s \right)\cdot {\bf \hat{F}}_{\mu\nu} + \left(\mathrm{Im\;} {\bf \hat{M}}_s \right)\cdot \left( \mathrm{Im\;} {\cal \hat{N}}^{-1} \cdot  {\bf \hat{G}}_{\mu\nu} -  \mathrm{Im\;} {\cal \hat{N}}^{-1} \cdot  \mathrm{Re\;} {\cal \hat{N}} \cdot {\bf \hat{F}}_{\mu\nu}\right)
\;.
\eea 
So the projected electric field strengths are a combination of the electric and magnetic fields. 
Another way to state the same property is that it projects the self-dual (\ref{fmMvecdfsd}) and anti self-dual (\ref{fmMvecdf}) parts of the fields differently
\be 
{\bf \hat{F}}^{-}_{{\bf \hat{M}_s},\mu\nu} = {\bf \overline{\hat{M}}_s} \cdot {\bf \hat{F}}^{-}_{\mu\nu} \;,\;\; {\bf \hat{F}}^{+}_{{\bf M},\mu\nu} = {\bf \hat{M}_s} \cdot {\bf \hat{F}}^{+}_{\mu\nu}  \;,
\ee 
and therefore does not preserve the electric field strengths
\be 
{\bf \hat{F}}_{\mu\nu} = {\bf \hat{F}}_{\mu\nu}^{+} + {\bf \hat{F}}_{\mu\nu}^{-} \;.
\ee 

The matter sector associated to the projection ${\bf \hat{M}}_t$ is simple to identify, it is the W-boson state in the microscopic Heterotic picture, and in the supergravity it is the state with charge ${\bf \hat{q}}^{(e)}$. What is the matter sector associated to the projection  ${\bf \hat{M}}_s$ ? To understand this we should first consider how many different charges are becoming light in the regime (\ref{hetstage1lim}). There is the BPS state with electric charge ${\bf \hat{q}}^{(e)}$ and there is the BPS state with magnetic charge ${\bf \hat{p}}^{(m)}$, as given in (\ref{chargetmi2trans}). There are therefore two charges becoming light, and it is natural to associated this to the two matter sectors. An objection to this would be that the two charges are electromagnetic duals, so both are charged under the same $U(1)$ vector multiplet, which is only one matter sector. However, they are not precisely dual. Recall that the magnetic state couples to the magnetic field strengths ${\bf \hat{G}}_{\mu\nu}$, while the electric state couples to the electric field strengths ${\bf \hat{F}}_{\mu\nu}$, and the two are related by (\ref{GNFminrel}), which reads
\be 
{\bf \hat{G}}_{\mu\nu}^+ = {\cal \hat{N}} \cdot {\bf \hat{F}}^{+}_{\mu\nu} \;\;,\;\;\; {\bf \hat{G}}_{\mu\nu}^- = \overline{{\cal \hat{N}}} \cdot {\bf \hat{F}}^{-}_{\mu\nu} \;.
\label{GNFminrelvf}
\ee
The light magnetic state couples to the magnetic field strength combination ${\bf \hat{p}}^{(m)} \cdot {\bf \hat{G}}_{\mu\nu}$, which is just $2\hat{G}_{2,\mu\nu}$. In order to relate the electric and magnetic fields, we should use (\ref{GNFminrelvf}), which for the magnetic state ${\bf \hat{p}}^{(m)}$ reads
\bea 
{\bf \hat{p}}^{(m)} \cdot {\bf \hat{G}}^+_{\mu\nu} &=& \frac{\log t}{2 \pi i} \;{\bf \hat{q}}^{\left(e\right)}\cdot {\bf \hat{F}}^+_{\mu\nu} -\frac{\log s}{2 \pi i} \;\left(4 \beta \bar{s} \right)\; {\bf \hat{\bar{q}}}^{(b)}\cdot {\bf \hat{F}}^+_{\mu\nu}\nn \\
&-& \frac{\log t}{2 \pi i} \; 4\beta \left(\begin{array}{c} \mathrm{Re\;}s \\ 2\;\mathrm{Im\;} s \\ 0 \end{array} \right) \cdot {\bf \hat{F}}^+_{\mu\nu} 
 + ... \;.
\label{sdepptoeemrelpl}
\eea 
The first term in (\ref{sdepptoeemrelpl}) tells us that indeed the state with electric charge ${\bf \hat{q}}^{\left(e\right)}$ is the leading component in the dual to the magnetic state ${\bf \hat{p}}^{(m)}$. But there are sub-leading components. Indeed the second term in (\ref{sdepptoeemrelpl}) is precisely the charge ${\bf \hat{q}}^{(b)}$. This identifies it as a part of the magnetic state ${\bf \hat{p}}^{(m)}$, or more precisely part of its electric dual. Its complex nature is capturing the fact that both ${\bf \hat{F}}_{\mu\nu}$ and ${\bf \hat{\tilde{F}}}_{\mu\nu}$ appear in the expression for ${\bf \hat{G}}_{\mu\nu}$. 


In the expression (\ref{sdepptoeemrelpl}) both ${\bf \hat{q}}^{(e)}$ and ${\bf \hat{q}}^{(b)}$ appear. There is a more informative way to understand the electric-magnetic duality which separates the two contributions. This is the duality in the projected matter subspaces, so after the projections by ${\bf \hat{M}}_t$ and ${\bf \hat{M}}_s$. Indeed, the form (\ref{MsMtproj}) acts naturally on magnetic charges from the left, this is because the symplectic inner product between charges that are purely electric and purely magnetic is simple $\left({\bf \hat{q}}^{(e)}\right)^T \cdot {\bf \hat{p}}^{(m)}$. So the projections act naturally on the magnetic charge ${\bf \hat{p}}^{(m)}$, which give projected charges
\bea
{\bf \hat{M}}_t \cdot {\bf \hat{p}}^{(m)}  &=& -2\;{\bf \hat{q}}^{(e)} + ... \;, \nn \\
{\bf \hat{M}}_s \cdot {\bf \hat{p}}^{(m)}&=& 2 \left(4 \beta \bar{s} \right)\;{\bf \hat{\bar{q}}}^{(b)}   + ... \;.
\label{MsMtprojonp}
\eea 

The projections (\ref{MsMtprojonp}) show that we should think of the different terms in  (\ref{sdepptoeemrelpl}) as corresponding to the two different matter sectors.  Indeed, from the general expression (\ref{NBformpreGtFMqe}) we have that ${\bf \hat{p}}^{(m)}$ yields a projection to the matter sector of the type II$_0$ locus, so associated to the $\log s$ part. The $\log t$ parts of (\ref{sdepptoeemrelpl}) instead correspond to the matter sector associated to $\hat{\bf M}_t$ and $\log t$. If we consider the leading terms for each logarithm, associated to the two matter sectors, we have that (\ref{NBformpreGtFMqe}) reads 
\bea 
{\bf \hat{p}}^{(m)} \cdot {\bf \hat{G}}^+_{\mu\nu} &=& \frac{\log t}{2 \pi i} \;{\bf \hat{q}}^{\left(e\right)}\cdot {\bf \hat{F}}^+_{\mu\nu} -\frac{\log s}{2 \pi i} \;\left(4 \beta \bar{s} \right)\; {\bf \hat{\bar{q}}}^{(b)}\cdot {\bf \hat{F}}^+_{\mu\nu}  + ... \;.
\label{sdepptoeemrelply}
\eea 

In order to compare the two sectors we should appropriately normalize the gauge field strengths, using the gauge kinetic matrices (\ref{NMdeftratc}) and (\ref{gaugeindNmim}). So let us define normalized field strengths ${\bf \hat{F}}^{(t),+}_{\mu\nu}$ and ${\bf \hat{F}}^{(s),+}_{\mu\nu}$ as
\begin{align}
{\bf \hat{F}}^+_{\mu\nu} \cdot {\cal \hat{N}}_{{\bf \hat{M}}_t}\cdot {\bf \hat{F}}^{+,\mu\nu} &={\bf \hat{F}}^{(t),+}_{\mu\nu} \cdot  {\cal \hat{N}}_{{\bf \hat{M}}_t}^{(t)} \cdot {\bf \hat{F}}^{(t),+}_{\mu\nu}=  -\frac{\log t}{2 \pi i}\;\;{\bf \hat{F}}^{(t),+}_{\mu\nu} \cdot  \left( \begin{array}{ccc} 
0 & 0 & 0 \\ 
0 & 0 & 0 \\ 
0 & 0 & 2 \\ 
\end{array}\right)\cdot {\bf \hat{F}}^{(t),+}_{\mu\nu}  \;, \nn \\
{\bf \hat{F}}^+_{\mu\nu} \cdot {\cal \hat{N}}_{{\bf \hat{M}}_s}\cdot {\bf \hat{F}}^{+,\mu\nu} &= {\bf \hat{F}}^{(s),+}_{\mu\nu} \cdot  {\cal \hat{N}}_{{\bf \hat{M}}_s}^{(s)} \cdot {\bf \hat{F}}^{(s),+}_{\mu\nu}= -\frac{\log t}{2 \pi i}\;\;{\bf \hat{F}}^{(s),+}_{\mu\nu} \cdot  \left( \begin{array}{ccc} 
2 & 4i & 0 \\ 
4i & -8 & 0 \\ 
0 & 0 & 0 \\ 
\end{array}\right) \cdot {\bf \hat{F}}^{(s),+}_{\mu\nu}  \;.
\label{fsgkmtfe}
\end{align} 
We therefore have explicitly
\be
{\bf \hat{F}}^{(t),+}_{\mu\nu}  =  {\bf \hat{F}}^{+}_{\mu\nu} \;,\;\; 
{\bf \hat{F}}^{(s),+}_{\mu\nu} = \left( \frac{\log s}{\log t}\right) \left(4\beta \bar{s} \right) {\bf \hat{F}}^{+}_{\mu\nu} \;\;.
\ee
It terms of the appropriately normalized field strengths we can then write (\ref{sdepptoeemrelply}) as
\bea 
{\bf \hat{p}}^{(m)} \cdot {\bf \hat{G}}^+_{\mu\nu} &=& -\frac{\log t}{2 \pi i} \left(-{\bf \hat{q}}^{\left(e\right)} \cdot {\bf \hat{F}}^{(t),+}_{\mu\nu}+ {\bf \hat{\bar{q}}}^{(b)}\cdot {\bf \hat{F}}^{(s),+}_{\mu\nu}\right)  + ... \;.
\label{sdepptoeemrelplyNm}
\eea 
This shows that, appropriately normalized, the two components of ${\bf \hat{p}}^{(m)} $, one along ${\bf \hat{q}}^{\left(e\right)} $ and one along ${\bf \hat{\bar{q}}}^{(b)}$ , are on an equal footing. 

The effective charge ${\bf \hat{\bar{q}}}^{(b)}$  is precisely the charge in (\ref{typeongenqb}). It is a type $\left(0,2\right)$ charge, which means that its chiral central charges (\ref{typencenchnspiis}) take the form
\bea 
Z_+ &=& -i\left(X^0\right)^{-1}e^{\frac{K}{2}} \;2\sqrt{2\pi}\; {\bf \hat{q}}^{(b)} \cdot {\bf \hat{X}}^{(0)}  + ... \;, \nn \\
Z_- &=& -i\left(X^0\right)^{-1}e^{\frac{K}{2}} \;2\sqrt{2\pi}\;{\bf \hat{\bar{q}}}^{(b)} \cdot {\bf \hat{X}}^{(2)} \;s^{2} + ... \;.
\label{typencenchnspiisqex}
\eea 
So $Z_-$ behaves as $s^2$. This is one extra power of $s$ relative to the chiral central charge $Z_-$ associated to the magnetic state ${\bf \hat{p}}^{(m)}$, which behaves as $s$. This additional suppression has a natural interpretation for the K-point physics studied in \cite{Hattab:2025aok}. In that setting, there were no integral charged states becoming light as $s$, so the magnetic states had a mass of order $s^0$. A similar interpretation would then assign an extra power of $s$ to the dual of the magnetic state, which would lead to a chiral central charge $Z_-$ behaving as $s$. This is indeed what was proposed in \cite{Hattab:2025aok}.  

\subsection{Finite-energy physics and threshold corrections}
\label{sec:sdepn}

The calculation of the gauge kinetic matrix around the type II point in moduli space yields its infrared form. There are two types of contributions to this infrared form. The first type are ultraviolet contributions which fix the values of the gauge couplings at high energies. The second are threshold corrections, which arise from integrating out light states. In this section we attempt to distinguish these two types of contributions. 

Let us write the gauge kinetic matrix, in the basis (\ref{rotamatb}), in terms of the parameters $s$ and $t$ explicitly 
\be 
    2 \pi i \;\mathcal{\hat{N}}_{\mathrm{II}}^H = - \frac{1}{2}\left( \begin{array}{ccc} 
1&0&0 \\ 
0&4&0 \\
0&0&1 
  \end{array} \right)\log\left(ts^4\right)\;
  +2\left( \begin{array}{ccc} 
0&0&0 \\ 
0&0&0 \\
0&0&1 
  \end{array} \right)\log s  + ...\;.
  \label{gkmstformrobas}
\ee 
In the type II$_1$ limit (\ref{hetlimitpert}), we have a well-understood perturbative Heterotic string description of the physics, and accordingly, we can distinguish the tree-level ultraviolet physics from the threshold correction. In this case, ultraviolet means the value of the gauge couplings at the Heterotic string scale. As studied in detail in section \ref{sec:hetdual}, the $\log\left(t s^4 \right)$ term in (\ref{gkmstformrobas}) corresponds to the ultraviolet tree-level contributions, while the $\log s$ term in (\ref{gkmstformrobas}) corresponds to a threshold correction from integrating out the light electric state of charge ${\bf \hat{q}}^{(e)}$ in (\ref{chargetmi2trans}). We can write this as
\be 
 \text{Pert. Het} \;:\;   2 \pi i \;\mathcal{\hat{N}}_{\mathrm{II}}^H = \underbrace{- \frac{1}{2}\left( \begin{array}{ccc} 
1&0&0 \\ 
0&4&0 \\
0&0&1 
  \end{array} \right)\log\left(ts^4\right)\;}_{\text{Classical ultraviolet}}
  +\underbrace{2\left( \begin{array}{ccc} 
0&0&0 \\ 
0&0&0 \\
0&0&1 
  \end{array} \right)\log s}_{\text{Quantum threshold from}\; {\bf \hat{q}}^{(e)}} + ...\;.
  \label{pertHetgkmstformrobas}
\ee 

It is worth explicitly stating a simple point regarding the threshold correction from the electric state with charge ${\bf \hat{q}}^{(e)}$. This charge appears in the projection to the matter sector (\ref{MsMtproj}), as the leading component of ${\bf \hat{M}}_t$. The projection ${\bf \hat{M}}_t$ is an exact projection orthogonal to the graviphoton. The charge ${\bf \hat{q}}^{(e)}$ by itself is therefore not exactly orthogonal to the graviphoton, since it is only the leading part of ${\bf \hat{M}}_t$. This can be seen explicitly by noting that its central charge is not identically zero. This means that integrating it out does yield a contribution to the graviphoton kinetic terms. This is crucial because the integrating out calculation should be performed as a background field calculation in a purely graviphoton background. This was the methodology used in \cite{Gopakumar:1998jq}, and in the follow-up works \cite{Dedushenko:2014nya,Hattab:2024ewk,Hattab:2024ssg}. The role of this background field calculation was also studied in the context of the K-point thresholds in \cite{Hattab:2025aok}. If the charge ${\bf \hat{q}}^{(e)}$ was exactly orthogonal to the graviphoton, it could not contribute to such a calculation. Specifically, it could not yield a threshold contribution to the prepotential. The overlap with the graviphoton, of order $s$, is what then leads to the prepotential contribution, which is of order $s^2 \log s$, and then to the gauge kinetic threshold in (\ref{pertHetgkmstformrobas}). 

Now we consider moving to the intermediate regime (\ref{hetstage1lim}). In transitioning to this regime, from the perturbative description, the magnetic state, with charge ${\bf \hat{p}}^{(m)}$ in (\ref{chargetmi2trans}), becomes lighter than the fundamental Heterotic string scale. Its contribution to the gauge kinetic function is therefore naturally transformed from what was considered an ultraviolet, or classical, one, to a threshold correction. 

Because now both the electric ${\bf \hat{q}}^{(e)}$ and magnetic ${\bf \hat{p}}^{(m)}$ states need to have a threshold interpretation, we cannot really treat them as fundamental local independent degrees of freedom. We should say that there is a threshold correction from a light sector, of which ${\bf \hat{q}}^{(e)}$ and ${\bf \hat{p}}^{(m)}$ are part of. We do not know how to fully calculate the threshold from this sector microscopically. Our proposed identification of threshold effects is therefore not certain. Nonetheless, the combination of the matter projections studied in section \ref{sec:sdepngravmatii0}, and of knowing the infrared structure of the gauge kinetic matrix, supports the following proposals. 

We first note that the gauge kinetic matrix is associated to the electric field strengths ${\bf \hat{F}}_{\mu\nu}$ and ${\bf \hat{\tilde{F}}}_{\mu\nu}$. The magnetic state instead couples to the magnetic field strength ${\bf \hat{G}}_{\mu\nu}$. As discussed in section \ref{sec:sdepngravmatii0}, see (\ref{sdepptoeemrelplyNm}), converting one to the other means that the magnetic state develops two components, given (at leading order) by the two matter projections (\ref{MsMtprojonp}). There is a component along the charge ${\bf \hat{q}}^{(e)}$ and one along ${\bf \hat{q}}^{(b)}$. We therefore can schematically expect threshold contributions from these two components, which in the perturbative Heterotic regime (type II$_1$ locus) were classical terms. We can denote this as
\be 
\text{Classical} \rightarrow \text{Threshold from\;} {\bf \hat{p}}^{(m)} 
\begin{array}{cc} 
\multicolumn{2}{l}{\;\;\;\;\;\text{Threshold from dual ${\bf \hat{q}}^{(e)}$ component}} \\ 
\nearrow  & \\ 
\searrow  & \\ 
\multicolumn{2}{l}{\;\;\;\;\;\text{Threshold from dual ${\bf \hat{q}}^{(b)}$ component}}  \\
\end{array} 
\;.
\label{schemspl}
\ee 

There is a natural identification of the gauge kinetic matrix components following the schematic expectation (\ref{schemspl}). We can consider interpreting the infrared gauge kinetic matrix as
\bea 
 \label{mixedHetgkmstformrobas}
2 \pi i \;\mathcal{\hat{N}}_{\mathrm{II}}^H = &- \underbrace{\frac{1}{2}\left( \begin{array}{ccc} 
1&0&0 \\ 
0&4&0 \\
0&0&0 
  \end{array} \right)\log t\;}_{\text{Classical ultraviolet}} 
  &-
  \underbrace{\frac{1}{2}\left( \begin{array}{ccc} 
0&0&0 \\ 
0&0&0 \\
0&0&1 
  \end{array} \right)\log t\;}_{\text{Classical ultraviolet}} 
  \quad\quad\quad  \nn \\  
    & & \nn \\
  \nn 
  &-
   \underbrace{ 2\left( \begin{array}{ccc} 
0&0&0 \\ 
0&0&0 \\
0&0&1 
  \end{array} \right)\log s\;}_{\text{Threshold from}\; {\bf \hat{q}}^{(e)}\;\subset \;{\bf \hat{p}}^{(m)}} 
  &+\underbrace{2\left( \begin{array}{ccc} 
0&0&0 \\ 
0&0&0 \\
0&0&1 
  \end{array} \right)\log s}_{\text{Threshold from}\; {\bf \hat{q}}^{(e)}} + ...\nn \\
    & & \nn \\
  &\;\;\;- 
  \underbrace{ 2\left( \begin{array}{ccc} 
1&0&0 \\ 
0&4&0 \\
0&0&0 
  \end{array} \right)\log s\;}_{\text{Threshold from}\; {\bf \hat{q}}^{(b)}\;\subset \;{\bf \hat{p}}^{(m)}}
 + &\;\;...  \;.
\eea

Let us consider each row in (\ref{mixedHetgkmstformrobas}) in turn. The two terms in the first row are proposed to be ultraviolet contributions, so classical from the perspective of the fundamental string scale. One could consider the possibility that the second term in the first row of (\ref{mixedHetgkmstformrobas}) could be a threshold correction. However, we find this unlikely. Most directly, there is no obvious sector with a mass going as $t$, that could lead to such a threshold. There is also a more subtle, but quite strong reason to expect that a splitting of the two terms in the top row of (\ref{mixedHetgkmstformrobas}) is not correct. 
The separation of the matrices as 
\be 
\left( \begin{array}{ccc} 
1&0&0 \\ 
0&4&0 \\
0&0&1 
  \end{array} \right) \rightarrow \left( \begin{array}{ccc} 
1&0&0 \\ 
0&4&0 \\
0&0&0 
  \end{array} \right) + \left( \begin{array}{ccc} 
0&0&0 \\ 
0&0&0 \\
0&0&1
  \end{array} \right) \;,
  \label{splitnondiag}
\ee 
is a natural one in terms of the leading form of the two matter sectors. However, we should not expect such a splitting in terms of the ultraviolet gauge couplings. Indeed, as discussed in section \ref{sec:factoriz}, the theory does not factorise in an integral basis. There is no splitting of the ultraviolet theory into a $U(1)\times U(1)$ sector and a further decoupled light $U(1)$. In particular, in such a splitting, the two sectors would not be mutually local. It is only a non-integral basis splitting. It is therefore not clear what such a splitting would mean in terms of the ultraviolet gauge couplings, which should control independent sectors. That is, if we chose to split the first two terms in (\ref{mixedHetgkmstformrobas}) into an ultraviolet gauge coupling and a threshold correction, then it would imply that the parameter $\log t$ only controls the gauge couplings of a $U(1) \times U(1)$ sector, which would be inconsistent with the absence of mutually-local factorization into such a sector. 

Consider the second row of (\ref{mixedHetgkmstformrobas}). The second term in that row is the threshold correction associated to the electric state with charge  ${\bf \hat{q}}^{(e)}$ , identified already in the perturbative Heterotic regime (\ref{pertHetgkmstformrobas}). The first term in the row is proposed to also be a threshold correction. It can be attributed schematically to the magnetic state ${\bf \hat{p}}^{(m)}$, and corresponds to the ${\bf \hat{q}}^{(e)}$ part of its threshold, in the sense of (\ref{sdepptoeemrelplyNm}), and as illustrated in (\ref{schemspl}). This threshold contribution cancels precisely against the perturbative electric one. This tells us that the theory behaves conformally along the direction of ${\bf \hat{q}}^{(e)}$, so that there is no net renormalisation associated to the sector whose mass scale is set by $s$. It  is naturally expected to be a strongly-coupled Conformal Field Theory (CFT) sector, involving both electric and magnetic dynamical states, and so having no local Lagrangian description. This is somewhat similar to Argyres-Douglas (AD) theories \cite{Argyres:1995jj}. 


The third row of (\ref{mixedHetgkmstformrobas}) is also proposed to be a threshold correction. It can be attributed schematically to the magnetic state ${\bf \hat{p}}^{(m)}$, and corresponds to the ${\bf \hat{q}}^{(b)}$ part of its threshold, in the sense of (\ref{sdepptoeemrelplyNm}), and as illustrated in (\ref{schemspl}). The matrix appearing is ${\bf \hat{B}}_s$, and this indeed corresponds to the natural form of a threshold correction from a charge ${\bf \hat{q}}^{(b)}$, as shown generally in (\ref{Bintermsofb}). In the rotated non-integral basis we can write this as
\be 
{\bf \hat{B}}_{IJ} = {\bf \hat{q}}_I^{(b)}\; {\bf \hat{\bar{q}}}_J^{(b)} + \hat{\bar{{\bf q}}}_I^{(b)}\; {\bf \hat{q}}_J^{(b)}  \;.
\label{Bintermsofbrr}
\ee 
The corresponding term in the gauge kinetic matrix, so the one in the third row of (\ref{mixedHetgkmstformrobas}), is then of the form (\ref{NBformb2}), which in the non-integral basis reads
\be 
\left({\cal \hat{N}}_{\mathrm{II}}^H\right)_{IJ} \supset -\frac{1}{2 \pi i} \Big( {\bf \hat{q}}_I^{(b)}\; {\bf \hat{\bar{q}}}_J^{(b)} + \hat{\bar{{\bf q}}}_I^{(b)}\; {\bf \hat{q}}_J^{(b)}\Big) \log s  \;.
\label{NBformb2rr}
\ee 
The term is matched onto the term in the prepotential of the threshold form, as in (\ref{FBformb2fgen}), involving the chiral central charges (\ref{typencenchnspiisqex}), so explicitly  
\be 
F \supset -\frac{i}{16 \pi^2}\;\frac{1}{2} \; \left(X^0\right)^2e^{-K} Z_+ Z_- \;\log \left( \left(X^0\right)^2e^{-K} Z_+ Z_- \right) + ... \;.
\label{FBformb2fgenrr}
\ee
In summary, the proposal for the threshold interpretation relies on the forms of the gauge kinetic matrix (\ref{NBformb2rr}) and prepotential (\ref{FBformb2fgenrr}), in relation to the complex charge ${\bf \hat{q}}^{(b)}$, which in turn features as the leading component of the projection of the light magnetic state to the matter sector associated with the $\log s$ directions, as in (\ref{sdepptoeemrelplyNm}). 

It is worth noting that the absence of the splitting (\ref{splitnondiag}) at the gauge group level lends further support to interpreting the third row of (\ref{mixedHetgkmstformrobas}) as a threshold correction. The alternative situation is that it should be an ultraviolet tree-level contribution. But there is no analogous classical $\log s$ term along the $\hat{F}^2_{\mu\nu}$ direction. At least not if we identify the second row of (\ref{mixedHetgkmstformrobas}) as a threshold correction. Therefore, it would again be a splitting of the ultraviolet gauge couplings in a way which is not compatible with the gauge group and mutual locality. 

Indeed, (\ref{mixedHetgkmstformrobas}) suggests that simply all the $\log s$ dependence in the infrared theory should be interpreted as infrared physics, so threshold corrections, while the $\log t$ dependence sets the ultraviolet classical values. 

This threshold correction interpretation is one of the most central aspects of the study. It is crucial because in the pure type II$_0$ limit it is the leading term in the gauge kinetic matrix and in the Kahler potential. This would mean that the leading divergences, including the divergent distance, associated to the type II$_0$ locus would be due to a threshold effect. So would be infrared, rather than ultraviolet, in nature. This was also the proposal for the one-parameter K-point setting in \cite{Hattab:2025aok}. We study this pure type II$_0$ regime next, in section \ref{sec:transs2}.

 \subsection{The $\mathrm{II}_1 \rightarrow \mathrm{II}_0$ transition : Stage 2}
\label{sec:transs2}

So far, we have considered the mixed  II$_1$-II$_0$ regime in parameter space (\ref{hetstage1lim}). As we keep decreasing $s$, we enter a new regime of interest where
\be 
\text{Pure II$_0$\;}:\; -\log s \rightarrow \infty  \;\;,\;\; -\log t = \text{constant} \gg 1 \;.
\label{hetstage2lim}
\ee 
This can be regarded as a pure type II$_0$ limit. Of course, we still keep $\log t$ large, so that the supergravity description is valid.

In the regime (\ref{hetstage2lim}), the terms (\ref{NBformb2rr}) in the gauge kinetic function, and (\ref{FBformb2fgenrr}) in the prepotential, are the leading terms. In particular, their associated term in the Kahler potential (\ref{Kahlerpotentypiif}) also becomes dominant, which we reproduce here for convenience 
\be 
\frac{\pi}{4} \;e^{-K} = -\log |t| - 4 \log |s| + ... \;.
\label{Kahlerpotentypiifrr} 
\ee 
The $\log s$ divergence in (\ref{Kahlerpotentypiifrr}) is what is responsible for the type II$_0$ locus being at infinite distance in moduli space. Within the framework, discussed in section \ref{sec:sdepn}, where this is associated to a threshold correction, that means that the infinite distance is emergent in the infrared. In this setting, the ultraviolet distance would be associated to the $\log t$ part of (\ref{Kahlerpotentypiifrr}). So the transition from the mixed regime (\ref{hetstage1lim}) to the pure type II$_0$ regime (\ref{hetstage2lim}), is proposed to be one where the infinite distance is transferred from an ultraviolet property to an emergent infrared one. 

From a perturbative Heterotic perspective, we have that the ratio of the Planck scale to the string scale behaves as in (\ref{mstomprel}), so explicitly we have
\be 
\frac{2 \pi^2 M_p^2} {\left(2 \pi \;m_s^H\right)^2} = -\log \left|t\right| - 4\log \left|s\right| + ... \;.
\label{mstomprelw2}
\ee
In the regime (\ref{hetstage2lim}), this diverges. However, the threshold interpretation assigns the $\log |s|$ part of (\ref{mstomprelw2}) to a threshold correction. In this picture, the actual string scale in the ultraviolet is controlled only by $\log t$, and stays finite (in Planck units). The string is weakly-coupled, but only by $\log t$, so the type II$_0$ locus is not a weak-coupling limit. 

We note that there are many similarities between this setting and the one-parameter K-point case studied in \cite{Hattab:2025aok}. For example, the embedding of the graviphoton is mixed electric and magnetic (\ref{gravlimtypIIhat}). Also a purely anti self-dual graviphoton background of the form (\ref{gravproj}), but with $W^+_{\mu\nu}=0$ , would require a Wick rotation to Euclidean space and then complex field-strengths. The discussion in \cite{Hattab:2025aok} on the Wick rotation and complex fields can be applied directly also in this setting, in the Heterotic context. Indeed, this was anticipated already in \cite{Hattab:2025aok}, through Heterotic M-theory duality. 

As stated in section \ref{sec:sdepn}, the interpretation of the last term in (\ref{mixedHetgkmstformrobas}) as a threshold correction is central, and can be motivated by various arguments, but we are not able to prove it in the absence of a microscopic understanding of the finite energy physics. That physics is non-perturbative from the Heterotic string perspective, and we have not been able to gain a sufficient understanding of it. It is natural to expect that in the pure type II$_0$ limit (\ref{mstomprelw2}), there should be some simplified effective finite energy description in terms of a light state of effective charge ${\bf \hat{q}}^{(b)}$, with an effective gauge coupling matrix for the normalized field strengths of the form (\ref{fsgkmtfe}).
We reproduce both here for convenience
\be 
{\bf \hat{q}}^{(b)} = \left( \begin{array}{c} 1 \\ -2 i \\ 0 \end{array}\right) \;\;,\;\; 
{\cal \hat{N}}_{{\bf \hat{M}}_s}^{(s)} =  -\frac{\log |t|}{2 \pi i}  \left( \begin{array}{ccc} 
2 & 4i & 0 \\ 
4i & -8 & 0 \\ 
0 & 0 & 0 \\ 
\end{array}\right) 
\;\; \;.
\label{lighteffthelogs}
\ee 
One reason to expect such a description is the simple form of the threshold correction associated to the state. The gauge kinetic matrix (\ref{lighteffthelogs}) has a negative eigenvalue and divergent real parts. This does not lead to a well-defined gauge sector. However, we can redefine
\be 
\hat{F}^{(s),1}_{\mu\nu} \rightarrow i \hat{F}^{(s),1}_{\mu\nu} \;.
\label{F1rot}
\ee 
This transforms the system (\ref{lighteffthelogs}) as
\be 
\hat{F}^{(s),1}_{\mu\nu} \rightarrow i \hat{F}^{(s),1}_{\mu\nu} \;\;:\;\; {\bf \hat{q}}^{(b)} \rightarrow \left( \begin{array}{c} 1 \\ 2 \\ 0 \end{array}\right) \;\;,\;\; 
{\cal \hat{N}}_{{\bf \hat{M}}_s}^{(s)} \rightarrow  -\frac{\log |t|}{2 \pi i}  \left( \begin{array}{ccc} 
2 & -4 & 0 \\ 
-4 & 8 & 0 \\ 
0 & 0 & 0 \\ 
\end{array}\right) 
\;\; \;.
\label{lighteffthelogsro}
\ee 
The gauge kinetic matrix is now well-behaved, with only imaginary components diverging and a positive eigenvalue. The charge ${\bf \hat{q}}^{(b)}$ now becomes integral. The rotation (\ref{F1rot}) also makes the graviphoton (\ref{gravlimtypIIhat}) embedding electric (the rotation should be applied to (\ref{gravWdef}) first). After such a rotation, a light effective description in terms of (\ref{lighteffthelogsro}) becomes completely standard. The threshold effect can be calculated in that framework, and after integrating out the state, the rotation back can be applied. 

Microscopically, this would only be part of the picture. Indeed, a central aspect of the threshold interpretation is that the two chiral central charges $Z_{\pm}$ (\ref{typencenchnspiisqex}) are not equal. This suggests some sort of an asymmetric setting, where the left and right moving sectors of the Heterotic string are treated quite differently. Indeed, the fact that the Heterotic string, defined by an asymmetry between the left and right-moving sectors, appears in the microscopic description is quite promising in this sense. The rotation (\ref{F1rot}) is likely, microscopically, to be related to a Wick-type rotation on the metric, as was discussed for the K-point case in \cite{Hattab:2025aok}.

There is a possible clue as to the microscopic structure that can be extracted from the threshold form by comparing (\ref{NBformb2rr}) with (\ref{N1loopvecto}). We can think of the factor of 2 in (\ref{N1loopvecto}) as corresponding to the $W^+$ and $W^-$ boson contributions, which are identical. In (\ref{NBformb2rr}) we do not have an overall factor of 2, and instead decompose the identical contributions into two complex terms. The sum of them yields a real gauge kinetic matrix. One therefore naturally expects some weakly-coupled description in terms of complex contributions from $W$-bosons.

In \cite{Monnee:2025msf} a microscopic interpretation of the regime (\ref{hetstage2lim}) was proposed. It was proposed that there is a non-perturbative transition involving the nucleation of spacetime-filling NS5 branes, such that the light $U(1)$ direction corresponds to the theory on these branes. The light charged states would then be wrapped E-string states. We have nothing new to add to the microscopic physics of this regime, other than that such a realisation appears in tension with the non-factorization of the gauge group. There was no attempt to consider the threshold effect of the light sector on the infrared theory in \cite{Monnee:2025msf}, and so it is not clear how to connect our picture with the proposed microscopic one. In fact, there is no clear contradiction between the two pictures from the threshold perspective, rather an open question as to what is the effect of integrating out the non-perturbative sector. The appearance of the E-string may well be relevant microscopically. A natural, though very speculative, idea is that the state of charge ${\bf \hat{q}}^{(b)}$ corresponds to a wrapped string state whose left-moving sector would be associated to the Heterotic string, and right-moving sector to an E-string. The former is massive, matching the central charge $Z_+$, while the latter is very light, matching the charge $Z_-$. 

\section{Discussion}
\label{sec:sum}

In this paper we studied various aspects of type II loci (in the classification of \cite{Grimm:2018ohb,Grimm:2018cpv}) in the moduli space of type IIB string theory compactified on Calabi-Yau manifolds to four dimensions. 

We first showed results on type II loci using the tools of Mixed Hodge Structures as introduced in \cite{Grimm:2018ohb,Grimm:2018cpv}. In particular, we showed that  there is always an electric symplectic frame where the prepotential vanishes in the type II limit. We also showed that for type II$_0$ loci defined by a limit $s \rightarrow 0$, it is possible to write the $\log s$ dependence of the prepotential and the gauge kinetic matrix in the form of a threshold correction from a state of effective complex charge ${\bf q}^{(b)}$.

We then discussed how to separate the graviphoton component from the matter vector multiplets.  We showed that for each modulus, there exists a matrix ${\bf M}$ which acts as a type of projection to a sector which is orthogonal to the graviphoton. It identifies pure matter sectors in the theory, and there is a natural pull-back of the gauge kinetic matrix to those sectors. The projection acts on the self-dual parts of the gauge fields, and in general does not respect electric and magnetic splitting. 

We studied a two parameter example Calabi-Yau, with moduli space parameters $s$ and $t$. The example is the prototypical one used for type II-Heterotic duality in \cite{Kachru:1995wm,Kachru:1995fv}, and was recently studied also in \cite{Blumenhagen:2025zgf,Monnee:2025ynn,Monnee:2025msf,Castellano:2026bnx}. We calculated the integral supergravity periods around large complex structure and around the intersection point between type II$_1$ and type II$_0$ loci. Using these, we reproduced in detail some of the known results on Heterotic duality, but expanded beyond. In particular, we calculated higher order corrections to the gauge kinetic matrix,  which are needed to determine the associated matter projection matrices ${\bf M}_t$ and ${\bf M}_s$. 

The perturbative Heterotic regime is dual to the type II$_1$ limit in moduli space. Moving from the II$_1$ regime to the II$_0$ regime passes through two regions. First, there is a regime where the perturbative Heterotic picture breaks down due to a non-perturbative state, a Kaluza-Klein monopole, becoming lighter than a the Heterotic string scale. We showed that there is a $\mathbb{Z}_2$ obstruction to factorization of the gauge group into this non-perturbative sector and a massive perturbative one, so that the strongly-coupled states can affect all of the gauge couplings. Further, upon entering this regime the graviphoton becomes a mixed embedding into the electric and magnetic fields. This means that one of the matter sectors orthogonal to it, associated to the projection matrix ${\bf M}_s$, must also be embedded into a mixture of electric and magnetic fields. 

When the monopole state becomes lighter than the string scale, it needs to be reinterpreted as a threshold contribution to the gauge kinetic function. We proposed to identify two of the leading terms in the gauge kinetic function as arising from threshold corrections associated to the monopole sector (more precisely to the whole light electric and magnetic sector). The two contributions corresponds to the two components of the magnetic state along the two matter sectors defined by the projection matrices ${\bf M}_t$ and ${\bf M}_s$. The former contribution cancelled precisely the electric state threshold, implying a conformal nature to the sector. The latter contribution matched a threshold correction from a state of charge ${\bf q}^{(b)}$.

The second regime is a pure type II$_0$ regime, where the threshold term associated to the ${\bf q}^{(b)}$ state becomes the dominant contribution to the gauge kinetic matrix. In that regime, we proposed that there is a weakly-coupled infrared description of the physics, like in Seiberg-Witten theory, in terms of the state of charge ${\bf q}^{(b)}$. More precisely, the associated matter sector requires a complex rotation so that its kinetic terms become positive. This rotation simultaneously rotates the charge ${\bf q}^{(b)}$ into a real integral one.

As explained in the introduction, and in more detail in \cite{Hattab:2025aok}, one implication of our results is evidence that infinite distance type II$_0$ loci are emergent in the infrared from threshold corrections. If this is indeed correct, as explained in \cite{Hattab:2025aok}, it is a counter example to the standard formulation of the Distance Conjecture \cite{Ooguri:2006in} and the Emergent String Conjecture \cite{Lee:2019wij}. This could have implications for the Swampland program, and potentially for phenomenology, such as large-field inflation models. It is important to keep in mind that that, in the absence of a full microscopic understanding, we cannot yet completely prove this picture. 

More generally, to the best of our knowledge, the type of theories that we have uncovered around type II$_0$ loci are qualitatively new. They bear similarities to Argyres-Douglas theories \cite{Argyres:1995jj}, but are much more exotic. They are a type of Seiberg-Witten like theory where the infrared theory is embedded into both electric and magnetic gauge fields in such a way that it cannot be made electric by an integral electromagnetic transformation. Remarkably, the theory does appear to have an electric weakly-coupled description in the infrared, but under a complex rotation of the gauge fields. This leads to the property that the weak-coupling limit is at infinite distance in moduli space, rather than the finite distance attributed to all known weakly-coupled infrared descriptions of strongly-coupled field theories. 

We did not develop the microscopic physics associated to type II$_0$ loci. From the Heterotic perspective, they correspond to a non-perturbative phase where a Kaluza-Klein monopole becomes lighter than the fundamental string. There are many similarities to the discussion of the microscopic physics of the K-point in \cite{Hattab:2025aok}. In particular, the fact that the graviphoton is embedded into both the electric and magnetic fields should map to a Wick rotation in the microscopic description. Indeed, already in \cite{Hattab:2025aok} it was proposed that the microscopic physics should be some sort of Wick rotation of a torus direction in the context of the Heterotic string. We do not have much new to add beyond these ideas. A microscopic candidate description for the type II$_0$ locus was proposed in \cite{Monnee:2025msf}. There it was proposed that the transition to the type II$_0$ locus from the type II$_1$ locus involves nucleation of an NS5-brane strongly-coupled sector. We discussed possible relations to our results.

\vspace{10pt}
{\bf Acknowledgements}
\noindent
We would like to thank Ofer Aharony, Timo Weigand and Max Wiesner for very useful discussions. This work was supported by the German Research Foundation through a German-Israeli Project Cooperation (DIP) grant ``Holography and the Swampland" and by the Israel planning and budgeting committee grant for supporting theoretical high energy physics.

\appendix

\section{The period vector around Large Complex Structure (LCS)}
\label{app:lcs}
\subsection{The Picard--Fuchs system and a local period basis}
\label{app:pf}

Near the large complex structure point $(z_1,z_2)=(0,0)$, the periods are power-series multiplied by $\log$ factors that solve the Picard--Fuchs system $\mathcal{L}_i \Pi = 0$ with
\begin{align}
\mathcal{L}_1
&=\theta_1^{2}\,(\theta_1-2\theta_2)
-8 z_1\,(6\theta_1+5)(6\theta_1+3)(6\theta_1+1)\,,
\\[4pt]
\mathcal{L}_2
&=\theta_2^{2}-z_2\,(2\theta_2-\theta_1+1)(2\theta_2-\theta_1)\,,
\qquad
\theta_i \equiv z_i\partial_{z_i}\,.
\end{align}
A convenient symplectic period vector is
\be 
\Pi_{\mathrm{LCS}}^A=\bigl(X^0,X^1,X^2,F_0,F_1,F_2\bigr)^T,
\qquad
F_\Lambda=\frac{\partial F^A}{\partial X^\Lambda}\,,
\ee
where $F^A$ is the prepotential. 
We take the following ansatz for the solutions
\bea
X^0 &=& 1 + f_0 \nn \;,\\
\frac{X^1}{X^0} &=& \frac{1}{2\pi i}  \Big( \log z_1 + f_1 \Big)\nn \;,\\
\frac{X^2}{X^0} &=& \frac{1}{2\pi i}  \Big( \log z_2 + f_2 \Big) \nn \;,\\
\frac{F_0}{X^0} &=& \frac{1}{\left(2\pi i\right)^3} \Big( \frac23 \left( \log z_1\right)^3 + \left( \log z_1\right)^2\log z_2 + 2f_1 \log z_1 \left(\log z_1 + \log z_2 \right) + f_2\left( \log z_1\right)^2\nn \\
&+& 2\left(f_3+f_4 \right) \log z_1  + f_3 \log z_2+ f_5 \Big) +\frac{13  }{12 \pi i} \left( \log z_1 + f_1 \right) \nn \\
&+&\frac{1}{2\pi i} \left( \log z_2 + f_2 \right) - \frac{63 i \zeta(3)}{2\pi^3}  \nn\;,\\
\frac{F_1}{X^0} &=& \frac{13}{6} - \frac{2}{\left(2\pi i\right)^2} \Big( \left(\log z_1\right)^2  + \log z_1 \log z_2 + \left(2 f_1+f_2\right) \log z_1  + f_1 \log z_2  + f_4\Big) \nn \;,\\
\frac{F_2}{X^0} &=& 1-\frac{1}{\left(2\pi i\right)^2} \Big( \left(\log z_1\right)^2  + 2 f_1 \log z_1 + f_3\Big)   \;,
\label{perveclcs}
\eea
where each $f_k(z_1,z_2)$ is a holomorphic power series in $(z_1,z_2)$ with vanishing constant term. This leads directly to an integral symplectic basis where the prepotential takes the form
\be 
\frac{1}{(X^0)^2}\;F^A  = -\frac23 \left(t_1\right)^3- \left(t_1\right)^2 t_2 + \frac{13}{6} t_1 + t_2 - \frac{252i\zeta(3)}{2\left(2\pi\right)^3} + {\cal O}\left(e^{-t_i}\right)\;,
\label{lcsprepo}
\ee 
and we define the mirror complexified Kahler moduli coordinates as
\be 
t_i = \frac{X^i}{X^0} \;.
\ee 
The prepotential (\ref{lcsprepo}) matches the correct mirror form \cite{Candelas:1993dm}.

We write down the first few terms in the $f_k$ expansions for completeness
\begin{align}
    f_0 &= 1+120 z_1+83160z_1^2+166320 z_1^2 z_2+\cdots \;,\\
    f_1 &=744 z_1-z_2+473652 z_1^2+240 z_1z_2-\frac{3}{2}z_2^2+1429704 z_1^2 z_2+240 z_1z_2^2+220680 z_1^2 z_2^2+\cdots \nn\;,\\
    f_2 &= 240 z_1 +2 z_2 + 220680 z_1^2  - 480 z_1 z_2+ 3 z_2^2 - 441360 z_1^2 z_2  - 
 480 z_1 z_2^2 - 441360 z_1^2 z_2^2+\cdots\nn\;,\\
    f_3 &=  2 z_2+553536 z_1^2 + 1488 z_1 z_2  + \frac{11}{2} z_2^2+ 4306104 z_1^2 z_2 + 24 z_1 z_2^2 - 1397412 z_1z_2^2+\cdots\nn\;,\\
    f_4 &= 1248 z_1 + 1884672 z_1^2 - 240 z_1 z_2 - z_2^2 + 1929096 z_1^2 z_2 + 
 120 z_1 z_2^2 + 110340 z_1^2 z_2^2+\cdots\nn\;,\\
    f_5 &= -4992 z_1 - 4 z_2 - 2305152 z_1^2 - 2976 z_1 z_2-\frac92 z_2^2 - 7287984 z_1^2 z_2  + 840 z_1 z_2^2 + 1376892 z_1^2 z_2^2+\dots\nn \;.
\end{align}

\subsection{The gauge kinetic matrix}

The gauge kinetic matrix $\mathcal{N}^A_{IJ}$ can be obtained by evaluating \eqref{genNexo}. We decompose 
\begin{equation}
t_i=x_i+i\; y_i\;\,,\qquad y_i>0\,,
\end{equation}
and present here only the polynomial parts in the $y_i$ , setting $x_i=0$ for clarity. The real part reads
\begin{align}
\Re\,\mathcal{N}^A_{01}
&=\frac{13}{6}-2 y_1^2-2 y_1 y_2
+\frac{8 y_1 (y_1+y_2)\big(4 \pi ^3 y_1^2 (2 y_1+3 y_2)-189 \zeta (3)\big)}
{16 \pi ^3 y_1^2 (2 y_1+3 y_2)-189 \zeta (3)}\;,
\nonumber\\
\Re\,\mathcal{N}^A_{02}
&=1-y_1^2+
\frac{16 \pi ^3 y_1^4 (2 y_1+3 y_2)-756 y_1^2 \zeta (3)}
{16 \pi ^3 y_1^2 (2 y_1+3 y_2)-189 \zeta (3)}\;,
\\
\Re\,\mathcal{N}^A_{00} &=0\;,
\qquad
\Re\,\mathcal{N}^A_{11}=0\;,
\qquad
\Re\,\mathcal{N}^A_{12}=0\;,
\qquad
\Re\,\mathcal{N}^A_{22}=0\;.
\nonumber
\end{align}
The imaginary part reads
\begin{align}
\Im\,\mathcal{N}^A_{00}
&=-\frac{4}{3} y_1^{3}-2 y_1^2 y_2+\frac{63 \zeta (3)}{2 \pi ^3}
+\frac{\big(4 \pi ^3 y_1^2 (2 y_1+3 y_2)-189 \zeta (3)\big)^2}
{48 \pi ^6 y_1^2 (2 y_1+3 y_2)-567 \pi ^3 \zeta (3)}\;,
\nonumber\\
\Im\,\mathcal{N}^A_{11}
&= 4y_1+2y_2-\frac{192 \pi ^3 y_1^2 (y_1+y_2)^2}
{16 \pi ^3 y_1^2 (2 y_1+3 y_2)-189 \zeta (3)}\;,
\nonumber\\
\Im\,\mathcal{N}^A_{12}
&= 2y_1-\frac{96 \pi ^3 y_1^3 (y_1+y_2)}
{16 \pi ^3 y_1^2 (2 y_1+3 y_2)-189 \zeta (3)}\;,\\
\Im\,\mathcal{N}^A_{22}
&=
-\frac{16 y_1^4}{\frac{32 y_1^3}{3}+16 y_1^2 y_2-\frac{63 \zeta (3)}{\pi ^3}}\;,\nn\\
\Im\,\mathcal{N}^A_{01}&=0\;,
\qquad
\Im\,\mathcal{N}^A_{02}=0
\;.\nonumber
\end{align}

The gauge kinetic matrix does not admit a unique extension to the boundary direction $(y_1,y_2)\to(+\infty,+\infty)$; its asymptotic form depends on the path of approach (equivalently, on the relative scaling of $y_1$ and $y_2$), and this leads to distinct qualitative coupling behavior in different regions of moduli space.
To make this dependence transparent, we drop the constant $\zeta(3)$ contributions and keep only the leading polynomial terms in the large-$y_i$ expansion. In that approximation,
\begin{align}
    \text{Im}\,\mathcal{N}^A = y_1^3\left( \begin{array}{ccc} 
    -\frac{2}{3}-\frac{y_2}{y_1} & 0 & 0\\
    0 & 0 & 0\\
    0 & 0 & 0 \end{array}\right)+y_1\left( \begin{array}{ccc} 
    0 & 0 & 0\\
    0 & -\frac{4+8\frac{y_2}{y_1}+6\frac{y_2^2}{y_1^2}}{2+3\frac{y_2}{y_1}} & -\frac{2}{2+3\frac{y_2}{y_1}}\\
    0 & -\frac{2}{2+3\frac{y_2}{y_1}} & -\frac{3}{2+3\frac{y_2}{y_1}} \end{array}\right)\;.
\end{align}
One distinguishes three qualitatively different regimes in this form depending on whether the ratio $y_2/y_1$ goes to $0$, a constant, or $+\infty$. In the last case where $y_2/y_1\rightarrow+\infty$ there is a further important qualitative split controlled by the ratio $y_2/y_1^2$. If $y_1\ll y_2\ll y_1^2$, the smallest eigenvalue remains parametrically large and the theory remains weakly coupled. If instead $y_2 = k\,y_1^2$ for some constant $k$ or $y_2\gg y_1^2$, the eigenvalues $\lambda_i$ of $\Im\,\mathcal{N}$ behave as
\begin{align}
\lambda_1 &= -y_1^2y_2  + ...\to -\infty,\nonumber\\
\lambda_2 &= -2y_2  + ...\to -\infty,\nonumber\\
\lambda_3 &= -\frac{y_1^2}{y_2} + ... \ \to\
\begin{cases}
-\dfrac{1}{k}\;, & y_2 =  k\,y_1^2\;,\\
0\;, & y_2\gg y_1^2\;,
\end{cases}
\end{align}
signaling strong coupling in this frame.\\
The symplectic transformation in \eqref{Hetmap} maps this frame to the Heterotic frame in which the limit $y_2\gg y_1^2$ admits a weakly coupled description. Denoting the transformed gauge kinetic matrix by $\mathcal{N}^H$ one finds at leading order
\begin{align}
\label{ggmHLCS}
    \mathcal{N}^H = -i\left( \begin{array}{ccc} 
    y_1^2y_2+\frac{2}{3}y_1^2+\frac{y_2}{y_1^2}+\frac{2}{3y_1} & \frac{3i}{2} & -\frac{y_2}{y_1^2}-\frac{2}{3y_1}\\
    \frac{3i}{2} & y_2+\frac{4}{3}y_1 & \frac{2i}{3}\\
    -\frac{y_2}{y_1^2}-\frac{2}{3y_1} & \frac{2i}{3} & \frac{y_2}{y_1^2}+\frac{2}{3y_1} \end{array}\right)\;.
\end{align}
Its eigenvalues satisfy in the limit $y_2 \;y_1^{-2} \rightarrow \infty$ the relation
\begin{equation}
\lambda_1^{H} \rightarrow \frac{\lambda_1}{2}\;,\qquad
\lambda_2^{H} \rightarrow  \lambda_2\;,\qquad
\lambda_3^{H} \rightarrow  \frac{2}{\lambda_3}\;,
\end{equation}
so the previously small/degenerating eigenvalue is exchanged for a parametrically large (weakly coupled) one in the Heterotic frame.

\section{The period vector around the Type II point}
\label{app:ii}

In this appendix we calculate the period vector locally around the type II locus. 

\subsection{Local solutions}

One can solve for six independent periods
\[
\Pi=(\Pi_1,\Pi_2,\Pi_3,\Pi_4,\Pi_5,\Pi_6)^T \;,
\]
as local series about the type II point, using coordinates $(s,t)$ as in \eqref{lociiw}. A convenient set of solutions is
\begin{align}
\pi\Pi_1
&=
1+\frac{5}{36}s^2+\frac{295}{7776}s^4\left(2+t	\right)+\dots,
\nonumber\\
\pi\Pi_2
&=
s^2+\frac{77}{216}(2+t)s^4+\cdots,
\label{eq:iiw_pis}
\\
\pi\Pi_3
&=
-s\Bigg(
1+\frac{23}{54}s^2+\frac{2689}{9720}s^4
+\Big(-\frac{1}{16}+\frac{23}{288}s^2\Big)t
-\frac{15}{1024}t^2
+\cdots\Bigg),\nonumber\\
i\pi^2\Pi_4
&=
\pi\Pi_3\big(\log(t/2^6)+3\big)
-s\Bigg(
1+\frac{161}{162}s^2+\frac{126383}{145800}s^4
+\Big(\frac{1}{16}-\frac{115}{288}s^2\Big)t
-\frac{1}{512}t^2+\cdots\Bigg)\;.
\nonumber \\
-2i\pi^2\Pi_5
&=
\pi\Pi_1\log(s^4t)+5+\frac{25}{36}s^2+\frac{2597}{11664}s^4
+\cdots,
\nonumber\\
-2i\pi^2\Pi_6
&=
\pi\Pi_2\log(s^4t)+s^2+\frac{787}{324}s^4
+\cdots \;.
\nonumber
\end{align}
This basis is related to the integral LCS basis $\Pi^{A}_{\mathrm{LCS}}$ by a constant transformation
\begin{equation}
\Pi = T\cdot\Pi^{A}_{\mathrm{LCS}}\;,\qquad
T=
\begin{pmatrix}
0 & -i\beta & 0 & 0 & 0 & \frac{\beta}{2}\\
0 & \frac{i}{\beta} & 0 & 0 & 0 & \frac{1}{2\beta}\\
1 & 0 & 0 & 0 & 0 & -\frac12\\
0 & 0 & 0 & 1 & 0 & 0\\
0 & \zeta_1 & -\beta & \frac{\beta}{2} & -\frac{i\beta}{2} & i\zeta_3\\
0 & \zeta_2 & -\frac{1}{\beta} & \frac{1}{2\beta} & \frac{i}{2\beta} & i\zeta_4
\end{pmatrix}\;.
\label{eq:T_iiw_to_lcs}
\end{equation}
Here
\begin{equation}
\beta=\frac{\Gamma(3/4)^4}{\sqrt{3}\,\pi^2}\;\;,\;\;\;
\zeta_3  = -\frac{i\zeta_1}{2}+\frac{5i\beta}{2\pi }+\frac{i\beta\log(1728)}{2\pi}\;,\;\;\;
\zeta_4 =\frac{i\zeta_2}{2}+\frac{5i}{2\pi \beta}+\frac{i\log(1728)}{2\pi \beta}\;,
\end{equation}
and
\begin{equation}
\zeta_1 \approx 2.251234070,\quad \zeta_2 \approx -23.36100861\;.
\end{equation}
In particular, the period vector in the Heterotic frame $\Pi^H=S^{HA}\cdot T^{-1}\cdot\Pi$, takes a nice form with all logarithmic terms confined to the magnetic periods
\begin{align}
    2\pi i \beta  X^0 &= i - is\beta\left(2-\frac{t}{8}-\frac{15}{512}t^2\right)+is^2\left(\beta^2+\frac{5}{36}\right)+\cdots\;,\\
     2\pi i \beta X^1 &= -1+s^2\left(\beta^2-\frac{5}{36}\right)+\cdots\;,\\
     2\pi i \beta X^2 &= 2i+2is^2\left(\beta^2+\frac{5}{36}\right)+\cdots\;,\\
    2\pi i F_0 &= (2 X^0-X^2) \log \left(t/2^6\right)+g_0\;,\\
    2\pi i F_1 &=-2 X^1 \log \left(s^4 t\right)+g_1 \;, \\
    2\pi i F_2 &= -\frac{X^2}{2} \log(s^4 t)-\left(X^0-\frac{X^2}{2}\right) \log \left(t/2^6\right)+g_2\;,
\end{align}
with holomorphic pieces
\begin{align}
\label{g012}
    g_0 &= \frac{1}{\pi}\left(-8s+\frac{st}{4}+\frac{47 st^2}{512}+\dots\right)\;,\\
    g_1 &= -2i\zeta_1-\frac{5i}{\pi \beta}-is^2\left(2\zeta_2+\frac{5}{18}\zeta_1+\frac{25}{36\pi \beta}-\frac{\beta}\pi{}\right)+\cdots\;,\\
    g_2 &=  2\zeta_4-\frac{5}{2\pi \beta}+\frac{s}{\pi}\left(4-\frac{t}{8}-\frac{47 t^2}{1024}\right)+s^2\left(2 \zeta_3-\frac{25}{72\pi \beta}-\frac{\beta}{2\pi}+\frac{5\zeta_4}{18}\right)+\cdots\;.
\end{align}

\subsection{A symplectic frame with a vanishing electric period}

There is a second natural symplectic frame about the type II point, where one of the electric periods vanishes. 
Applying the block-diagonal integer symplectic transformation $R$ ,
\begin{equation}
R=
\begin{pmatrix}
-1&0&1&0&0&0\\
0&1&0&0&0&0\\
2&0&-1&0&0&0\\
0&0&0&1&0&2\\
0&0&0&0&1&0\\
0&0&0&1&0&1
\end{pmatrix},
\end{equation}
brings the period vector to a frame in which one electric period becomes parametrically small on the type II$_0$ locus. In this frame,
\begin{align}
2\pi i\,\beta\,X^0
&=
i+2i\beta s-\frac{i\beta}{8}st
+i s^2 \beta^2+\frac{5 i}{36}s^2
-\frac{15i\beta}{512}st^2+\cdots,
\nonumber\\
2\pi i\,\beta\,X^1
&=
-1+s^2\Big(\beta^2-\frac{5}{36}\Big)+\cdots\;,
\nonumber\\
2\pi i\,\beta\,X^2
&=
2 i \beta s\Big(-2+\frac{t}{8}+\frac{15}{512}t^2+\dots\Big)\;,
\\[4pt]
2\pi i\,F_0
&=-(2X^0+X^2)\log(s^4t)+g_0+2g_2\;,\nonumber\\
2\pi i\,F_1
&=-2X^1\log(s^4t)+g_1\;,\nonumber\\
2\pi i\,F_2
&=\frac{X^2}{2}\log(t/2^6)-\Big(X^0+\frac{X^2}{2}\Big)\log(s^4t)+g_0+g_2\;.\nonumber
\end{align}

\bibliographystyle{jhep}
\bibliography{Higuchi}

\end{document}